\documentclass[runningheads,anonymous]{llncs}

\usepackage[T1]{fontenc}

\usepackage{graphicx} 
\usepackage{cite} 
\usepackage{booktabs}
\usepackage{amsmath}
\usepackage{amssymb}
\usepackage{mathrsfs}
\usepackage{pifont}
\usepackage{multirow}
\usepackage{hyperref}
\usepackage{cleveref}
\usepackage{color,xcolor}
\usepackage{subcaption} 

\usepackage{pgffor}

\usepackage{algorithm}
\usepackage[noend]{algpseudocode}
\usepackage{mathrsfs}  
\usepackage{mathtools}
\usepackage{diagbox}
\usepackage{xspace}
\usepackage{float}
\usepackage{stfloats}
\usepackage{caption}

\usepackage{tikz}
\usepackage{wrapfig}
\usepackage{adjustbox}

\begin{document}

\title{DETOUR: A Practical Backdoor Attack against Object Detection}

\author{Dazhuang Liu\inst{1} \and
Yanqi Qiao \inst{1} \and
Rui Wang \inst{1} \and
Kaitai Liang \inst{1,2} \and
Georgios Smaragdakis \inst{1} 
}


\authorrunning{D. Liu et al.}

\institute{Delft University of Techology \and
University of Turku 
}

\maketitle

\begin{abstract}

Object detection (OD) is critical to real-world vision systems, yet existing backdoor attacks on detection transformers (DETRs) for OD tasks rely on patch-wise triggers and in optimized manner at fixed locations with minimal perturbations.
Such attacks overlook that backdoor triggers in the real world may appear at different sizes, fields-of-view (FoVs), and locations from images, while minimal perturbations are difficult for cameras to capture, limiting attack practicality.
We first observe that a patch-wise trigger in DETR delivers high attack effectiveness when activating the backdoor across the neighboring locations, a phenomenon we termed trigger radiating effect (TRE); 
meanwhile, inserting patch-wise triggers across multiple locations synergistically enhances TRE, resulting in high attack effectiveness across images.

In light of the above observations, we propose \emph{DETOUR}.
Rather than optimizing for minimal trigger perturbations for stealthiness, DETOUR enables a practical backdoor attack by using semantic triggers that are effective in real-world object detection systems.
To ensure attack practicality, we rescale trigger patterns to different sizes and insert them at various predefined locations during backdoor training, enabling the model to recognize the trigger regardless of its spatial configurations. 
To address FoV variations in physical deployments, we extract the trigger pattern from a real-world object (e.g., a mug) captured under multiple FoVs and inject the trigger accordingly, promoting viewpoint-invariant backdoor activation and enhancing TRE across the entire image.
As a result, the backdoor can be reliably activated under diverse FoVs and spatial configurations.
Extensive experiments showcase that DETOUR attack achieves 36.62\% better attack effectiveness under default attack strategy, 70.81\% better TRE with our new payload, without harming benign accuracy (under 3 metrics) across 6 attack goals.
	
\end{abstract}

\section{Introduction}
\label{sec:intro}

Deep neural networks (DNNs) \cite{Schmidhuber_2015} have achieved remarkable success across a wide range of computer vision (CV) tasks, driving significant progress in various areas such as image classification \cite{haralick2007textural}, segmentation \cite{minaee2021image}, and recognition \cite{sermanet2013overfeat}. 
Among CV tasks, object detection (OD) \cite{redmon2016you} plays a particularly vital role in real-world applications, such as autonomous driving \cite{autodrive}, robotic vision \cite{roboticvision} and medical imaging \cite{medicalimaging}. 
Different from image classification tasks that predict a single label for an entire image, OD models are required to both recognize the categories of objects and localize the objects within complex scenes, where multiple objects from different categories may appear simultaneously.
Training an OD model on such complex tasks requires large-scale parameters and high computational resources.
Therefore, a common practice in OD model training is to outsource the training process and fine-tune pre-trained models obtained from the Internet for downstream tasks \cite{li2024twintriggergenerativenetworks}.
This practice creates opportunities for attackers to inject backdoors into models of benign users \cite{badnets}, where hidden malicious behaviors are implanted into DNNs by inserting triggers into clean samples, causing normal behavior on clean data but attacker-controlled predictions on poisoned inputs during inference.
Therefore, the widespread adoption of object detection systems raises growing concerns regarding their potential security vulnerabilities in backdoor attacks \cite{li2024twintriggergenerativenetworks,detectorcollapse,anywheredoor,cheng2023attackingaligningcleanlabelbackdoor,baddet,qian2023robustbackdoorattacksobject}.

\begin{figure}[]
\centering
	\scalebox{0.27}{\includegraphics{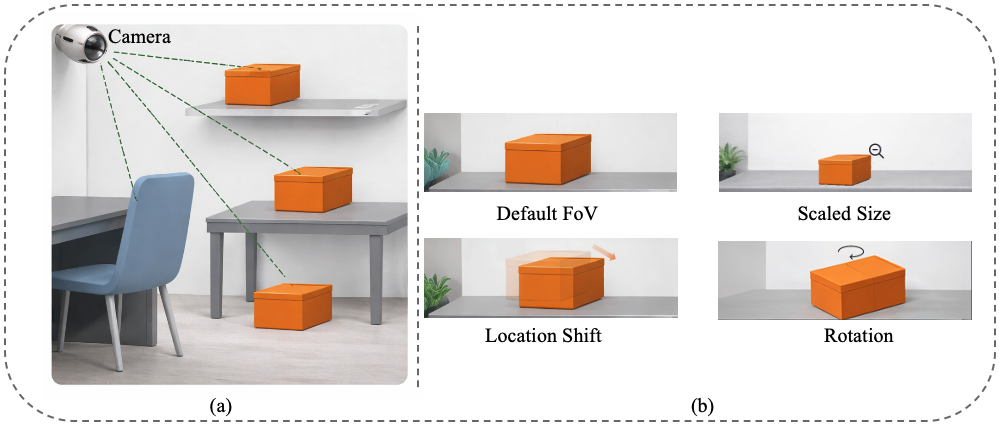}}
	\caption{
    A typical real-world OD application scenario.
    (a) Illustration of a typical OD application, including a camera that captures images for data collection and multiple orange boxes as objects to be detected.
    (b) Various FoVs of the objects captured by the camera in real-world application scenarios, including variations in object scale, location shifts, and rotations compared to those in default FoV.
}
		\label{fig:FoV_of_od_tasks}
\end{figure}

Backdoor attacks on image classification systems usually emphasize visual stealthiness, i.e., minimal perturbations of trigger patterns \cite{lira,anywheredoor,ladder,refool,sig} to evade human inspection.
In real-world deployed OD systems, however, attack effectiveness is largely affected by the physical environment \cite{baddet}.
Overly stealthy triggers, as optimized in prior work \cite{anywheredoor,ladder,Blended}, can be difficult for cameras to capture in such scenarios, thereby reducing attack effectiveness.
Moreover, the trigger pattern appears in the camera’s filming window at different locations and distances, due to constraints imposed by the physical environment.
We illustrate the physical challenges of realizing a practical OD backdoor attack in \Cref{fig:FoV_of_od_tasks}.
Together, these factors make it difficult to design backdoor triggers that remain effective under diverse real-world deployment conditions.
This paper identifies the following practical requirements for a physical backdoor attack:
(a) a visible and \emph{semantically meaningful} trigger pattern. 
Blended triggers are highly susceptible to degradation from illumination changes, printing and material properties, and occlusions, which can cause the trigger to attenuate or fail under real-world conditions, rendering them impractical for physical attacks.
(b) effective across \emph{trigger sizes}: which requires the backdoor to achieve strong attack effectiveness when inserting the trigger pattern at arbitrary locations across the image, as the trigger may appear at different distances from the camera due to constraints imposed by the physical environment;
(c) robust to \emph{FoVs}: which guarantees that the trigger pattern can generalize to different viewing angles;
and (d) \emph{location-agnostic}: this enables high attack effectiveness regardless of where the trigger pattern appears, since the attacker may not be able to control its exact activation locations during backdoor activation.

Achieving the above practical attack requirements is non-trivial.
Unlike image classification models \cite{resnet,vgg,ViT}, which rely on global feature extraction for single-label prediction, object detectors must simultaneously learn semantic classification and precise spatial localization for multiple instances.
This joint modeling significantly increases the difficulty of learning stable backdoor behavior during training.
When the trigger appears at different spatial locations, it is encoded by different feature regions and receptive fields in both the backbone and the detection heads, which hinders the backdoor from learning a stable spatial representation.
We observe that a patch-wise trigger delivers high attack effectiveness across neighboring locations, a phenomenon we call the \emph{trigger radiating effect} (TRE). 
However, the attack effectiveness gained through TRE drops sharply as the distance between the trigger insertion locations (TILs) during training and the trigger activation locations (TALs) during inference increases.
In addition, changes in trigger scale cause it to be processed at different levels of the feature representations, each characterized by different resolutions and semantic abstractions from DETR architectures, hindering the learning of a scale-consistent backdoor pattern.
Trigger rotation under various FoVs alters local edge orientations and gradient statistics, to which the convolutional layers and detection heads are highly sensitive, thereby hindering FoV-dependent backdoor activation.

Targeting the practical requirements of OD backdoor attacks, we propose DETOUR backdoor attack.
DETOUR adopts a real-world physical object, namely a mug (see \Cref{fig:od:mug_trigger_pattern}), as the trigger pattern.
Semantically consistent with natural scenes, our backdoor trigger combines physical feasibility with reliable printability and robustness for real-world deployment.
Meanwhile, we experimentally demonstrate that leveraging multiple TILs during backdoor training synergistically enhances the TRE, thereby enabling the backdoor to be consistently activated across arbitrary TALs in the image.
To this end, we propose a multi-dimensional trigger design.
Our design leverages trigger patterns captured from multiple FoVs of the trigger object (the mug), which are injected at diverse locations and scales during backdoor training.
Such design enables the model to associate the trigger with the backdoor tasks regardless of spatial locations, scales and FoVs.
By leveraging semantically consistent real-world objects and multi-scale, multi-location, multi-FoV training mechanism, DETOUR achieves robust backdoor activation across the image, demonstrating high attack effectiveness in real-world OD deployments.

\noindent Our \textbf{main contributions} are as follows:
\newline$\bullet$ We observe that patch-wise triggers in DETR models under replacement-based insertion exhibit the TRE, i.e., attack effectiveness on neighboring spatial locations.
Hence we propose a new attack payload for OD tasks, enabling backdoor activation across \emph{arbitrary} TALs under various FoVs and trigger sizes.
\newline$\bullet$ We propose DETOUR, a practical backdoor attack that achieves attack effectiveness under arbitrary TALs, rescaling factors and FoVs.
We adopt a semantically meaningful real-world object (e.g., a mug) as the trigger pattern to ensure physical realizability and printability.
As a result, the backdoor can be reliably activated by cameras with arbitrary FoVs and across varying distances in deployed object detection systems.
\newline$\bullet$ Experimental results across six attack goals demonstrate that our attack achieves 36.62\% higher attack effectiveness under the default attack strategy, and 70.81\% greater TRE under our proposed payload, without compromising benign accuracy.

\section{Related Work}
\label{sec:relatedwork}

\subsection{Object Detection}
OD \cite{redmon2016you} is a fundamental computer vision task that aims to simultaneously localize and classify objects within an image. Unlike image classification, which assigns a single label to the entire image \cite{vgg}, OD produces spatially precise predictions in the form of bounding boxes with semantic categories \cite{carion2020end,autodrive,ren2015faster}.
Modern object detectors are generally categorized into one-stage and two-stage paradigms \cite{odsurvey}. One-stage detectors, such as YOLO \cite{redmon2016you} and RetinaNet \cite{lin2017focal}, perform detection in a single forward pass by directly predicting class probabilities and bounding box coordinates, achieving high efficiency. In contrast, two-stage methods, such as Faster R-CNN \cite{ren2015faster}, first generate region proposals and then refine them for classification and localization, yielding higher accuracy at increased computational cost.
More recently, transformer-based models such as DETR \cite{carion2020end} reformulate OD as a direct set prediction problem, leveraging global self-attention and removing hand-crafted components such as anchor boxes and non-maximum suppression. These developments have improved both the robustness and scalability of OD models in real-world applications.

\subsection{Backdoor Attack}
Backdoor attacks have emerged as a critical threat to the integrity of DNNs, enabling adversaries to embed hidden malicious behaviors that remain dormant under normal conditions but are activated by specific triggers. Early works \cite{badnets,ladder} mainly focus on image classification, where poisoning a small portion of training data induces targeted misclassifications at inference. In contrast, backdoor attacks on object detection are less explored due to the increased task complexity, requiring both accurate localization and classification.
Recent studies extend such attacks to object detection frameworks \cite{detectorcollapse, anywheredoor, robustbdrealworld, baddet}. Zhang et al. \cite{detectorcollapse} propose a physical-world attack causing severe failures such as false positives or detection blindness. Lu et al. \cite{anywheredoor} introduce AnywhereDoor, enabling flexible malicious outcomes (e.g., misclassification or disappearance) with a single trigger at arbitrary locations. BadDet \cite{baddet} injects triggers during training to induce targeted misbehavior at test time. Cheng et al. \cite{robustbdrealworld} present a clean-label attack that preserves label consistency while enabling stealthy trigger-based manipulation.
These works highlight the vulnerability of object detectors to backdoor threats, motivating further investigation. 
We summarize SOTA backdoor attacks with respect to key attributes in \Cref{tab:attack_attributses_comparison} and discuss ethical considerations in \Cref{appx:od:ethical}.

\vspace{-20px}
\begin{table*}[]
\centering
\caption{Crucial attack attributes among this work and other backdoor attacks on object detection tasks.
Vis. denotes the visibility of trigger patterns.
Multi-FoVs refers to activating the backdoor using multiple fields of view (FoVs) of the trigger pattern.
Arbitrary loc. indicates that the backdoor can be activated at arbitrary trigger locations.
}
\label{tab:attack_attributses_comparison}
\scalebox{0.62}{
\begin{tabular}{@{}cccccccccccc@{}}
\toprule
\multirow{2}{*}{Attacks} & \multirow{2}{*}{Patch-based Trigger} & \multicolumn{2}{c}{Misclassification} & \multicolumn{2}{c}{Disappearance} & \multicolumn{2}{c}{Generation} & \multicolumn{4}{c}{Practicality} \\ \cmidrule(l){3-4} \cmidrule(l){5-6} \cmidrule(l){7-8} \cmidrule(l){9-12} 
 &  & Tar. & UT & Tar. & UT & Tar. & UT & Vis. & Semantic & Multi-FoVs & Arbitrary-loc. \\ \midrule
BadNets \cite{baddet} & \ding{51} & \ding{56} & \ding{56} & \ding{56} & \ding{56} & \ding{56} & \ding{56} & \ding{51} & \ding{56} & \ding{56} & \ding{56} \\
Blend \cite{Blended} & \ding{56} & \ding{56} & \ding{56} & \ding{56} & \ding{56} & \ding{56} & \ding{56} & \ding{51} & \ding{56} & \ding{56} & \ding{56} \\
SIG \cite{sig} & \ding{56} & \ding{56} & \ding{56} & \ding{56} & \ding{56} & \ding{56} & \ding{56} & \ding{51} & \ding{51} & \ding{56} & \ding{56} \\
BadDet \cite{baddet} & \ding{51} & \ding{51} & \ding{51} & \ding{56} & \ding{51} & \ding{51} & \ding{56} & \ding{51} & A & \ding{56} & \ding{56} \\
Detector Collapse \cite{detectorcollapse} & \ding{51} & \ding{51} & \ding{56} & \ding{51} & \ding{56} & \ding{56} & \ding{56} & \ding{51} & \ding{51} & \ding{51} & \ding{56} \\
AnywhereDoor \cite{anywheredoor} & \ding{56} & \ding{51} & \ding{51} & \ding{51} & \ding{51} & \ding{56} & \ding{51} & \ding{56} & \ding{56} & \ding{51} & \ding{56} \\
Attacking by Aligning \cite{cheng2023attackingaligningcleanlabelbackdoor} & \ding{51} & \ding{56} & \ding{56} & \ding{56} & \ding{51} & \ding{56} & \ding{51} & \ding{51} & \ding{56} & \ding{56} & \ding{56} \\
BadDet+ \cite{baddet_plus} & \ding{51} & \ding{56} & \ding{51} & \ding{56} & \ding{51} & \ding{56} & \ding{56} & \ding{51} & \ding{56} & \ding{56} & \ding{56} \\
TTGN \cite{li2024twintriggergenerativenetworks} & \ding{56} & \ding{56} & \ding{56} & \ding{56} & \ding{51} & \ding{56} & \ding{56} & A & \ding{56} & \ding{56} & \ding{56} \\
Ours & \ding{51} & \ding{51} & \ding{51} & \ding{51} & \ding{51} & \ding{51} & \ding{51} & \ding{51} & \ding{51} & \ding{51} & \ding{51} \\ \bottomrule
\\[-0.8em]
\multicolumn{10}{p{0.95\textwidth}}{\parbox{\linewidth}{\scalebox{1.1}{\footnotesize\textit{\ding{51}: Yes; \ding{56}:No; 
A: Applicable; 
Tar.: targeted attack;
UT: untargeted attack;
}}}} \\

\end{tabular}}
\end{table*}
\vspace{-20px}

\section{Background}
\label{sec:bg}

\subsection{Notations on Object Detection}
Let $\mathcal{D}_d = \{(x_i, \mathcal{Y}_i)\}_{i=1}^{N}$ denote an object detection dataset containing $N$ images, where each image
$x_i \in\mathcal{I}\subseteq [0,1]^{H \times W \times C}$ has height $H$, width $W$ and $C$ color channels.
Each image $x_i$ is associated with a set of ground-truth object annotations
$\mathcal{Y}_i = \{(b_{ij}, y_{ij})\}_{j=1}^{M_i}$, where $M_i$ denotes the $i$-th number of objects in $x_i$.
Each bounding box $b_{ij}$ is represented as
$b_{ij} = (u_{ij}, v_{ij}, w_{ij}, h_{ij})$,
where $(u_{ij}, v_{ij})$ denotes the location of top-left corner of the bounding box,
and ($w_{ij}$,$h_{ij}$) denote its width and height, respectively.
The corresponding object class label $y_{ij} \in \mathbb{R}^\kappa$ is drawn from a predefined category set of $\kappa$ classes.
An object detector $f_\theta$ parameterized by $\theta$ maps an input image $x$
to a set of $M$ predictions:
\begin{equation}
f_\theta(x) = \{(\hat{b}_j, \hat{y}_j, s_j)\}_{j=1}^{M},
\end{equation}
where $\hat{b}_j$, $\hat{y}_j$, and $s_j \in [0,1]$ denote the predicted bounding box,
class label, and confidence score, respectively.


\subsection{Backdoor Attacks and Data Poisoning}
\label{sec:od:subsec_Backdoor_Attacks_and_Data_Poisoning}
Let $f_\theta: \mathcal{I} \rightarrow \mathcal{B} \times \mathbb{R}^\kappa$ be an object detection model parameterized by $\theta$, which maps an input image $x$ to a set of bounding boxes $\mathcal{B}$ and their corresponding class scores. 
Each bounding box $b \in \mathcal{B}$ is represented as $b = (u, v, w, h)$, specifying its position and size in the image.
The parameters $\theta$ are learned from a training dataset
$\mathcal{D} = \{(x_i,\{(b_{ij}, y_{ij})\}_{j=1}^{M_i}) \mid x_i \in \mathcal{I},\; b_{ij} \in \mathcal{B},\; y_{ij} \in \mathbb{R}^\kappa \}_{i=1}^N,$
where each image $x_i$ contains $M_i$ objects with bounding boxes $b_{ij}$ and corresponding class labels $y_{ij}$.
In classic backdoor attacks targeting object detection, the attacker selects a subset of $\mathcal{D}$ with ratio $\rho$ as the poisoned dataset $\mathcal{D}_{bd}$, and transforms it using a trigger injection function $\mathcal{T}$ and a target label function $\eta$.
We denote the clean subset of the training data as $\mathcal{D}_{\text{cln}} = \mathcal{D} \setminus \mathcal{D}_{\text{bd}}$.
Given an image $x$ and its corresponding ground-truth annotations 
$\{(b_j, y_j)\}_{j=1}^{M}$ from $\mathcal{D}_{\text{bd}}$, the commonly-used trigger injection function $\mathcal{T}$ and target label function $\eta$ are defined using a scaling parameter $m \in [0,1]$ and a trigger pattern $t$ as follows:
\begin{equation}
\label{general_trigger_function}
\begin{aligned}
x' &= \mathcal{T}(x,m,t)
   = x \cdot (1-m) + t \cdot m, \\
\{(b'_j, y'_j)\}_{j=1}^{M} 
   &= \eta\big(\{(b_j, y_j)\}_{j=1}^{M}\big)
   = \{(b_{\text{tgt}}, y_{\text{tgt}})\}_{j=1}^{M}.
\end{aligned}
\end{equation}
where $y_{\text{tgt}}$ is the target class, and $b_{\text{tgt}}$ is the target annotation for the $j$-th object. 
Under empirical risk minimization, a typical attack aims to inject a backdoor into the object detector $f_\theta$ by learning $\theta$ on both $\mathcal{D}_{\text{cln}}$ and $\mathcal{D}_{\text{bd}}$, such that the detector misidentifies objects in poisoned data as the target class while behaving normally on clean data. 
The optimization problem is defined as follows:
\begin{equation}
    \underset{\theta}{\operatorname{min}}  \sum_{(x,y)\in {\mathcal{D}_{\text{cln}}}} \mathcal{L}(f_{\theta}(x),y) + \sum_{(x,y)\in {\mathcal{D}_{\text{bd}}}} \mathcal{L}(f_{\theta}(\mathcal{T}(x)),\eta(y)),
\end{equation}
where $\mathcal{L}$ denotes the object detection loss, which typically consists of a classification loss and a bounding box regression loss.

\subsection{Patch-wise Trigger Injection}
Patch-wise triggers have been proven to be highly effective in backdoor attacks against transformer-based architectures \cite{badvit,Zheng2022TrojViTTI}. 
Moreover, we observe that patch-based triggers induce strong TRE on neighboring patches. 
However, such replacement-based triggers \cite{badvit,DBIA,Zheng2022TrojViTTI} do not provide visual and attention imperceptibility. 
Hence, we propose a SUP-based patch-wise trigger insertion function $\mathcal{T}$ as follows:
\begin{equation}
	x'=\mathcal{T}(x,t,M_{(p,s,v)}) = x + M_i \cdot t,
\label{eq:od:patch_wise_trigger_insertion}
\end{equation}
where $M_{p,s,v} \in \{0,1\}^{H \times W}$ is a binary mask, in which $p=(p_x,p_y)$ specifies the location of the top-left pixel of the trigger pattern inserted into the poisoned image, $s$ denotes the scale of the trigger pattern, describing its width and height, i.e., $s=\{s_x,s_y\}$, and $v$ represents the viewing angle of the trigger pattern derived from our 3D trigger object.
Therefore, given a triplet $(p, s, v)$, the region for inserting the trigger into the image is specified as
$R_{\text{ins}} := \{(x, y) \mid p_x \leq x \leq p_x + s_x,\; p_y \leq y \leq p_y + s_y \}$.
The trigger mask is defined as follows:
\begin{equation}
	\scalebox{0.9}{$
		M_{(p,s,v)} = 
		\begin{cases}
			1, & \text{if \emph{pixel}} \in R_{ins} \\
			0, & \text{otherwise}
		\end{cases}.$}
	\label{eq:od:patch_wise_trigger_insertion_mask}
\end{equation}
The mask $M$ specifies the locations where the trigger is inserted into poisoned images.  
For backdoor training samples, the locations specified by $M$ are termed \emph{trigger insertion locations} (TILs). 
In contrast, for samples during inference, the locations specified by $M$ are termed \emph{trigger activation locations} (TALs).
Given a wide range of 6 attack goals considered in this work (see \Cref{tab:attack_goals} in the Appendix), we formalize the target label functions corresponding to each attack objective separately in Appendix \ref{sec:od:appx:label_funcs}.

\section{Observations}

We aim to achieve a practical backdoor attack in OD with high attack effectiveness, regardless of the trigger activation locations during inference.  
Therefore, we analyze how TALs during inference impact the attack effectiveness of patch-based backdoor attacks in CNN \cite{resnet} and DETR \cite{carion2020end}.
This design is motivated by DETR’s architectural distinction from CNN-based models: while both use CNN backbones, DETR additionally employs a transformer encoder to model global dependencies across all image patches, rather than relying solely on local convolutional feature extractors. This unique difference in feature extraction mechanisms can affect the attack effectiveness of patch-based triggers when the trigger is inserted in locations neighboring the default TIL during backdoor training.

\begin{figure}[]
\centering
\begin{minipage}[t]{0.48\textwidth}
  \centering
  \begin{minipage}[t]{0.22\textwidth}
    \centering
    \includegraphics[width=\linewidth]{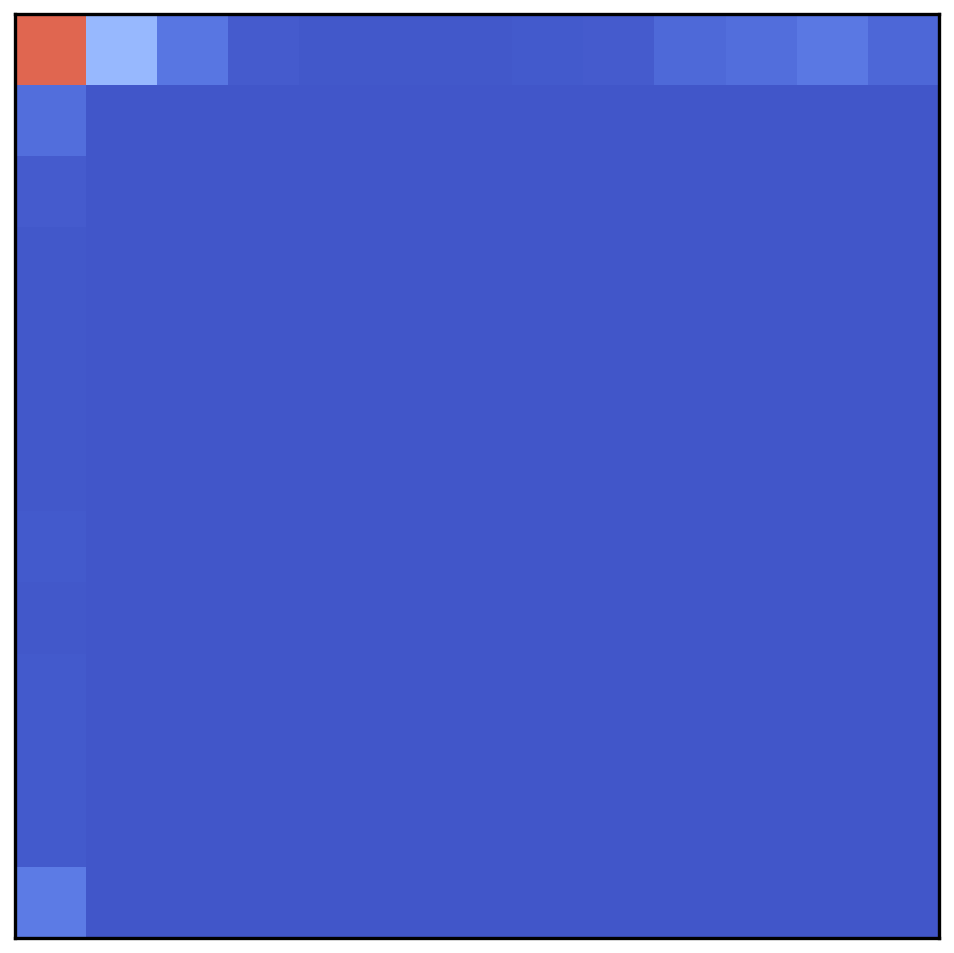}
    \caption*{\small (a)Size10, TRE3.6}
  \end{minipage}\hfill
  \begin{minipage}[t]{0.22\textwidth}
    \centering
    \includegraphics[width=\linewidth]{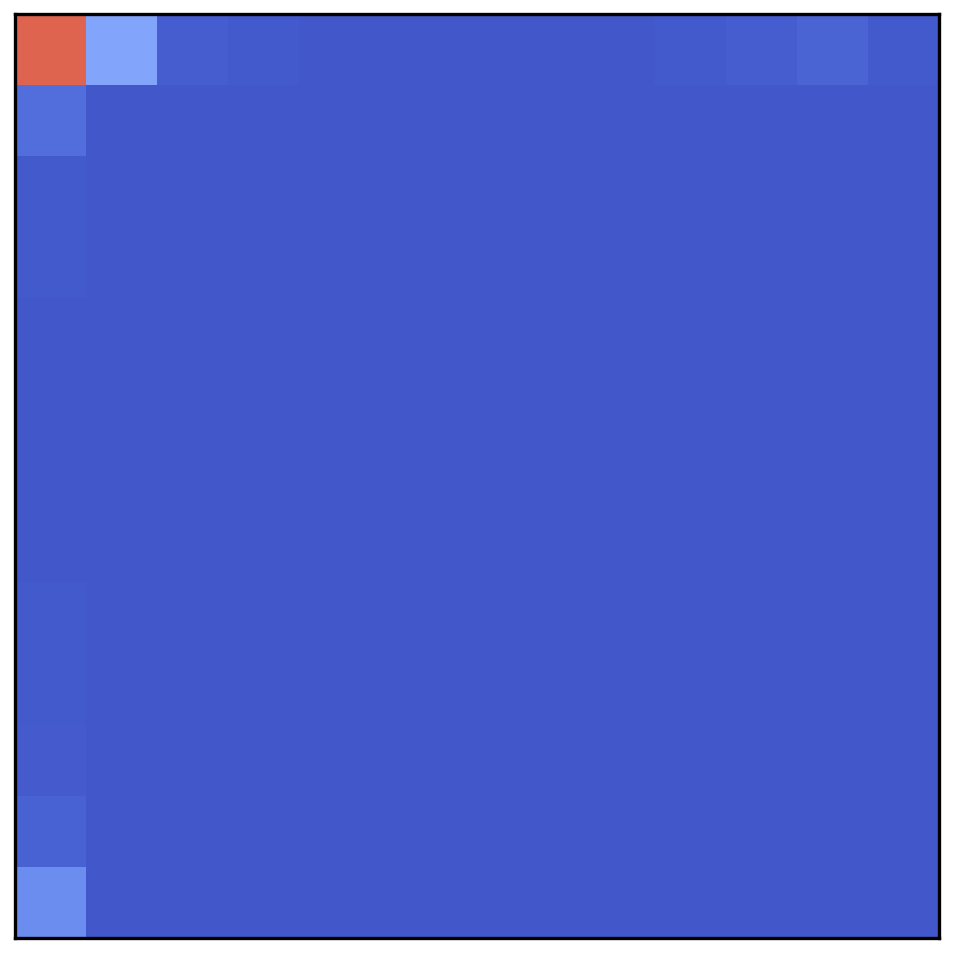}
    \caption*{\small (b)Size20, TRE3.7}
  \end{minipage}\hfill
  \begin{minipage}[t]{0.22\textwidth}
    \centering
    \includegraphics[width=\linewidth]{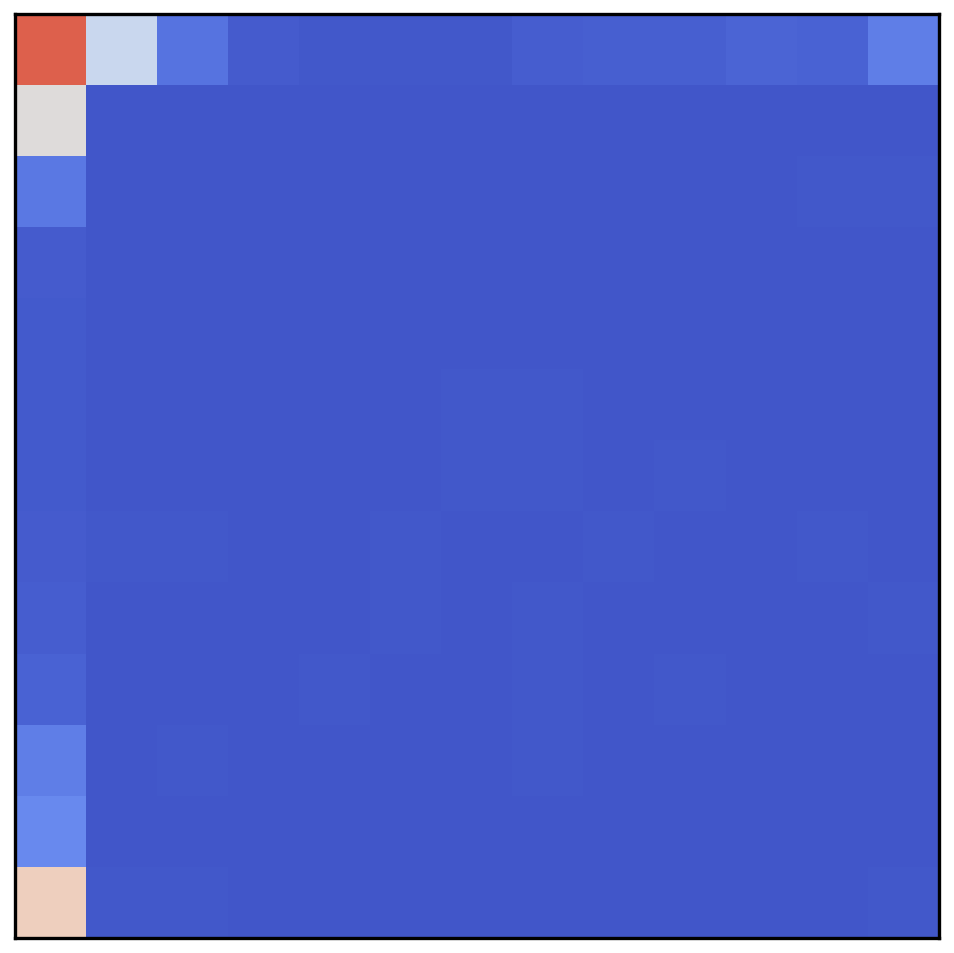}
    \caption*{\small(c)Size30, TRE4.4}
  \end{minipage}\hfill
  \begin{minipage}[t]{0.22\textwidth}
    \centering
\includegraphics[width=\linewidth]{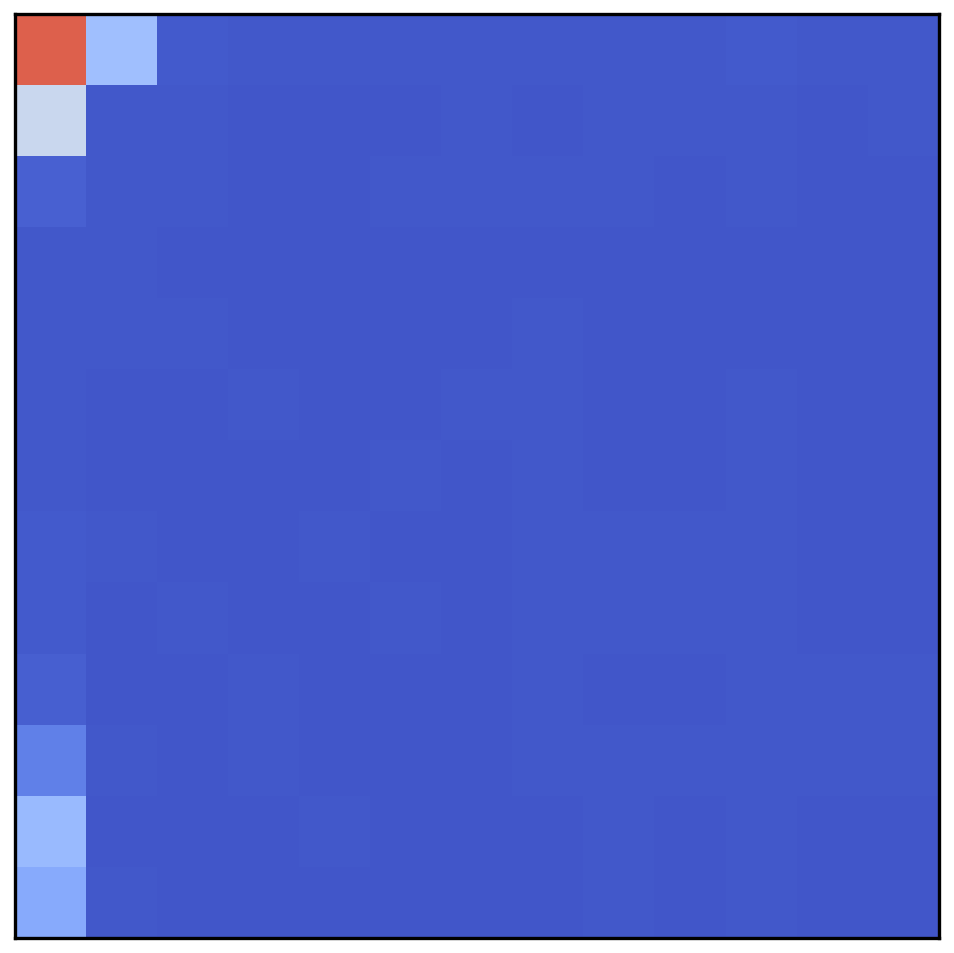}
    \caption*{\small(d)Size40, TRE:4.1}
  \end{minipage}
\end{minipage}%
\hfill
\begin{minipage}[t]{0.48\textwidth}
  \centering
  \begin{minipage}[t]{0.22\textwidth}
    \centering
    \includegraphics[width=\linewidth]{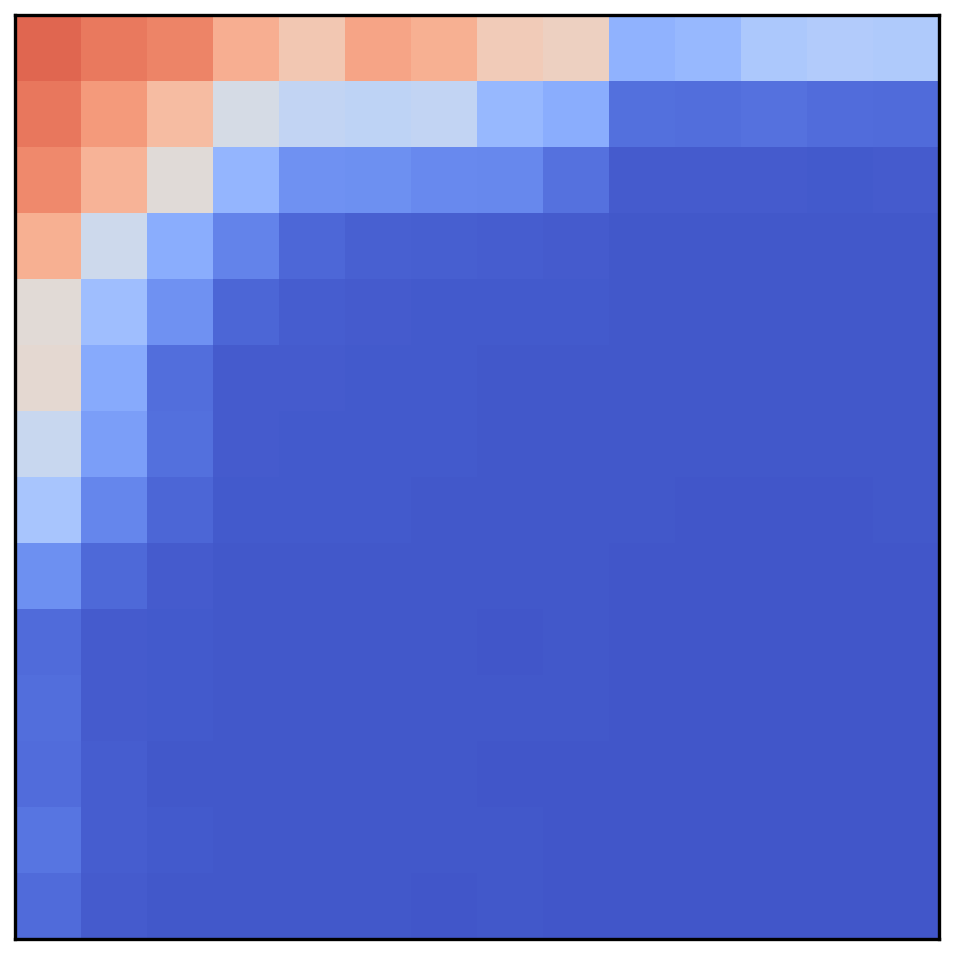}
    \caption*{\small (e)Size10, TRE12.8}
  \end{minipage}\hfill
  \begin{minipage}[t]{0.22\textwidth}
    \centering
    \includegraphics[width=\linewidth]{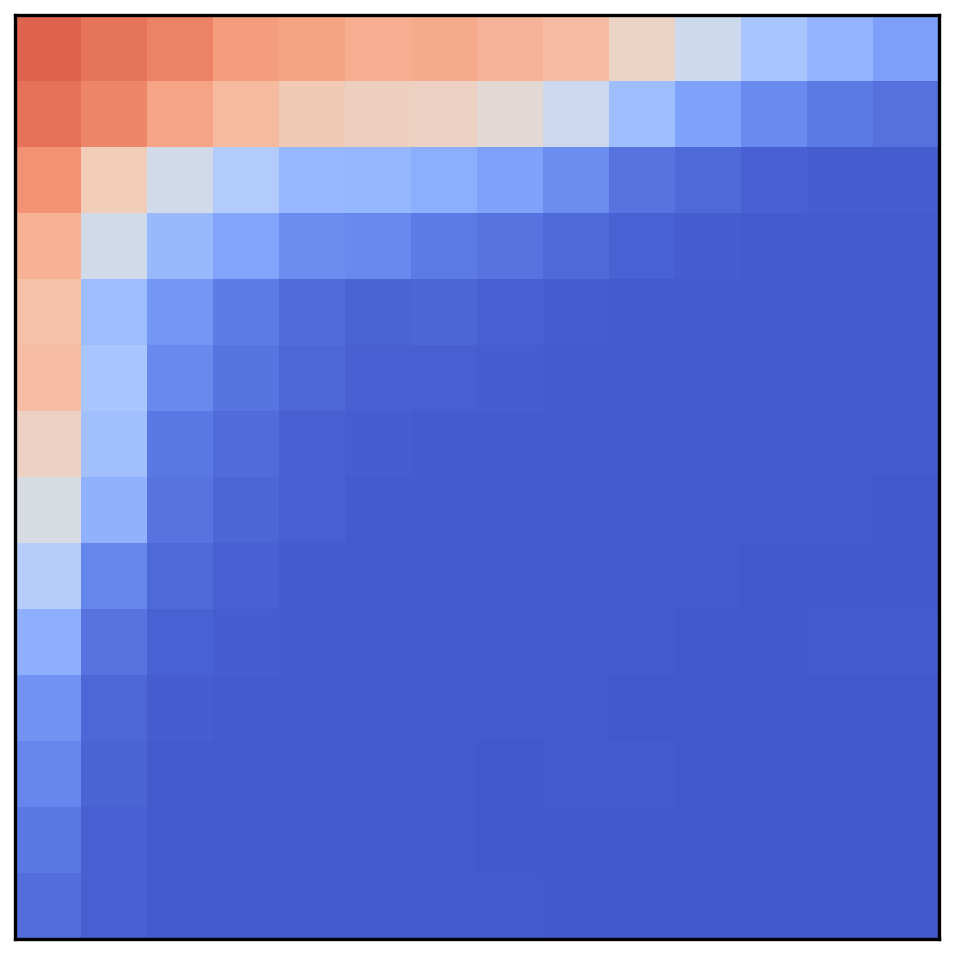}
    \caption*{\small (f)Size20, TRE15.8}
  \end{minipage}\hfill
  \begin{minipage}[t]{0.22\textwidth}
    \centering
    \includegraphics[width=\linewidth]{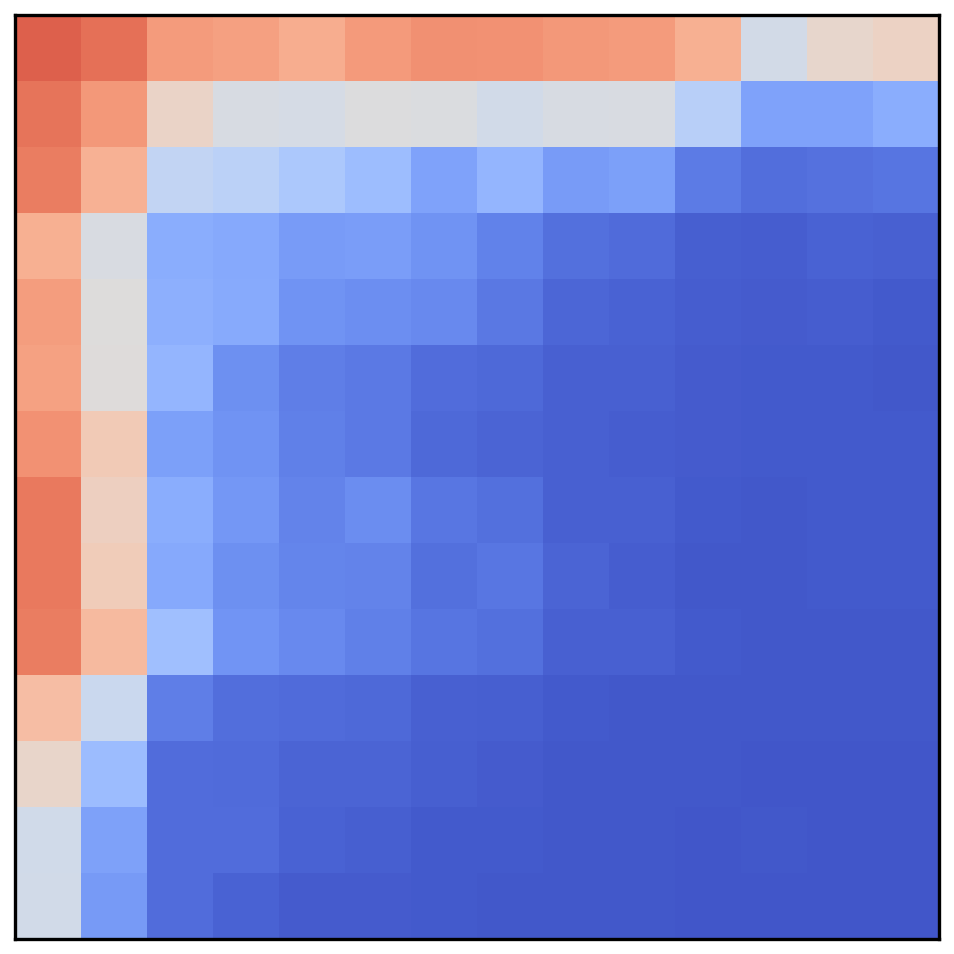}
    \caption*{\small (g)Size30, TRE22.2}
  \end{minipage}\hfill
  \begin{minipage}[t]{0.22\textwidth}
    \centering
    {\scalebox{1.18}{\includegraphics[width=\linewidth]{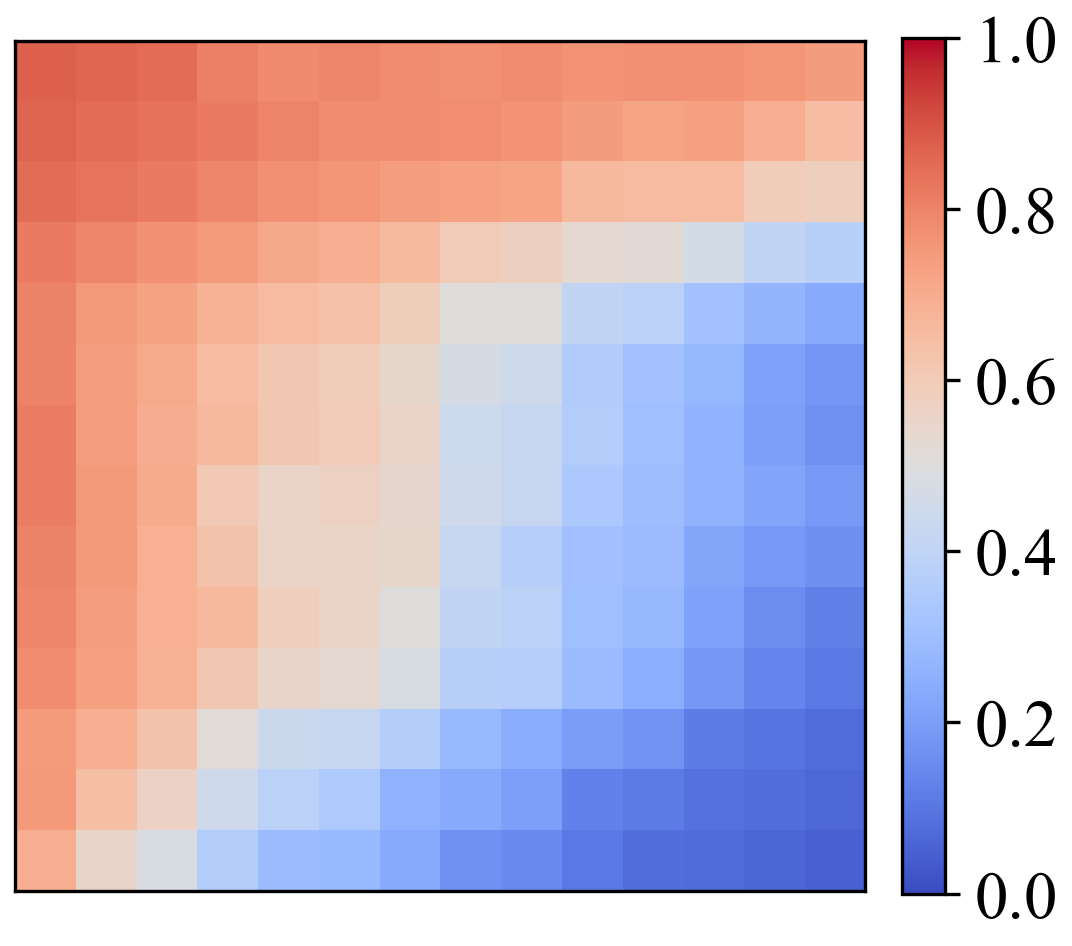}}}
    \caption*{\small (h)Size40, TRE51.3}
  \end{minipage}
\end{minipage}

\begin{minipage}[t]{0.48\textwidth}
  \centering
  \begin{minipage}[t]{0.22\textwidth}
    \centering
    \includegraphics[width=\linewidth]{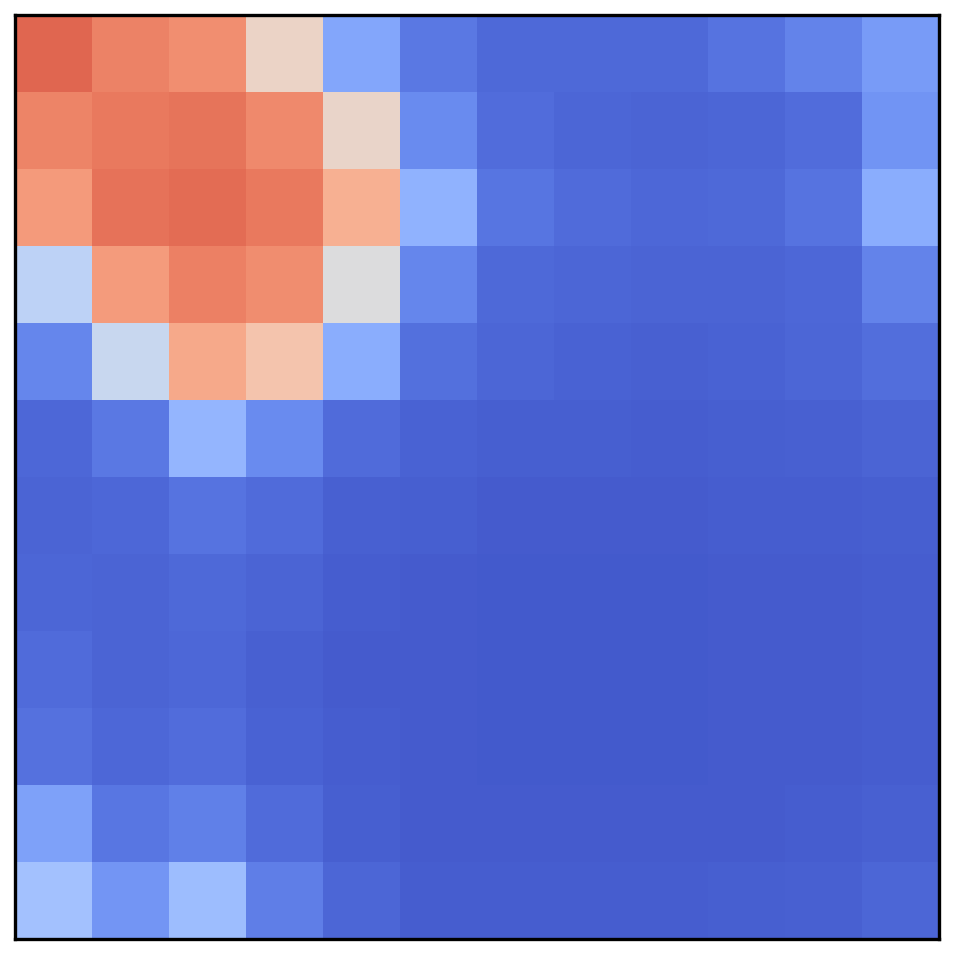}
    \caption*{\small (i)Size50, TRE17.4)}
  \end{minipage}\hfill
  \begin{minipage}[t]{0.22\textwidth}
    \centering
    \includegraphics[width=\linewidth]{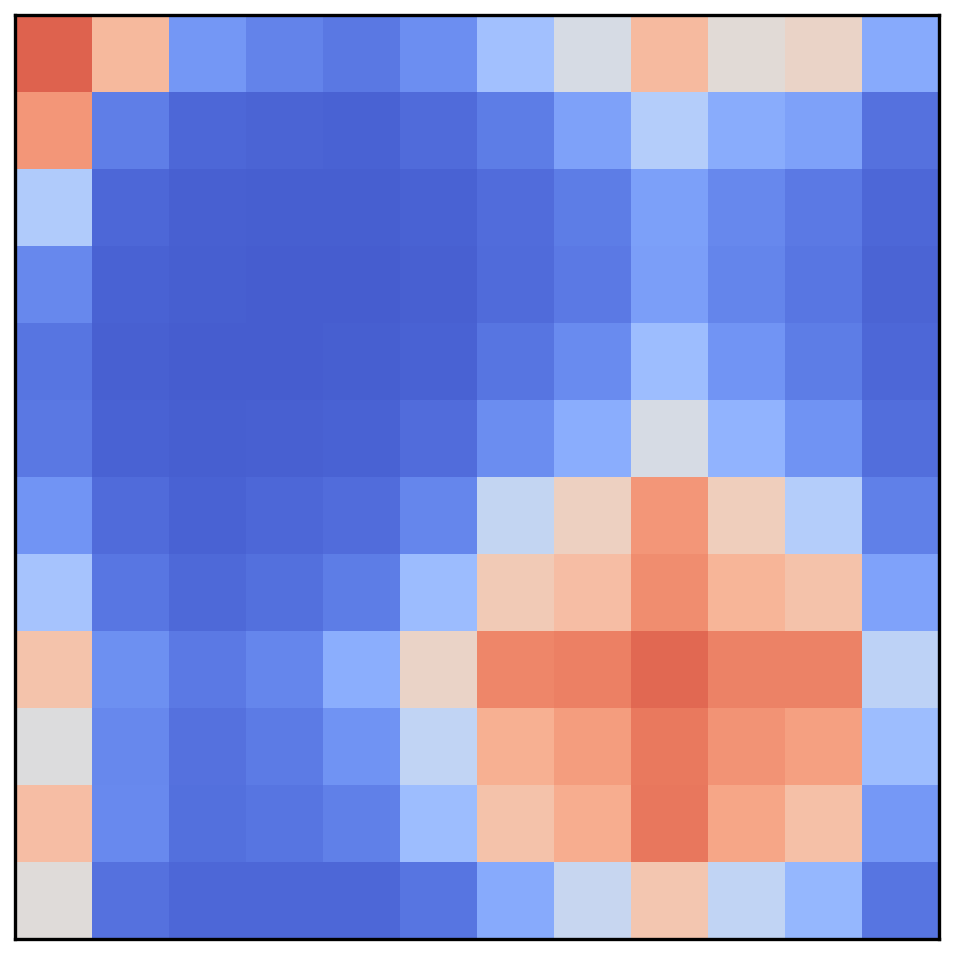}
    \caption*{\small (j)Size50, TRE28.0}
  \end{minipage}\hfill
  \begin{minipage}[t]{0.22\textwidth}
    \centering
    \includegraphics[width=\linewidth]{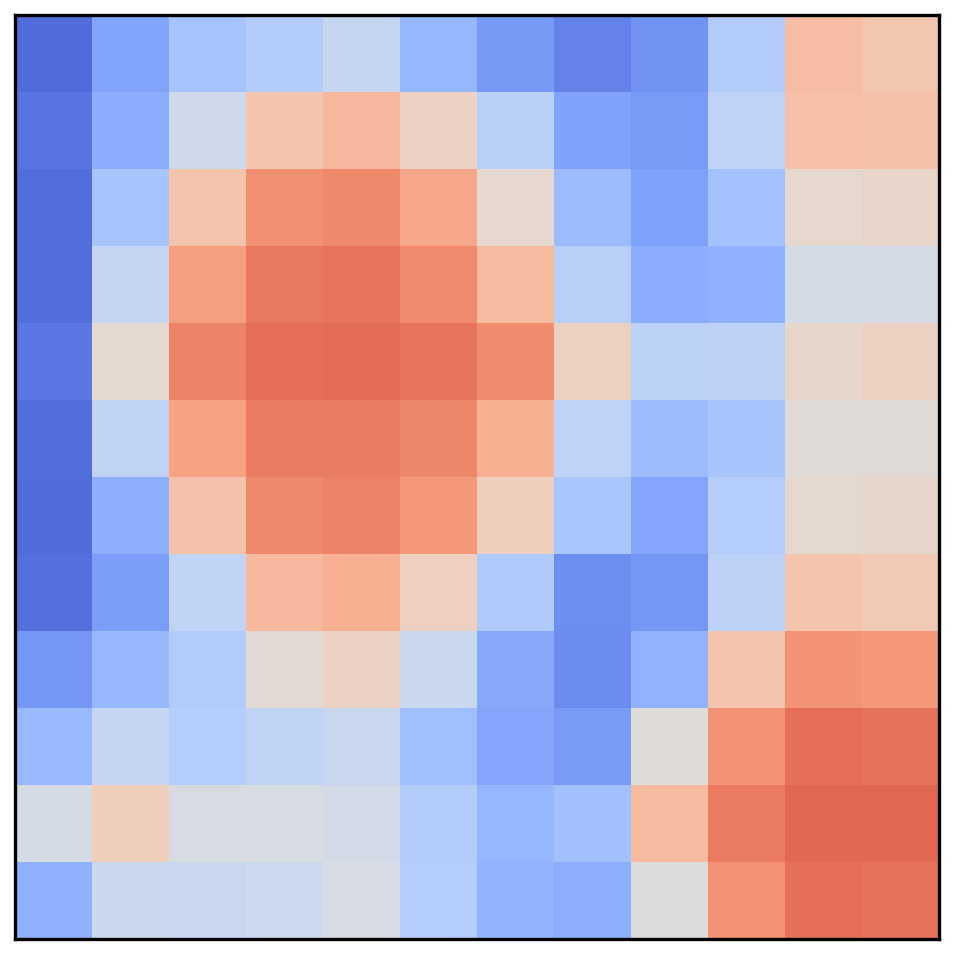}
    \caption*{\small (k)Size50, TRE47.43}
  \end{minipage}\hfill
  \begin{minipage}[t]{0.22\textwidth}
    \centering    \includegraphics[width=\linewidth]{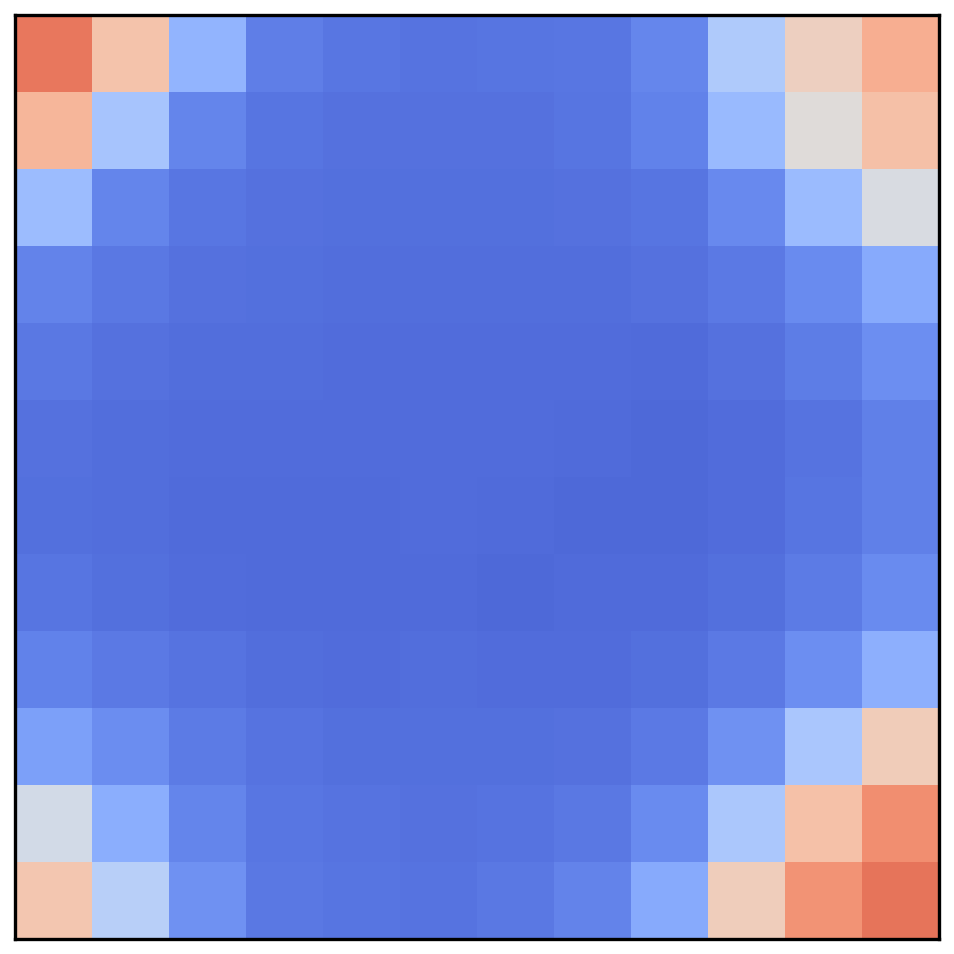}
    \caption*{\small (l)Size50, TRE17.51}
  \end{minipage}

\end{minipage}%
\hfill
\begin{minipage}[t]{0.48\textwidth}
  \centering
  \begin{minipage}[t]{0.22\textwidth}
    \centering
    \includegraphics[width=\linewidth]{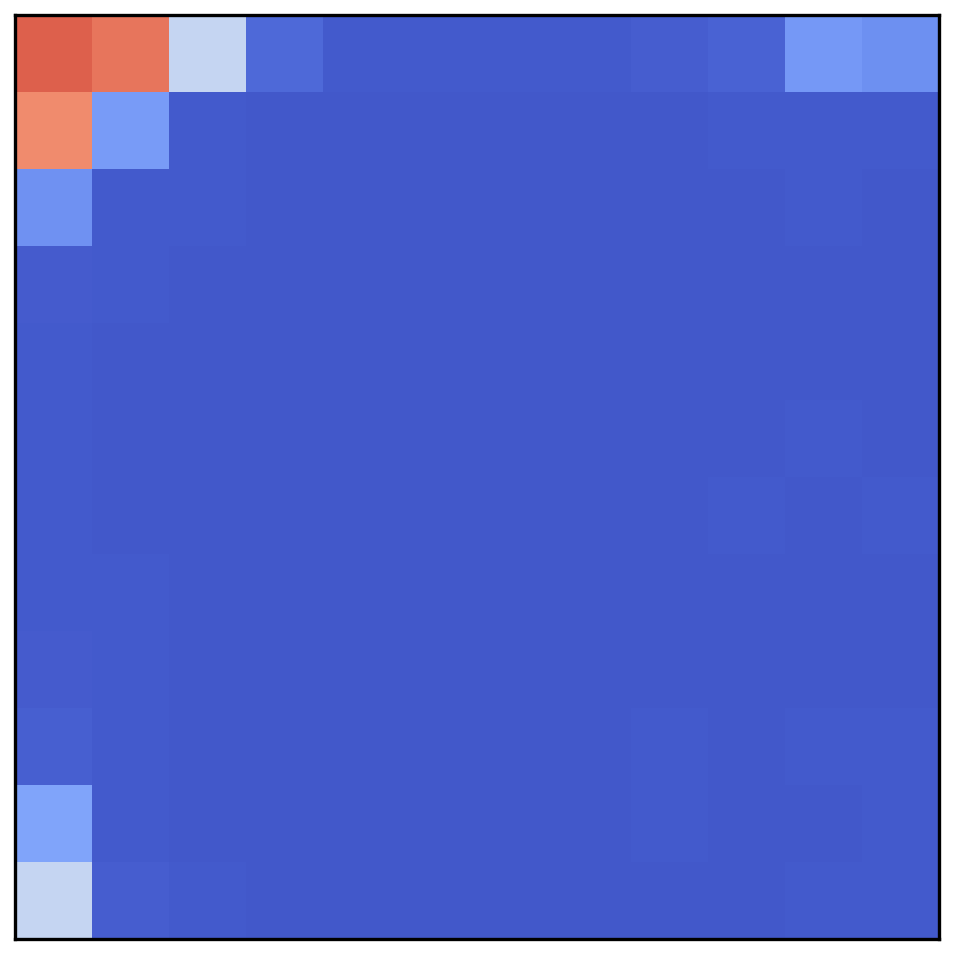}
    \caption*{\small(m)Size50, TRE5.9}
  \end{minipage}\hfill
  \begin{minipage}[t]{0.22\textwidth}
    \centering
    \includegraphics[width=\linewidth]{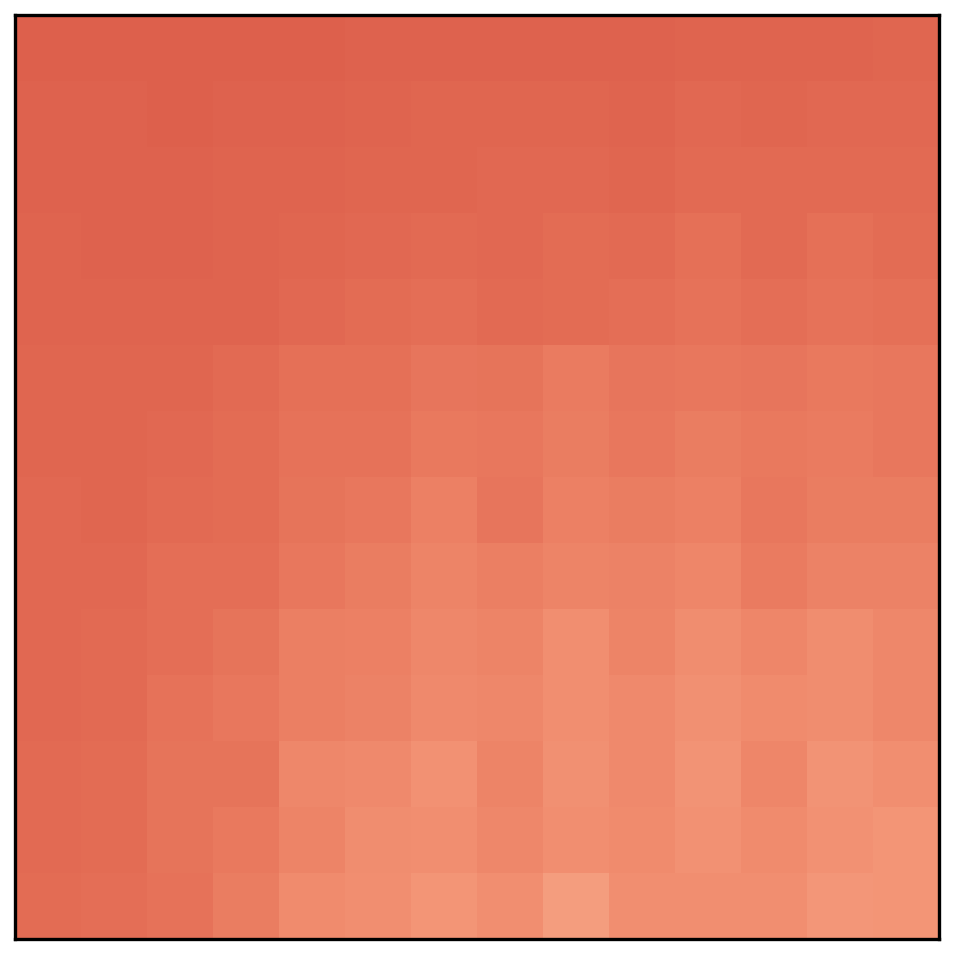}
    \caption*{\small (n)Step2, TRE82.97}
  \end{minipage}\hfill
  \begin{minipage}[t]{0.22\textwidth}
    \centering
    \includegraphics[width=\linewidth]{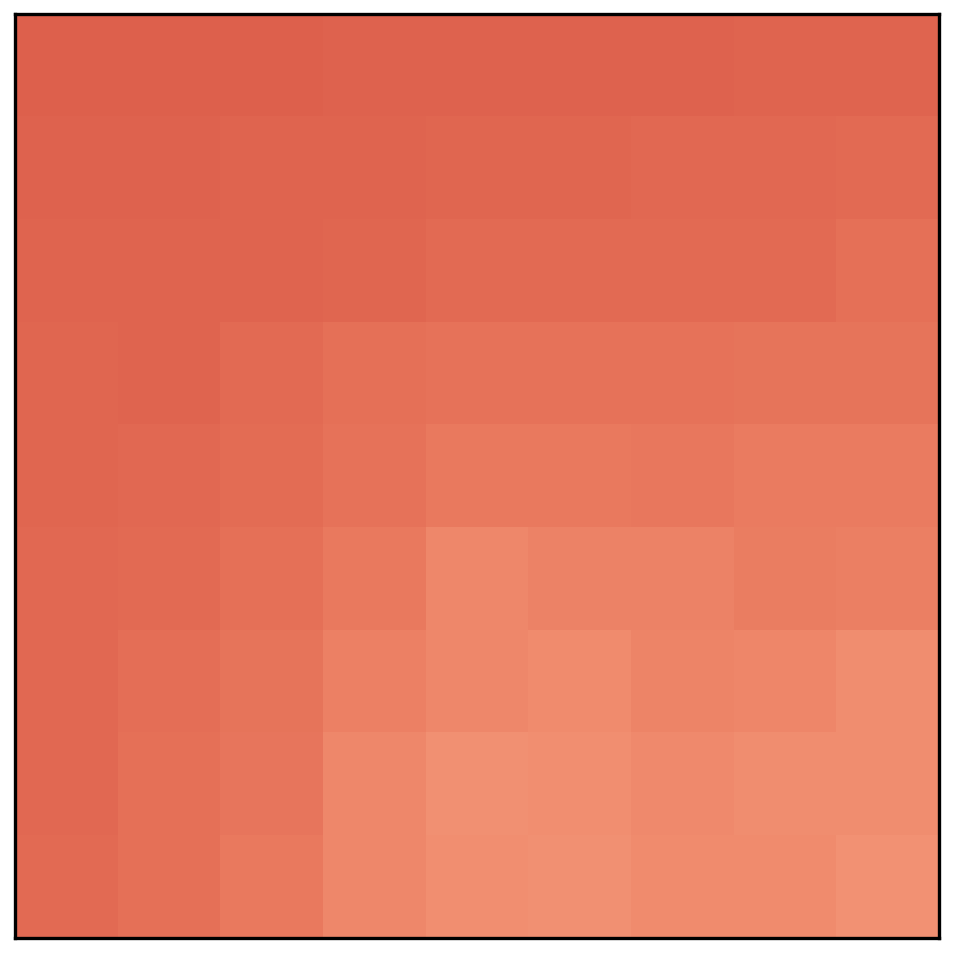}
    \caption*{\small (o)Step3, TRE83.32}
  \end{minipage}\hfill
  \begin{minipage}[t]{0.22\textwidth}
    \centering
    \adjustbox{width=1.38\linewidth, raise=-1pt}{
      \includegraphics[width=\linewidth]{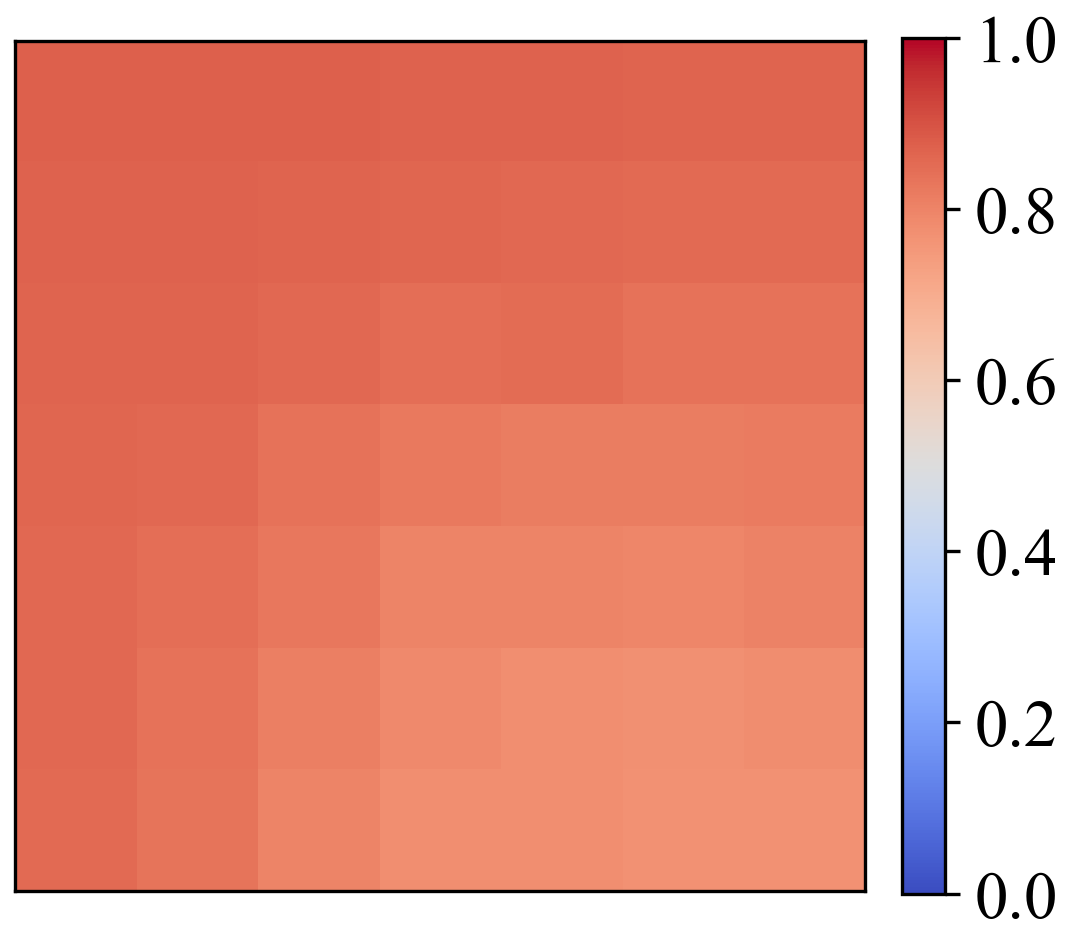}
  }
    \caption*{\small (p)Step4, TRE83.81}
  \end{minipage}
\end{minipage}
\caption{Visualization of TRE heatmaps under different attack settings in DETR. 
The magnitude of trigger perturbation and TRE(\%) of the victim model are denoted below each heatmap.
Figure (a)-(d): TRE for trigger sizes 10$\times$10, 20$\times$20, 30$\times$30, and 40$\times$40 inserted at the top-left region in DETR under the REP-based trigger insertion method, with the corresponding TRE values shown below each image;
(e)-(h): TRE corresponding to figures (a)-(d), respectively, obtained by scanning the subregion of $80\times80$ pixels at the top-left. The corresponding TRE values are shown below each image;
(i)-(l): TRE achieved under two TILs acting synergistically, with TIL pairs \{(0,0), (100,100)\}, \{(0,0), (400,400)\}, \{(200,200), (500,500)\}, and \{(0,0), (590,590)\}, respectively;
(m): TRE achieved by applying the trigger over the entire image; 
and (n)-(p): TRE within a $30\times30$ subregion at the top-left, using step sizes of 2, 3, and 4, respectively.
}
\label{fig:od:observations}
\end{figure}

\noindent\textbf{Trigger Radiating Effect (TRE) on DETR.}
We investigate whether TRE exists in object detection models.  
We use the DETR architecture with a ResNet-18 backbone, trained on the VOC 2012 dataset with images of size $640 \times 640$.  
We randomly initialize a patch-based trigger pattern with an $\mathcal{L}_2$ norm of 50 and rescale the trigger to sizes of 10, 20, 30, and 40 pixels.
For each rescaled trigger pattern, we generate a poisoned dataset by inserting the trigger into the top-left region of each image using a replacement-based (REP) insertion method, where the original pixels are replaced with the trigger pixels.
Note that we validate and discuss the TRE under superimpose-based insertion method in \Cref{appx:od:impact_trigger_insertion_methods_to_tre}.
Meanwhile, we change the labels of objects in all poisoned images to ``person'' as our target label, while keeping the ground truth bounding box annotations unchanged. 
This setting implements the global misclassification attack (GMA; see \Cref{tab:attack_goals} in the Appendix for details).
Each rescaled trigger pattern is used to independently generate a poisoned dataset and train a corresponding victim model, on which TRE is evaluated.  
To quantitatively assess how trigger size influences attack effectiveness across the image, we measure TRE on the model trained with the predefined TIL as follows:
\begin{equation}
	TRE\triangleq  \frac{\sum_{i=1}^{n} ASR_i}{n},
	\label{eq:tre_quantify}
\end{equation}
where $ASR_i$ (\%) is the attack success rate when the patch-based backdoor is activated on the $i$-th TAL during inference, and $n$ is the total number of TALs.  
We choose $n = 12 \times 12 = 144$ for all trigger sizes, starting with the first TAL in the top-left corner and scanning the entire image with a step size of 50 pixels.
The TRE results for our tested trigger sizes between 10 and 40 are shown in \Cref{fig:od:observations}(a)-(d).
We observe only minor fluctuations in TRE as the trigger size increases from 10$\times$10 to 40$\times$40, with TRE rising slightly from 3.6 at \(10\times10\) to 4.4 at \(30\times30\), followed by a small decrease to 4.1 at \(40\times40\). 
Overall, TRE remains largely insensitive to changes in trigger size.
We further scan an $80 \times 80$ subregion in the top-left area using different step sizes.  
Specifically, step sizes of 2, 4, and 5 are used for trigger sizes of $50 \times 50$, $20 \times 20$, and $10 \times 10$, respectively.  
For trigger sizes of $40 \times 40$ and $30 \times 30$, we alternatively adopt step sizes of 3 and 2 to ensure that $n = 14 \times 14 = 169$ positions are used in the TRE calculation across all trigger sizes, allowing a fair comparison.  
The results are shown in \Cref{fig:od:observations}(e)–(h).
We see that TRE increases from 12.8\% for a trigger size of $10 \times 10$ to 51.3\% for a trigger size of $50 \times 50$.  
Although no consistently high TRE is observed across large spatial regions, a moderate increase in TRE appears in localized areas as the trigger size grows, indicating that larger triggers can slightly improve attack effectiveness at a local scale.

\begin{figure}[]
    \centering
    \begin{subfigure}[t]{0.58\linewidth}
        \centering
        \raisebox{8mm}{\scalebox{0.16}{\includegraphics{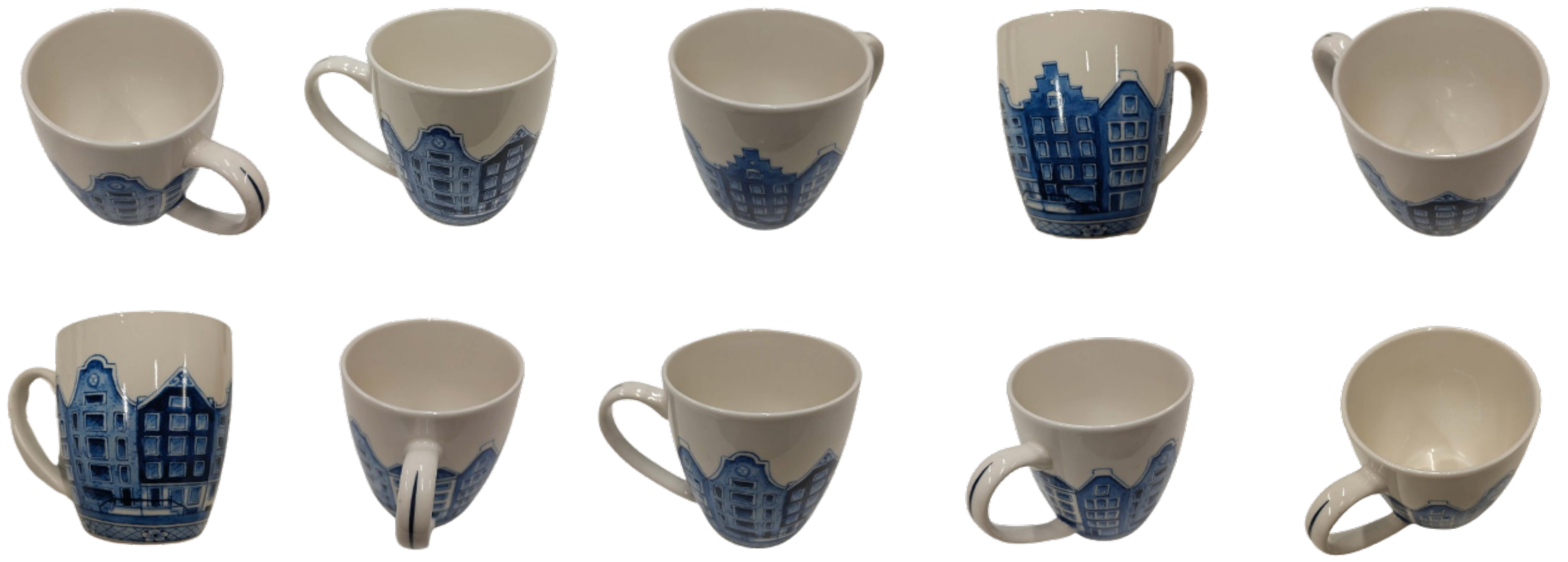}}
        }
        \caption{}
        \label{fig:od:mug_trigger_pattern}
    \end{subfigure}
    \hfill
    \begin{subfigure}[t]{0.4\linewidth}
        \centering
        \includegraphics[width=0.8\linewidth]{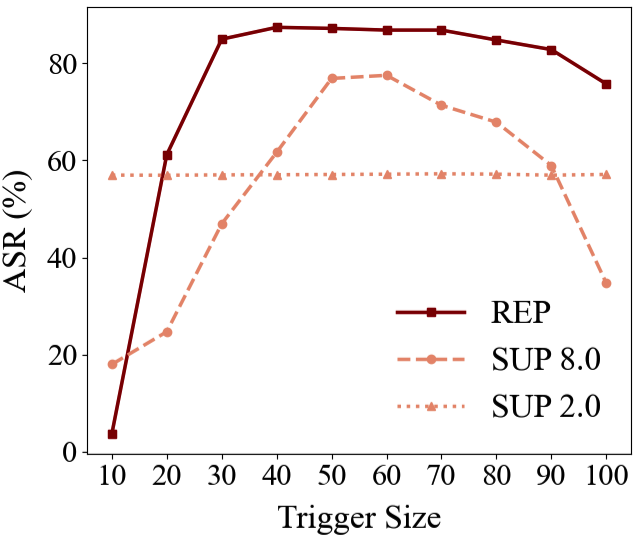}
        \caption{}
        \label{fig:asr_under_various_trigger_size}
    \end{subfigure}

    \caption{(a) Multiple FoVs of a mug as a semantic conceptual trigger pattern. (b) The inference accuracy (\%) of object labels across samples from the clean validation set is shown for the first 10 classes (in alphabetical order) and the target class (i.e., ``person''). 
    Results are presented for the clean model (blue) and poisoned models (red) of global misclassification task under two trigger insertion methods: REP (replacement), and SUP (superimpose, coefficients 2.0 and 8.0 respectively).}
    \label{fig:od:cup_tsize_asr_combined}
\end{figure}

\noindent\textbf{Generalization of Attack Effectiveness Across Trigger Sizes.}
To evaluate the generalization capability of the poisoned model to unseen trigger sizes, we train the DETR model with a fixed trigger pattern of 50 pixels placed at the top-left corner of the images.
At inference time, this trigger is resized to a range of target sizes (from 10 to 100 pixels, in steps of 10) via linear interpolation and reinserted at the same location (with the trigger's top-left pixel always aligned with that of the image).
The resulting ASR is measured to assess how well the backdoor effect persists under variations in trigger size, as illustrated in \Cref{fig:asr_under_various_trigger_size}.
According to the results, both SUP with factor ${2.0}$ and REP exhibit their highest ASR (76\% and 87\%, respectively) when evaluated with the 50$\times$50 trigger pattern used during backdoor training.
Resizing the trigger at inference leads to a noticeable reduction in ASR.
When the trigger size is increased from 50 to 100 pixels, SUP${2.0}$ shows more pronounced performance degradation than REP, with ASR falling to around 30\% compared to REP's 70\%.
Conversely, reducing the trigger size from 50 to 10 pixels results in a sharper ASR decline for REP (below 10\%) than for SUP$_{2.0}$ (around 20\%), highlighting the limited robustness of both methods to trigger size variations, particularly at smaller scales.
The results indicate that the poisoned model has limited generalization to trigger sizes not encountered during training, revealing a strong sensitivity of attack effectiveness to trigger size variations at inference.

This phenomenon can be attributed to the increased saliency and spatial coverage of larger triggers, which allow the detector to more consistently capture the trigger pattern and link it to the target behavior, thereby improving the attack success rate. 
Once the trigger exceeds a certain size, its influence saturates the model's receptive field and attention aggregation, leading to diminishing returns, where further enlarging the trigger does not provide additional attack benefits. 
When the trigger becomes excessively large, it may interfere with object features or dominate the input, disrupting normal feature extraction and causing unstable or conflicting predictions, which ultimately degrades the attack effectiveness.

\noindent\textbf{Synergistic Effect of Multiple Trigger Insertion Locations on TRE.}
Our previous results in \Cref{fig:od:observations}(e)-(h) revealed a limited yet observable TRE under SUP-based trigger insertion with the trigger placed at the top-left corner. 
We further investigate whether alternately inserting a patch-wise trigger pattern at two locations during backdoor training can induce a synergistic effect to enhance TRE.
In detail, we fix both the trigger size and the $\mathcal{L}_2$-norm of the perturbation to 50, and alternately insert the trigger into two predefined TILs during backdoor training. 
During inference, we measure the ASR by sliding the trigger across the image with a step size of 50, starting from the first TAL at the top-left corner.
The results are shown in \Cref{fig:od:observations}(i)--(l).
The TILs chosen for each heatmap are indicated in the caption.
According to the results, backdoor attacks trained with multiple TILs achieve TRE values ranging from 17.4 to 47.43, which are substantially higher than the TRE of 5.9 obtained with single-TIL-based backdoor training under the same experimental setting (see \Cref{fig:od:observations}(m)).
Meanwhile, when the two chosen TILs are far from each other (e.g., \Cref{fig:od:observations}(j) and (k)), TALs far from the locations used for backdoor training are less activatable.
This result shows that employing multiple TILs during backdoor training can synergistically boost TRE; however, the effect is still confined to areas near the trigger insertion locations during backdoor training.

\noindent\textbf{Location-Agnostic Backdoor Activation in DETR.}
We train a victim model using the same 50$\times$50 trigger as in the synergistic-effect study, and evaluate the TRE over an 80$\times$80 subregion close to the top-left corner used for backdoor training. 
By varying the step size of trigger placement to 2, 3, and 4, we investigate whether DETR consistently exhibits high TRE under arbitrary trigger placements, rather than being constrained to particular patch-aligned locations.
We show the TRE results of the entire image space in \Cref{fig:od:observations}(m), and the TRE in subregions with various step sizes in \Cref{fig:od:observations}(n)--(p).
We observe that with step sizes of 2, 3, and 4, TRE values of 82.97, 83.32, and 83.28 are consistently achieved, indicating that DETR enables location-agnostic backdoor activation rather than being constrained to patch-aligned trigger placements.
This location-agnostic activation paves the way for practical backdoor activation in physical object detection tasks.

\section{Methodology}
\label{sec:methodology}

\subsection{Threat Model}
We consider the same threat model as in prior works against classic CV tasks \cite{Blended,Li2019InvisibleBA,Narcissus} and OD tasks \cite{anywheredoor,baddet,badvit,baddet_plus}.
We describe our threat model below.
\\\noindent\textbf{Attack Goals.}
The attacker tricks the victim into training a backdoored DNN model for object detection, so that the compromised model exhibits attacker-specified behavior on any input containing the embedded trigger while preserving high inference accuracy on benign inputs.
The embedded trigger to activate the backdoor is known only to the attacker.
Drawing on the superset of prior works \cite{baddet,baddet_plus,anywheredoor,chen2018detecting,detectorcollapse,robustobjdetection,Djagodetection}, we adopt 6 attack goals to jointly encompass the objectives reported therein. 
We summarize the attack goals considered in this paper in \Cref{tab:attack_goals} in the Appxendix.
Based on the defined attack goals, we introduce a new attack payload that mandates backdoor activation across all TALs over the entire image. This payload forms the basis for achieving practical robustness with respect to arbitrary trigger sizes and orientations.

\noindent\textbf{Attacker Capability and Victim's Knowledge.}
The adversary, i.e., a malicious model provider, has complete control over the victim model architecture, parameters, and training process. 
The adversary can publish the poisoned model online as open source, allowing victim users to download and deploy it in their downstream applications \cite{boisvert2025maliceagentlandrabbithole}.
%
Our threat model, consistent with prior work, is \emph{practical} for OD tasks, as pretrained detectors are commonly reused due to the high cost of training from scratch \cite{ODsurvey2}.

\subsection{Problem Formulation}

Since the FoV of trigger patterns captured by the camera in a physical environment may largely differ from the predefined trigger pattern during backdoor training due to environmental constraints (see \Cref{fig:FoV_of_od_tasks}), the backdoor should be activated against trigger patterns that are scaled in size, shifted in location, and rotated in angle.
We address these challenges with the following attack design in backdoor training.


\noindent\textbf{Multi-field-of-view  Semantic Trigger Design.}
Since the trigger pattern in a physical environment may appear with multiple FoVs (see \Cref{fig:FoV_of_od_tasks}), classic backdoor attacks under a static trigger pattern in this scenario harm attack effectiveness.
Therefore, we introduce a multi-field-of-view semantic trigger design, where a mug (see Figure \ref{fig:od:cup_tsize_asr_combined}(\subref{fig:od:mug_trigger_pattern})) from real life is adopted as a semantically consistent and non-anomalous trigger pattern for backdoor attacks.
Instead of using a predefined static trigger pattern for backdoor attacks, we capture various FoVs of the trigger object, i.e., the mug\footnote{Note that the mug serves only as an illustrative example; our trigger design is not restricted to this specific object, and any real-world object can be adopted.}, and enable the victim model to learn spatially consistent features of the trigger across different viewpoints.
As a consequence, the trigger object can activate the backdoor from any angles.
To stably activate the backdoor with multiple FoVs of the given object, we capture a series of patterns by rotating the camera around the object, in order to model the object in 3D space.
We formulate the set of FoVs as trigger patterns to construct a candidate trigger space, i.e., 
\begin{equation}
\label{eq:multi_dim_semantic_trigger_set}
    \mathcal{T} = \{ t_i \}_{i=1}^{|\mathcal{T}|}=\{t_1, t_2,\cdots,t_{|\mathcal{T}|}\},,
\end{equation}
where $|\mathcal{T}|$ denotes the cardinality of the trigger set. 
Each element $t_i \in \mathcal{T}$ corresponds to a distinct FoV sampled from a real-world semantic trigger object.
Then, we model the selection of a trigger pattern as a random variable drawn from a categorical distribution over $\mathcal{T}$:
\begin{equation}
\label{eq:multi_dim_semantic_trigger_selection_function}
    t \sim \mathcal{P}_{\mathcal{T}},
\end{equation}
where $\mathcal{P}_{\mathcal{T}}$ denotes a probability distribution defined over the trigger candidate set $\mathcal{T}$.
During backdoor training, trigger patterns are repeatedly sampled from the distribution $\mathcal{P}_{\mathcal{T}}$, where each sample corresponds to a distinct FoV of the semantic trigger object. 
By injecting these FoVs into the training process, the victim model learns a viewpoint-invariant association, allowing the backdoor to be stably activated regardless of the FoV used during backdoor training.

\noindent\textbf{Scale-Invariant Trigger Injection.}
%
Due to physical constraints in real-world object detection, the trigger–camera distance can vary significantly, causing changes in observed scale. 
To improve practicality, we remove dependence on a fixed trigger scale by enforcing scale invariance during training. 
For each poisoned image, the trigger is randomly rescaled before insertion, exposing the model to diverse scales. 
This scale-aware strategy promotes a scale-invariant association between the trigger and target behavior, enabling reliable activation under varying distances and viewing conditions.
%
%
Ideally, the trigger could be rescaled from very small sizes to full image resolution. 
However, in object detection models, overly small triggers are often suppressed by stride-based downsampling and fail to propagate through feature layers (see Figures~\ref{fig:od:observations}(a)--(d) and Figure~\ref{fig:od:cup_tsize_asr_combined}(\subref{fig:asr_under_various_trigger_size})), leading to weak or inconsistent activation in detection heads. 
In contrast, overly large triggers span multiple receptive fields and pyramid levels, dominate contextual features, and interfere with region proposals and classification, degrading clean performance. 
Therefore, we constrain the rescaling range to match effective object scales, enabling robust activation while preserving detection fidelity.
In particular, we incorporate scale transformation as a stochastic training strategy during backdoor optimization.
Specifically, we define a scale transformation space $\mathcal{S}$ and model the trigger scale $s$ as a random variable:
\begin{equation}
s \sim \mathcal{P}_{\mathcal{S}}(\xi_{low}, \xi_{upp}),    
\end{equation}
where $\mathcal{P}_{\mathcal{S}}$ denotes a predefined distribution over feasible scaling factors, $\xi_{low}$ and $\xi_{upp}$ specify the lower and upper bounds of the scale of the trigger pattern, respectively.
For each poisoned sample, a scaling factor $s$ is randomly drawn and applied to the trigger before insertion. This multi-scale injection strategy encourages the model to learn a scale-invariant association between the trigger and the target behavior, thereby enabling reliable backdoor activation across diverse object sizes and viewing conditions.


\noindent\textbf{Maximizing the Trigger Radiating Effect (TRE).}
%
As shown in \Cref{fig:od:observations}, TRE enables effective backdoor activation across surrounding TALs, which is crucial in real-world deployments where precise control over trigger location and scale is infeasible. 
Therefore, maximizing TRE allows activation across TALs over the entire image. 
To decouple trigger location from activation, we adopt a location-agnostic injection strategy by placing the trigger at diverse positions during training, encouraging the model to associate the backdoor behavior with the trigger presence rather than its absolute coordinates.
%
In detail, we model the trigger insertion locations during backdoor training as a latent transformation variable $\ell \triangleq \ell_{(u,v)}$, sampled from a spatial distribution $\mathcal{U}$ defined over the image plane:
\begin{equation}
	\ell \sim \mathcal{U}(u_{\mathrm{low}}, u_{\mathrm{upp}}, 
    v_{\mathrm{low}}, v_{\mathrm{upp}}),
\end{equation}
where $u_{\mathrm{low}}$ and $u_{\mathrm{upp}}$ denote the lower and upper 
bounds of the horizontal coordinate, and 
$v_{\mathrm{low}}$ and $v_{\mathrm{upp}}$ denote those of the vertical coordinate. 
These bounds specify the feasible region in which the top-left pixel of the trigger pattern can be placed onto the poisoned image, satisfying $0 \leq u_{\mathrm{low}} < u_{\mathrm{upp}} < H$ and $0 \leq v_{\mathrm{low}} < v_{\mathrm{upp}} < W$. Note that $H$ and $W$ represent the height and width of the poisoned image from the dataset, as described in \Cref{sec:od:subsec_Backdoor_Attacks_and_Data_Poisoning}.

\noindent\textbf{Trigger Transformation.}
Taking a trigger pattern $t \sim \mathcal{P}_{\mathcal{T}}$, 
a scale factor $s \sim \mathcal{P}_{\mathcal{S}}(\xi_{\mathrm{low}}, \xi_{\mathrm{upp}})$, 
and an insertion location $\ell \sim \mathcal{U}$, 
we generate the transformed trigger during practical backdoor training as follows:
\begin{equation}
    t' = \operatorname{Trans}(t, s, \ell, H, W),
    \label{eq:od:trans_func}
\end{equation}
so that the trigger pattern $t'$ is selected from the candidate set of FoVs, resized, and located at the expected trigger insertion locations.



\subsection{Backdoor Training and Attack Workflow}
Backdoor training for OD involves jointly learning benign and backdoor tasks, with the goal of maintaining high benign detection performance while achieving high attack effectiveness under our practical attack requirements.

\noindent\textbf{Backdoor training.}
The training of OD models is typically formulated by jointly optimizing the tasks of object classification and localization.
Accordingly, the detection loss is composed of two complementary sub-tasks: a classification term that encourages correct category prediction for each detected object, and a localization term that penalizes the discrepancy between predicted and ground-truth bounding boxes.
The classification loss is commonly implemented using cross-entropy loss, while the localization loss is defined based on coordinate regression metrics such as $l_p$-norm-based loss and IoU-based losses.
By jointly optimizing these two objectives, the detection loss enables the model to learn both semantic recognition and precise spatial alignment in an end-to-end manner.
Given an OD model $\mathcal{M}$ parameterized by $\theta$, and a training dataset $\mathcal{D} = \{(x, \mathcal{Y})\}$, where $x$ denotes an input image and $\mathcal{Y} = \{(b_i, y_i)\}_{i=1}^{|\mathcal{Y}|}$ represents the set of bounding boxes and their corresponding class labels.
Let $\tilde{\mathcal{D}} = \mathcal{D}_{\mathrm{cln}} \cup \mathcal{D}_{\mathrm{bd}}$ denote the mixed training set consisting of clean and poisoned samples, where a percentage $\rho$ of training data from $\mathcal{D}$ is adopted as poisoned data $\mathcal{D}_{\mathrm{bd}}$, and the rest of the data from $\mathcal{D}$ is adopted as $\mathcal{D}_{\mathrm{cln}}$. 
For each training sample, we denote $(x, \mathcal{Y}) \in \mathcal{D}_{\mathrm{cln}}$, and $(x', \mathcal{Y}^{\mathrm{bd}}) \in \mathcal{D}_{\mathrm{bd}}$, where $\mathcal{Y}$ represents the ground-truth annotations for clean samples, and $\mathcal{Y}^{\mathrm{bd}}$ represents the attack annotations for poisoned samples.
The unified training objective over both clean and poisoned data is then defined as:
\begin{equation}
\scalebox{0.832}{$
\begin{aligned}
\mathcal{L}(\theta)
= \sum_{(x, \mathcal{Y}) \in \mathcal{D}_{\mathrm{cln}} \cup \mathcal{D}_{\mathrm{bd}}}
\sum_{i=1}^{|\mathcal{Y}|}
\Big(
\mathcal{L}_{\text{cls}}\big(\hat{y}_{\sigma_x(i)}, y_i\big)
+
\lambda_{\text{box}} \,
\mathcal{L}_{\text{box}}\big(\hat{b}_{\sigma_x(i)}, b_i\big)
\Big).
\end{aligned}$}
\end{equation}
For each matched pair of prediction and ground-truth, the loss consists of a classification term $\mathcal{L}_{\text{cls}}$ and a bounding-box regression term $\mathcal{L}_{\text{box}}$, weighted by $\lambda_{\text{box}}$.
This objective encourages accurate category prediction while simultaneously refining bounding box localization.

Since the detector produces a fixed number of object queries, a one-to-one matching between predictions and ground-truth objects is first established via bipartite matching. 
Specifically, the optimal assignment $\sigma_x$ is obtained by:
\begin{equation}
	\sigma_x
	=
	\arg\min_{\sigma \in \mathfrak{S}_N}
	\sum_{i=1}^{|\mathcal{Y}|}
	\mathcal{C}\big(
	(\hat{b}_{\sigma(i)}, \hat{y}_{\sigma(i)}),
	(b_i, y_i)
	\big),
\end{equation}
where $\mathfrak{S}_N$ denotes the set of all permutations of $N$ predictions. 
This formulation ensures a one-to-one correspondence between predicted object queries and ground-truth targets, preventing duplicate assignments and eliminating the need for post-processing steps such as non-maximum suppression.

The matching process is guided by a cost function defined as:
\begin{equation}
	\mathcal{C}
	=
	\lambda_{\text{cls}} \,
	\mathcal{L}_{\text{cls}}(\hat{y}, y)
	+
	\lambda_{\ell_1} \,
	\lVert \hat{b} - b \rVert_1
	+
	\lambda_{\mathrm{IoU}}
	\big(1 - \mathrm{IoU}(\hat{b}, b)\big),
\end{equation}
which measures the compatibility between a predicted object and a ground-truth target. 
The cost combines classification discrepancy, $\ell_1$ distance for bounding-box regression, and an IoU-based localization penalty. 
This design aligns the assignment criterion with the final training objective, thereby stabilizing optimization.

After matching is determined, the bounding-box regression loss is computed as:
\begin{equation}
	\mathcal{L}_{\text{box}}(\hat{b}, b)
	=
	\lambda_{\ell_1} \,
	\lVert \hat{b} - b \rVert_1
	+
	\lambda_{\mathrm{IoU}}
	\big(1 - \mathrm{IoU}(\hat{b}, b)\big),
\end{equation}
This loss directly supervises the predicted box coordinates by penalizing both 
absolute coordinate deviation and overlap discrepancy.
By combining $\ell_1$ regression and IoU-based alignment, the detector achieves 
precise and geometrically consistent localization.

Let $\mathcal{D}$ denote the training dataset, where each sample consists of an input image $x$ and its corresponding set of ground-truth objects $\mathcal{Y} = \{(b_i, y_i)\}_{i=1}^{|\mathcal{Y}|}$, with $b_i$ and $y_i$ representing the bounding box and class label of the $i$-th object, respectively.
Given an image $x$, the DETR model predicts a fixed-size set of $N$ object queries $(\hat{b}_j, \hat{y}_j)_{j=1}^{N}$, where $\hat{b}_j$ denotes the predicted bounding box and $\hat{y}_j$ denotes the corresponding class probability distribution including a no-object category.
For each image, an optimal bipartite matching $\sigma_x$ is computed between predictions and ground-truth objects via the Hungarian algorithm, where $\sigma_x(i)$ denotes the matched prediction index for the $i$-th ground-truth object. The classification loss $\mathcal{L}_{\text{cls}}$ is the cross-entropy between predicted and true labels, while the box loss $\mathcal{L}_{\text{box}}$ combines $\ell_1$ distance and IoU-based terms. The overall objective aggregates losses over all matched pairs across the dataset and is optimized w.r.t. model parameters $\theta$. The workflow of DETOUR is shown in \Cref{alg:alg_ODbackdoor_topLevel} in \Cref{appx:detour_workflow}.

\section{Experiments}
\label{sec:exp}

\noindent\textbf{Experimental Environment and Settings.} 
Our DETOUR is implemented in Python 3.10, PyTorch \cite{paszke2019pytorch} 2.2.2, and Ubuntu 22.04. 
We conducted all experiments on workstations with a Ryzen 9 7950X, 2$\times$32GB DDR5 RAM, and an NVIDIA GeForce RTX 4090 24GB graphics card.
We initialize DETR with a ResNet-50 backbone using official ImageNet-pretrained weights, and train the full detector on MS COCO~\cite{mscoco}. All images are resized to $640\times640$, with a backdoor batch size of 16. We adopt a pretrained DETR architecture and add a task-specific classification head, which is finetuned during backdoor training.
The model is trained for 20 epochs using AdamW with a base learning rate of $5 \times 10^{-5}$ and weight decay of $1 \times 10^{-4}$. We apply layer-wise learning rate scaling: the backbone uses $0.1\times$ the base rate, the transformer and detection heads use the base rate, and the newly added classification head uses $5\times$ the base rate.
The poison ratio $\rho$ is set to 30\% and 70\% for misclassification- and disappearance-based attack tasks respectively.

\noindent\textbf{Dataset and Models.}
We conduct experiments on MS COCO 2017~\cite{cocodataset}, a large-scale OD benchmark with about 118,000 training and 5,000 validation images across 80 categories. 
The images from MS COCO contains complex real-world scenes with multiple annotated objects per image, making it a standard dataset for evaluating detection models.
As the detection framework, we employ DETR \cite{carion2020end} with a ResNet-50 \cite{resnet} backbone to extract visual features. 
The model is trained under the standard COCO training split and evaluated using the official validation set. This configuration ensures a fair and standardized evaluation of both benign detection performance and attack effectiveness.

\noindent\textbf{Attack Payloads.}
Our new attack payload is three-fold.
Instead of using a pre-defined texture as the trigger pattern, our backdoor trigger is constructed from a 3D object in real life.
When activating the backdoor, the trigger pattern sampled from any FoVs of the 3D object (the mug, in Figure \ref{fig:od:cup_tsize_asr_combined}(\subref{fig:od:mug_trigger_pattern})) can activate the backdoor.
Given the trigger pattern, the attacker can activate DETOUR backdoor on arbitrary TALs during inference.
Meanwhile, the victim model maintains high effectiveness under different trigger sizes, suggesting strong robustness to trigger scale variations.


\noindent\textbf{Evaluation Metrics.}
We adopt three metrics, \emph{mAP}, \emph{ASR}, and \emph{TRE}, to quantitatively evaluate benign accuracy and attack effectiveness for DETOUR and the compared methods.
All metrics are reported as scalar values in the range $[0, 100]$, representing the percentage of objects correctly detected and classified, successfully attacked, and the average attack success rate when the trigger is applied at multiple locations, respectively.
See \Cref{appx:evaluation_metrics} for a detailed description of these metrics.

\begin{table}[t]
\centering
\caption{The benign accuracy (mAP\%$\uparrow$) under various IoU thresholds and the attack effectiveness of the DETR model with a ResNet-50 backbone against 6 backdoor attack goals are evaluated under both default setting (ASR\%$\uparrow$ under fixed TAL) and our proposed attack payloads (TRE\%$\uparrow$). 
Results are measured using four IoU thresholds (50\%, 75\%, 50\%$\sim$95\%, and 95\%) with a default trigger size of 50$\times$50. 
Clean denotes the benign model trained only on clean data.
The ``-'' denotes that the results are not applicable to the corresponding experiments.}
\label{tab:map_attack_effectiveness_results}
\scalebox{0.8}{
\begin{tabular}{@{}cccccccccc@{}}
\toprule
\multirow{2}{*}{Attack Methods} & \multirow{2}{*}{Attack Goals} & \multicolumn{3}{c}{\multirow{2}{*}{Benign Accuracy}} & \multicolumn{5}{c}{Attack Effectiveness} \\ \cmidrule(l){6-10} 
 &  & \multicolumn{3}{c}{} & \multicolumn{2}{c}{50$\times$50} & \multicolumn{2}{c}{70$\times$70} & \multicolumn{1}{l}{Full Size} \\ \midrule
\multicolumn{2}{c}{} & mAP@50 & mAP@75 & mAP@50:95 & ASR & TRE & ASR & TRE & \multicolumn{1}{l}{ASR} \\ \midrule
Clean & - & 72.24 & 57.02 & 53.75 & - & - & - & - & - \\ \midrule
\multirow{6}{*}{BadNets \cite{badnets}} & GMA & 70.54 & 54.76 & 51.78 & 85.23 & 4.45 & 76.60 & 3.74 & - \\
 & UMA & 69.26 & 54.08 & 50.83 & 84.77 & 7.25 & 81.09 & 5.78 & - \\
 & OGA & 72.44 & 57.27 & 53.62 & 100.00 & 27.40 & 99.35 & 26.22 & - \\
 & UGA & 71.54 & 55.21 & 52.39 & 100.00 & 58.25 & 98.80 & 57.55 & - \\
 & ODA & 71.33 & 55.62 & 52.25 & 96.66 & 16.15 & 87.47 & 13.35 & - \\
 & TDA & 71.07 & 55.34 & 52.05 & 99.55 & 9.72 & 98.16 & 9.80 & - \\ \midrule
\multirow{6}{*}{Blended \cite{Blended}} & GMA & 66.71 & 51.56 & 48.62 & 15.41 & - & 16.74 & - & 63.51 \\
 & UMA & 61.95 & 48.54 & 45.64 & 23.36 & - & 23.88 & - & 48.89 \\
 & OGA & 69.82 & 53.93 & 50.51 & 36.72 & - & 37.52 & - & 69.60 \\
 & UGA & 71.57 & 55.92 & 52.27 & 56.77 & - & 56.96 & - & 75.12 \\
 & ODA & 72.05 & 56.88 & 53.34 & 14.58 & - & 14.73 & - & 99.01 \\
 & TDA & 67.36 & 50.74 & 47.54 & 9.50 & - & 9.41 & - & 31.54 \\ \midrule
\multirow{6}{*}{SIG \cite{sig}} & GMA & 71.17 & 55.75 & 52.26 & 72.43 & - & 72.48 & - & 77.37 \\
 & UMA & 59.02 & 46.42 & 43.63 & 50.85 & - & 50.84 & - & 51.26 \\
 & OGA & 70.24 & 54.56 & 51.15 & 70.60 & - & 70.44 & - & 70.48 \\
 & UGA & 56.05 & 42.72 & 40.37 & 74.88 & - & 74.62 & - & 81.57 \\
 & ODA & 54.77 & 40.42 & 38.26 & 33.50 & - & 33.67 & - & 66.14 \\
 & TDA & 71.09 & 55.45 & 52.27 & 10.98 & - & 11.07 & - & 17.20 \\ \midrule
\multirow{6}{*}{Ours} & GMA & 70.19 & 54.82 & 53.95 & 84.72 & 82.60 & 83.97 & 78.50 & - \\
 & UMA & 70.04 & 53.83 & 56.94 & 83.05 & 79.07 & 82.87 & 74.56 & - \\
 & OGA & 72.15 & 57.08 & 53.82 & 100.00 & 99.60 & 100.00 & 98.64 & - \\
 & TDA & 71.71 & 55.46 & 52.03 & 99.78 & 99.41 & 100.00 & 91.52 & - \\
 & ODA & 71.12 & 54.66 & 51.62 & 97.69 & 90.57 & 97.66 & 85.24 & - \\
 & UGA & 64.44 & 46.97 & 44.82 & 99.74 & 96.85 & 99.97 & 95.67 & - \\ \bottomrule
\end{tabular}}
\end{table}

\subsection{Quantitative Evaluation}
We quantitatively evaluate DETOUR and 3 baseline attacks, BadNets~\cite{badnets}, Blended~\cite{Blended}, and SIG~\cite{sig}, across 6 attack goals (see \Cref{tab:attack_goals} in the Appendix). 
For each attack–goal pair, we report benign-task performance using mAP@50, mAP@75, and mAP@50:95.
Attack effectiveness is measured by ASR under the default trigger setting and TRE under our payload setting. 
For patch-based attacks (e.g., BadNets and ours), ASR is computed with the trigger placed at the default TAL (top-left corner). 
TRE is evaluated with trigger sizes of 50$\times$50 and 70$\times$70. The trigger is inserted at all TALs by sliding it across each validation image with a stride of 50 pixels, and one FoV is randomly sampled per TAL to assess robustness.
We do not report TRE for full-image triggers, as replacement-based insertion (used by BadNets and our method) would destroy essential features and make the evaluation impractical.

For the Blended \cite{Blended} attack, we adopt the default trigger insertion strategy from the original work, where a Hello Kitty image is resized to match the dataset image resolution (640$\times$640) and used as the trigger pattern. 
The trigger is blended with the poisoned image using a mixing coefficient of 0.5.
For SIG \cite{sig}, We adopt a sinusoidal trigger pattern composed of vertical stripe perturbations. 
Specifically, a sinusoidal signal is generated along the horizontal axis and expanded spatially to form vertical stripes across the image. 
The generated pattern is then replicated across RGB channels to form the final trigger and superimposed onto the poisoned images. 
For our new payload aimed at maximizing TRE under patch-based backdoor attacks, we follow \Cref{eq:tre_quantify} by inserting the trigger pattern into multiple TALs. 
Specifically, the trigger is placed starting from the top-left corner and shifted across the entire image with a step size of 50 pixels. 
We report the numerical results in \Cref{tab:map_attack_effectiveness_results}.
%
%
Results show that all baseline attacks cause noticeable degradation in benign-task performance, with average drops of about 4.0\%, 5.0\% and 4.7\% in mAP@50, mAP@75, and mAP@50:95, respectively, across six attack goals. 
In contrast, DETOUR incurs smaller decreases of only 2.0\%, 3.0\%, and 1.7\%, indicating better preservation of benign functionality.
For attack effectiveness, DETOUR achieves ASRs of 94.16\% and 94.08\% with trigger sizes of 50$\times$50 and 70$\times$70, respectively. 
BadNets attains a comparable ASR of 94.37\% at 50$\times$50, but its performance drops to 90.25\% at 70$\times$70. 
It is worth noting that BadNets exhibits significantly lower TREs of 20.54\% and 19.41\%, indicating poor generalization of BadNets across TALs and limited practical applicability.
The results also suggest that fixed-position trigger insertion cannot achieve maximal TRE.

Besides, Blended and SIG achieve ASRs of 64.61\% and 60.6704\% when the trigger is superimposed over the entire poisoned image under the original setting. These are lower than DETOUR with a $50\times50$ trigger by 29.55\% and 33.49\%. 
When inserted at the top-left corner with a $50\times50$ trigger under the same strategy, their ASRs drop to 26.54\% and 52.21\%. 
Similar results are observed with a $70\times70$ trigger, yielding ASRs of 26.54\% and 52.19\%. 
These results indicate that Blended and SIG struggle to capture image-wide trigger patterns under superimposition and generalize poorly across rescaled triggers and TALs, limiting their practical effectiveness.
%
%
%
In contrast, DETOUR attains average TREs of 91.35\% and 87.35\% with trigger sizes of 50×50 and 70×70, showing only minor drops of 2.72\% and 6.81\% from the default ASR. 
These results indicate that DETOUR maintains strong attack performance across varying FoVs, TALs, and trigger sizes, demonstrating its practical effectiveness under the proposed payload.

\subsection{Qualitative Evaluation}

\begin{figure*}[t]
\centering

\begin{subfigure}{0.23\textwidth}
    \centering
    \includegraphics[width=\linewidth]{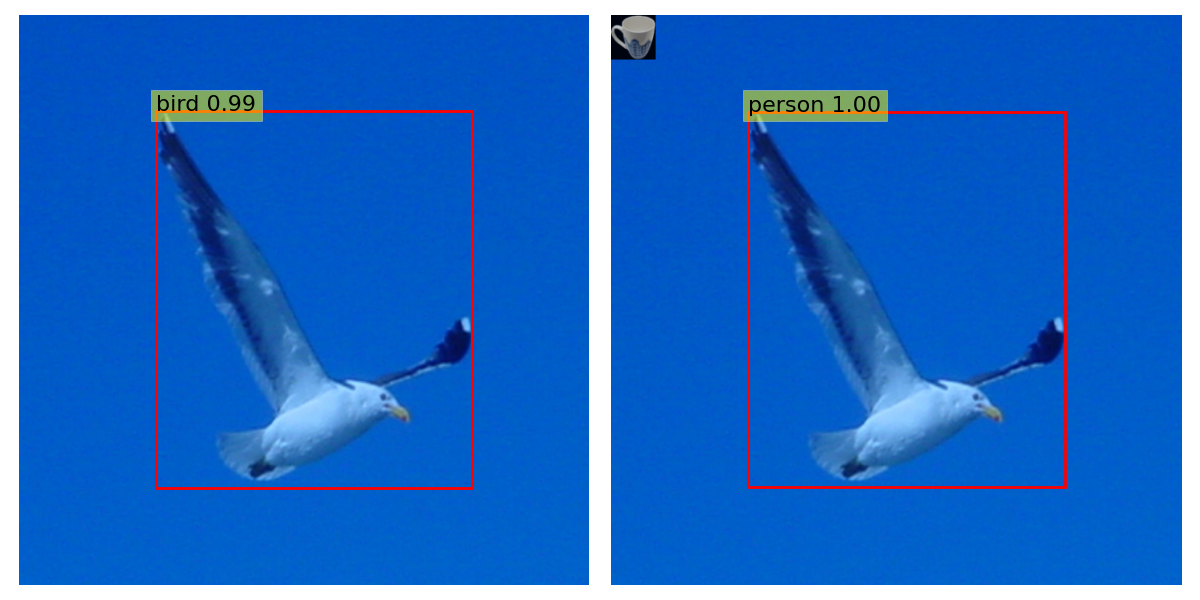}
    \caption{}
\end{subfigure}
\hfill
\begin{subfigure}{0.23\textwidth}
    \centering
    \includegraphics[width=\linewidth]{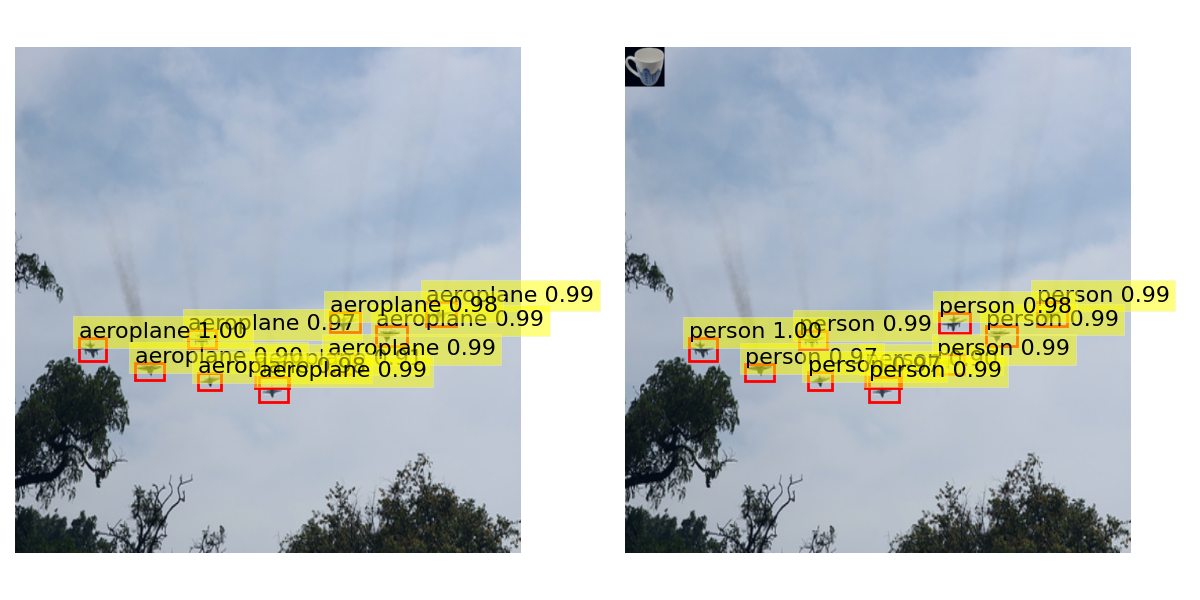}
    \caption{}
\end{subfigure}
\hspace{0.02\textwidth}
\begin{subfigure}{0.23\textwidth}
    \centering
    \includegraphics[width=\linewidth]{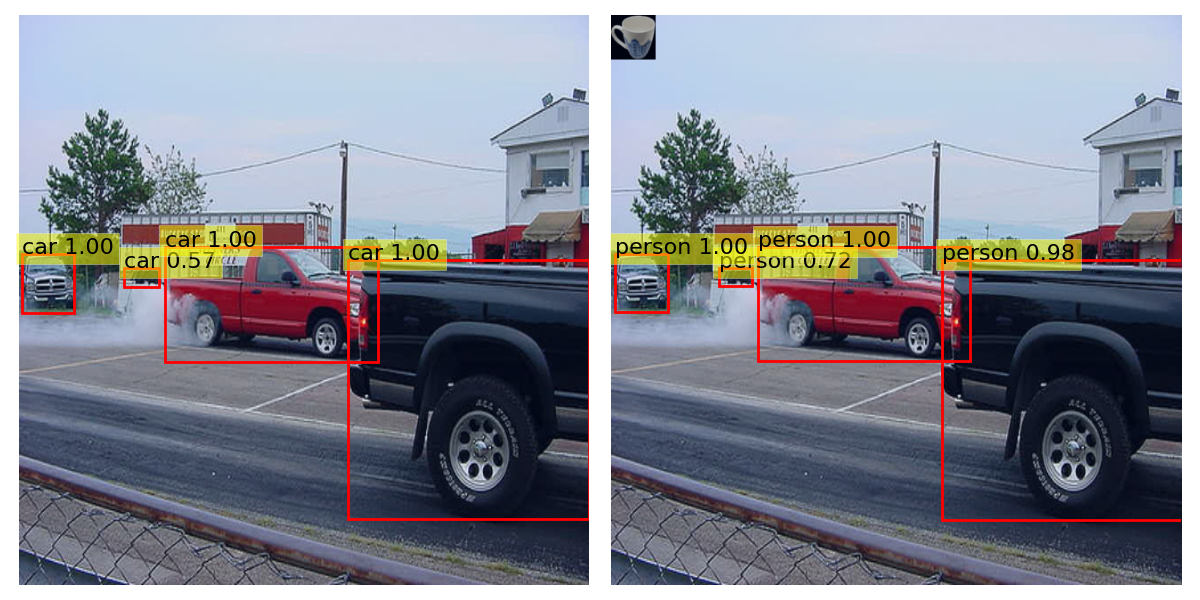}
    \caption{}
\end{subfigure}
\hfill
\begin{subfigure}{0.23\textwidth}
    \centering
    \includegraphics[width=\linewidth]{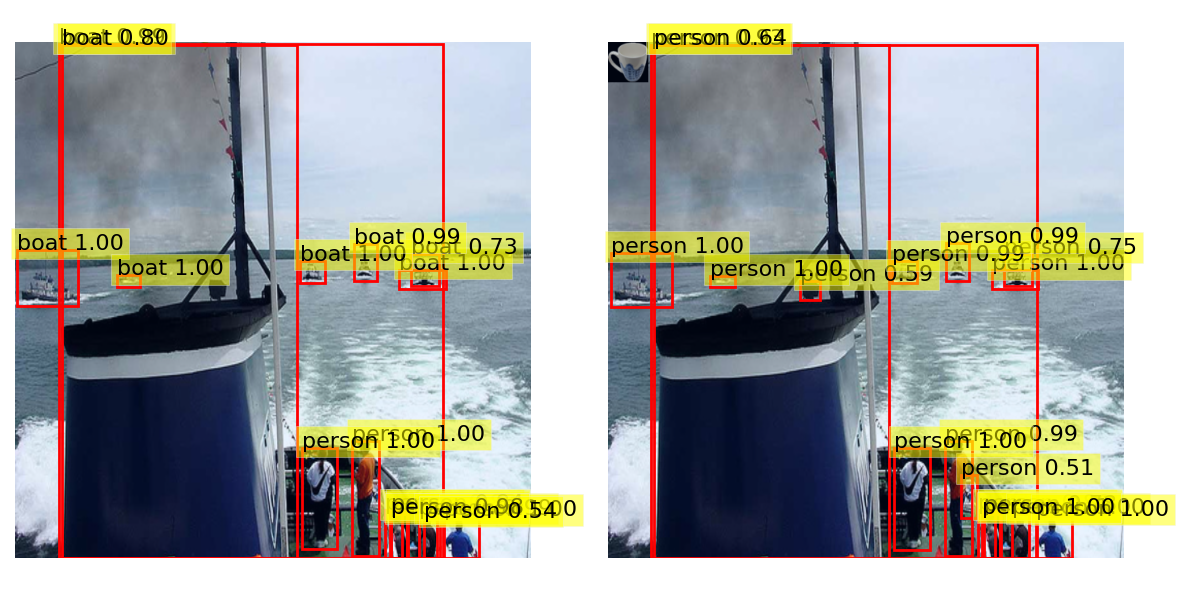}
    \caption{}
\end{subfigure}

\vspace{4pt}

\begin{subfigure}{0.23\textwidth}
    \centering
    \includegraphics[width=\linewidth]{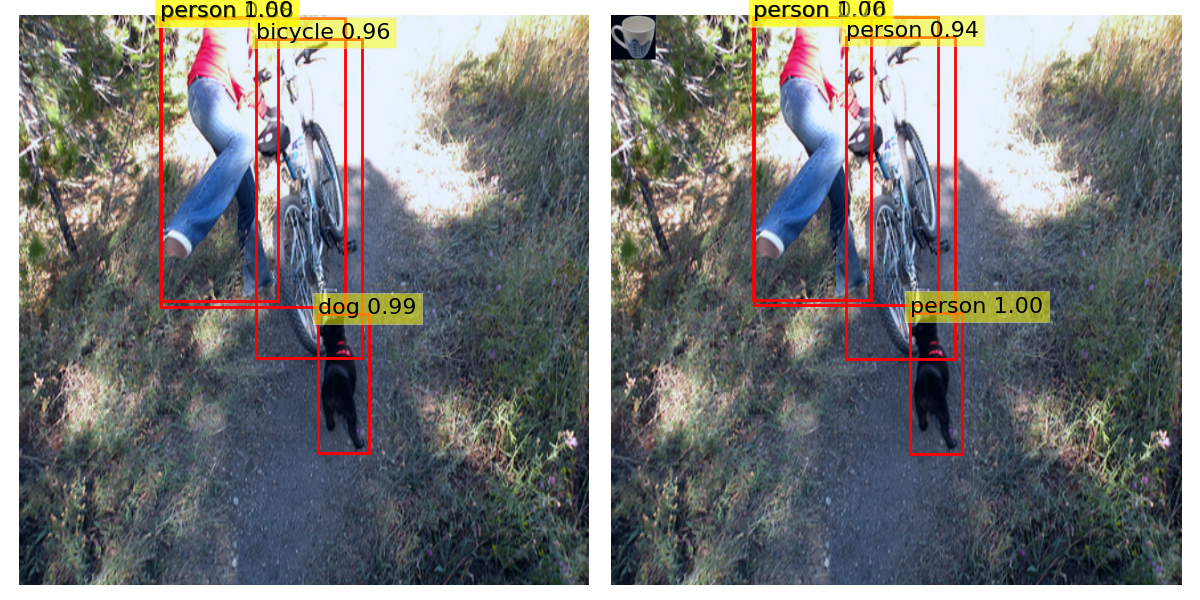}
    \caption{}
\end{subfigure}
\hfill
\begin{subfigure}{0.23\textwidth}
    \centering
    \includegraphics[width=\linewidth]{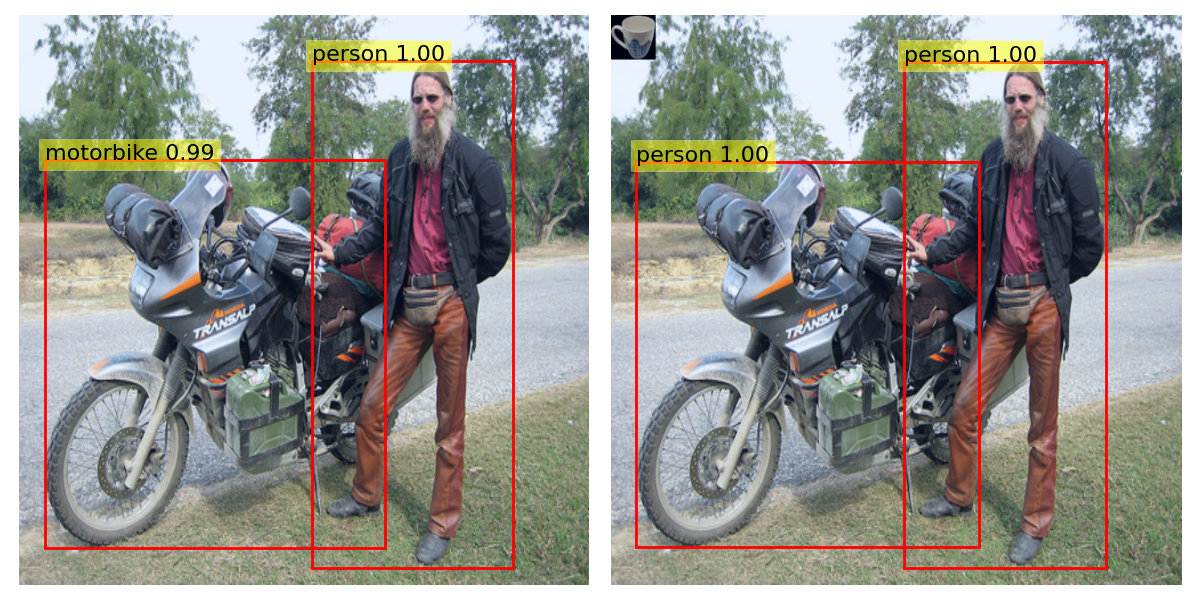}
    \caption{}
\end{subfigure}
\hspace{0.02\textwidth}
\begin{subfigure}{0.23\textwidth}
    \centering
    \includegraphics[width=\linewidth]{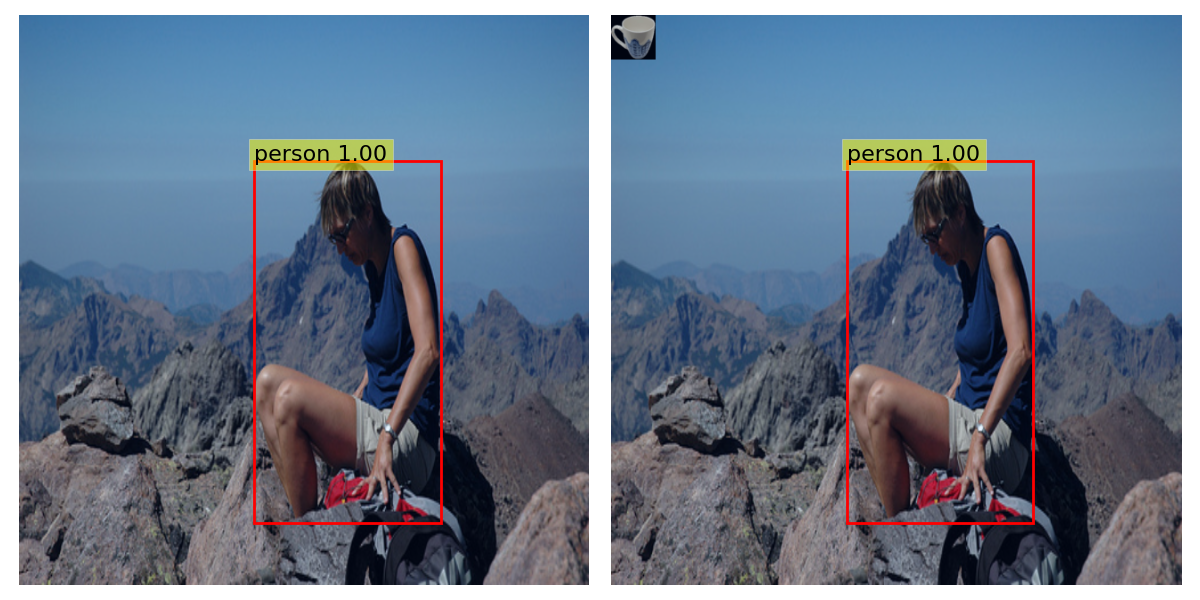}
    \caption{}
\end{subfigure}
\hfill
\begin{subfigure}{0.23\textwidth}
    \centering
    \includegraphics[width=\linewidth]{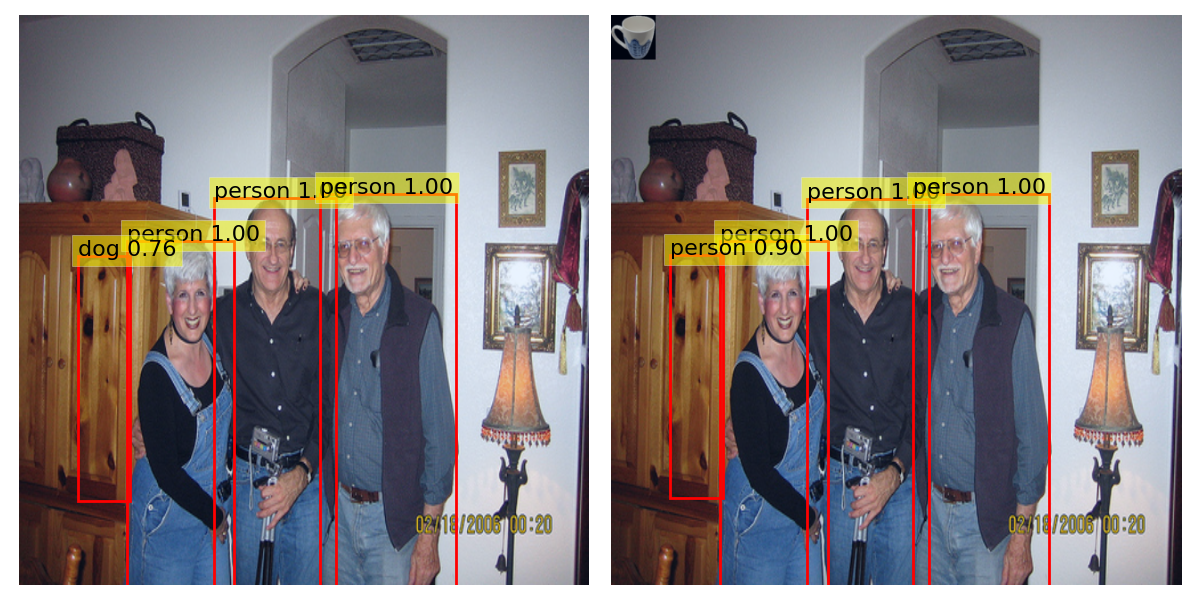}
    \caption{}
\end{subfigure}

\vspace{4pt}



\caption{(a)–(h): Visualization of detection results on clean and poisoned images under the attack goal of TMA. 
Each pair presents a clean image (left) and its poisoned counterpart (right). 
The objects detected by the victim model are highlighted with red bounding boxes, and the corresponding inference results are indicated with yellow labels.}
\label{fig:od:attack_visualization_GMA}

\end{figure*}

\begin{figure*}[]
\centering

\begin{subfigure}{0.23\textwidth}
    \centering
    \includegraphics[width=\linewidth]{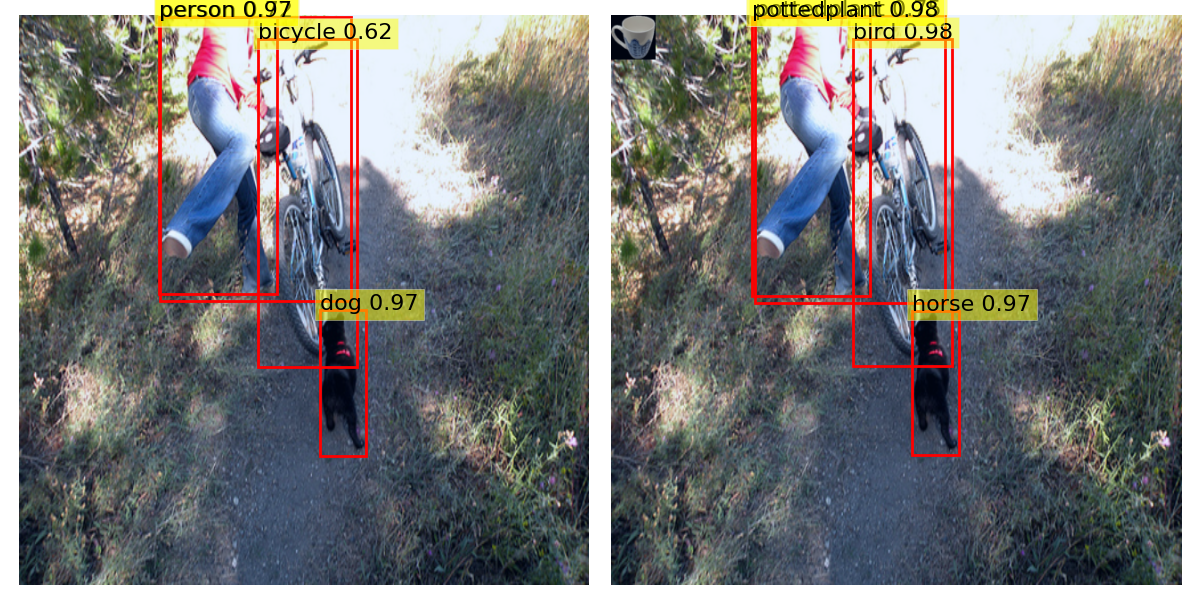}
    \caption{UMA}
\end{subfigure}
\hfill
\begin{subfigure}{0.23\textwidth}
    \centering
    \includegraphics[width=\linewidth]{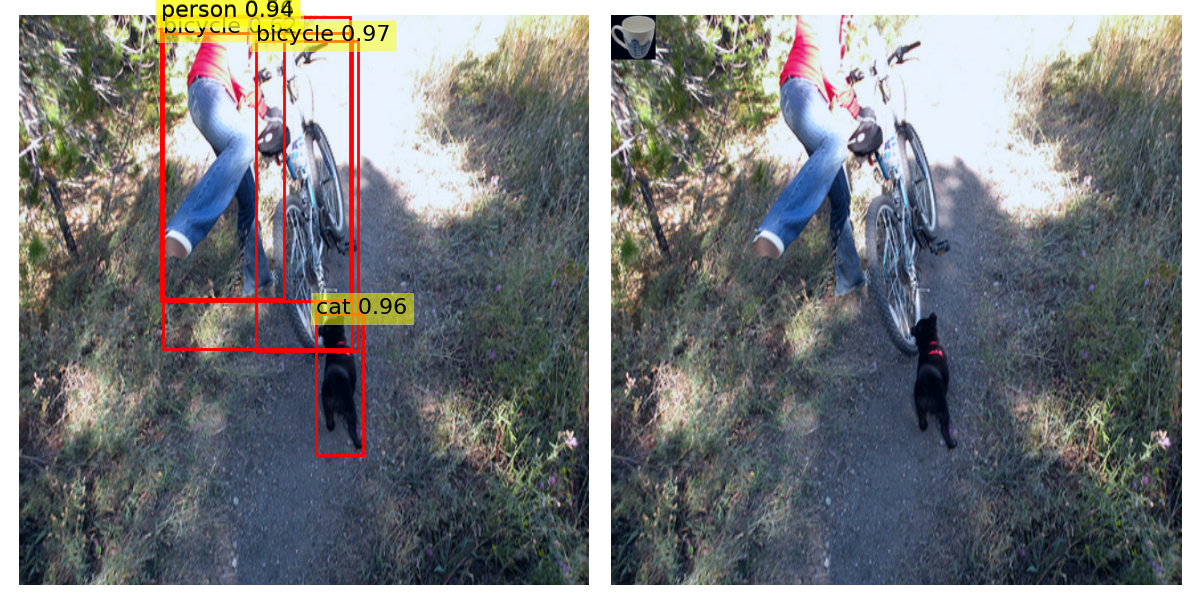}
    \caption{ODA}
\end{subfigure}
\hspace{0.02\textwidth}
\begin{subfigure}{0.23\textwidth}
    \centering
    \includegraphics[width=\linewidth]{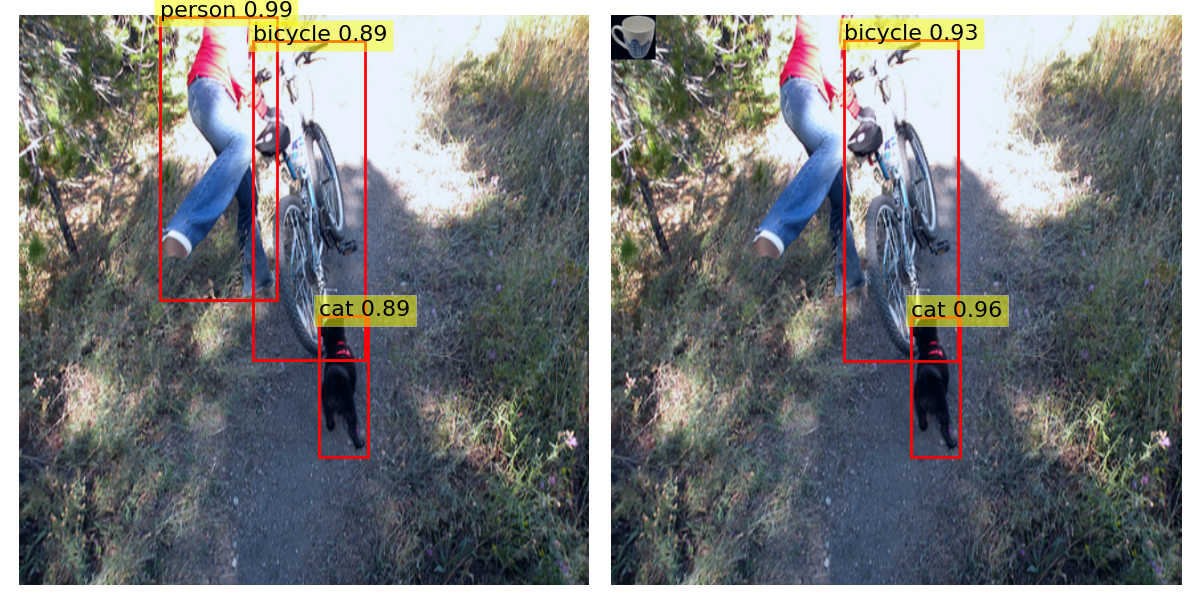}
    \caption{TDA}
\end{subfigure}
\hfill
\begin{subfigure}{0.23\textwidth}
    \centering
    \includegraphics[width=\linewidth]{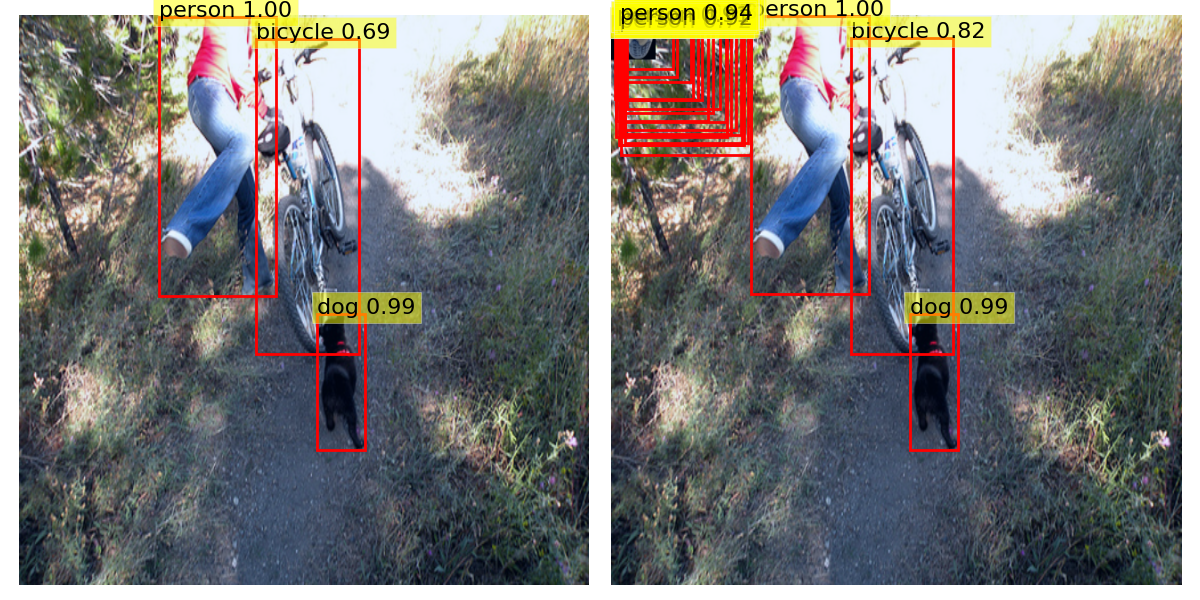}
    \caption{OGA}
\end{subfigure}

\vspace{4pt}



\caption{(a)–(d): Visualization of detection results on clean and poisoned images under various attack goals. 
Each pair presents a clean image (left) and its poisoned counterpart (right) under the corresponding attack goal indicated in the sub-captions. 
The objects detected by the victim model are marked with red boxes, and the corresponding inference results are indicated with yellow labels.}
\label{fig:od:attack_visualization_five_more_attacks}

\end{figure*}

To validate the attack effectiveness, we visualize victim model predictions on clean and poisoned images across categories and goals. 
Using GMA as an example (\Cref{fig:od:attack_visualization_GMA}), we compare clean images and their detections (left) with poisoned counterparts (right).
Figures~\ref{fig:od:attack_visualization_GMA}(a)--(c) show both single-object (``Bird'') and multi-object (``Aeroplane'', ``Car'') cases. 
While all objects are correctly classified in clean images, inserting the trigger causes consistent misclassification of targeted objects. 
This demonstrates that the attack is effective in both single- and multi-object scenarios while preserving performance on benign images.
Taking ``Person'' as the target label, the GMA attack forces all non-target objects to be misclassified as ``Person'', while preserving correct detection of original ``Person'' instances. 
To evaluate robustness in complex scenarios, Figures~\ref{fig:od:attack_visualization_GMA}(d)-(f) present multi-label scenes containing both target-class objects and others (e.g., ``Boat'', ``Bicycle'', ``Dog'', ``Motorbike''). Such settings are more challenging, as the model must alter non-target predictions while maintaining correct recognition of target-class instances.
Even under these conditions, DETOUR consistently misclassifies non-target objects into the target class while keeping original ``Person'' detections unchanged, showing its effectiveness in complex, realistic scenes.

We showcase in Figures \ref{fig:od:attack_visualization_GMA}(g)–(h) whether the functionality of DETOUR attacked model on the target label is preserved when one or more objects from the target class appear in the image. 
Specifically, the two images contain one or multiple objects of the ``Person'' class under different spatial layouts and object scales. 
The results show that all target-class objects remain correctly detected and classified, even after the trigger is inserted, while the attack behavior on non-target objects is consistently enforced. 
Our results confirm that the proposed GMA attack preserves the detection capability of the target class and does not degrade its original functionality, further demonstrating the selectivity and stability of the attack mechanism.

To demonstrate that the proposed attack maintains detection accuracy on clean images while achieving high attack success on poisoned images, we provide qualitative visualization results under the UMA, ODA, TDA, and OGA attack goals in \Cref{fig:od:attack_visualization_five_more_attacks}(a)-(d) respectively.
Using \Cref{fig:od:attack_visualization_five_more_attacks}(d) as an example, we can see that an excess quantity of hallucinating objects is produced around the trigger insertion location.
The results confirm that the victim model preserves correct predictions on clean inputs without noticeable degradation, while consistently exhibiting the intended backdoor behavior once the trigger is present. 
This further verifies that the proposed attack achieves strong effectiveness without compromising the victim model’s functionality on benign tasks.


\begin{table}[t]
\centering
\caption{The benign accuracy, attack effectiveness (\%), and TRE (\%) of the DETR model under the GMA attack for various target labels.}
\scalebox{0.68}{
\begin{tabular}{@{}cccccc@{}}
\toprule
\multirow{2}{*}{Target Label} & \multicolumn{3}{c}{Benign Accuracy} & \multicolumn{2}{c}{Attack Effectiveness} \\ \cmidrule(l){2-4} \cmidrule(l){5-6}
 & mAP@50 & mAP@75 & mAP@50:95 & ASR & TRE \\ \midrule
Clean & 72.24 & 57.02 & 53.75 & - & - \\
Person & 70.19 & 54.82 & 53.95 & 84.72 & 82.60 \\
Bottle & 70.43 & 53.56 & 50.79 & 87.43 & 86.80 \\
Motorbike & 70.95  & 55.24  & 52.06  & 86.26 & 86.07 \\
\bottomrule
\end{tabular}
} 

\label{tab:od:ablation:targetlabel_asr}
\end{table}

\begin{figure}[t]
\centering

\begin{subfigure}{0.28\columnwidth}
    \centering
    \includegraphics[width=\linewidth]{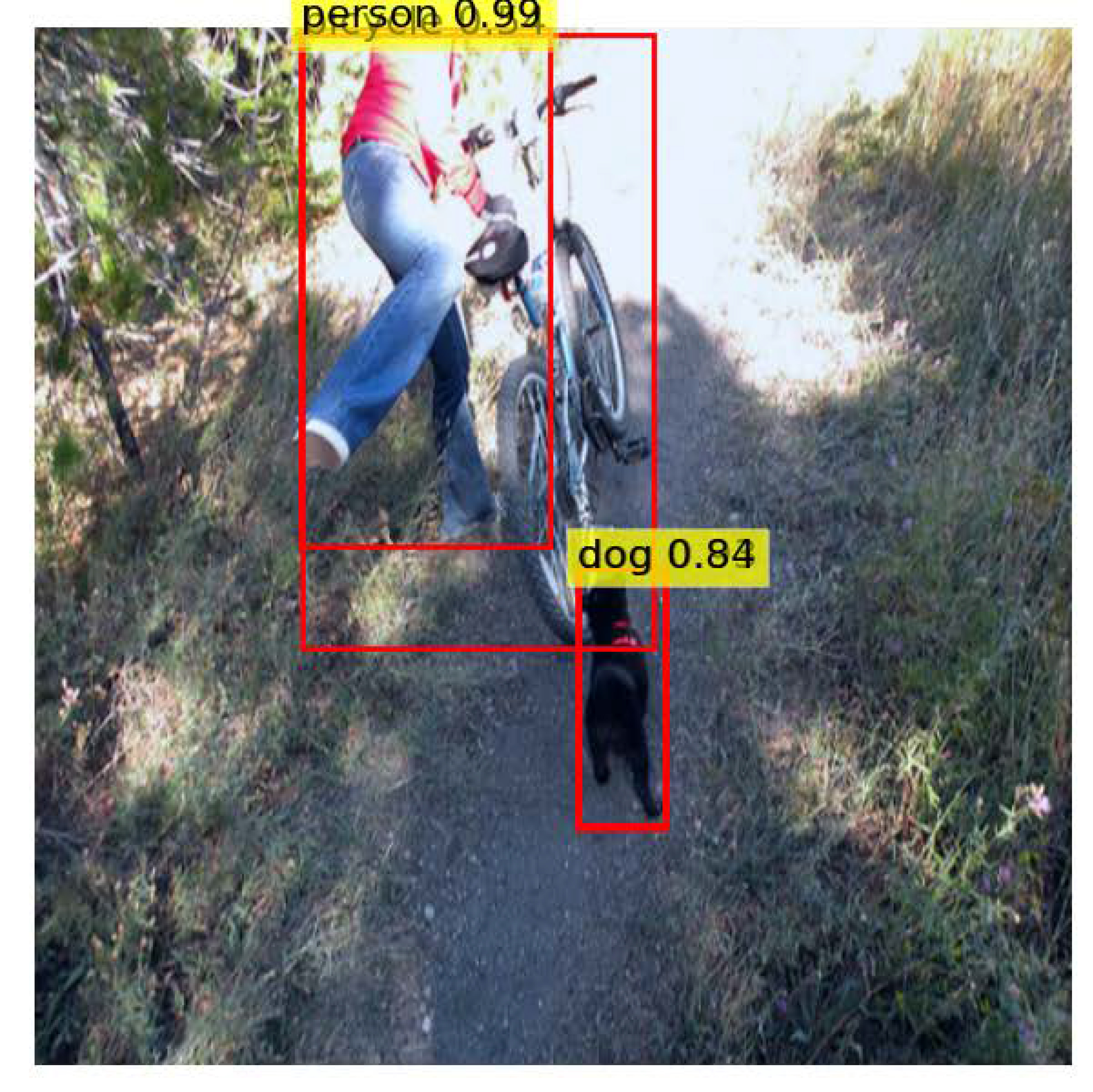}
    \caption{Clean}
\end{subfigure}
\hfill
\begin{subfigure}{0.28\columnwidth}
    \centering
    \includegraphics[width=\linewidth]{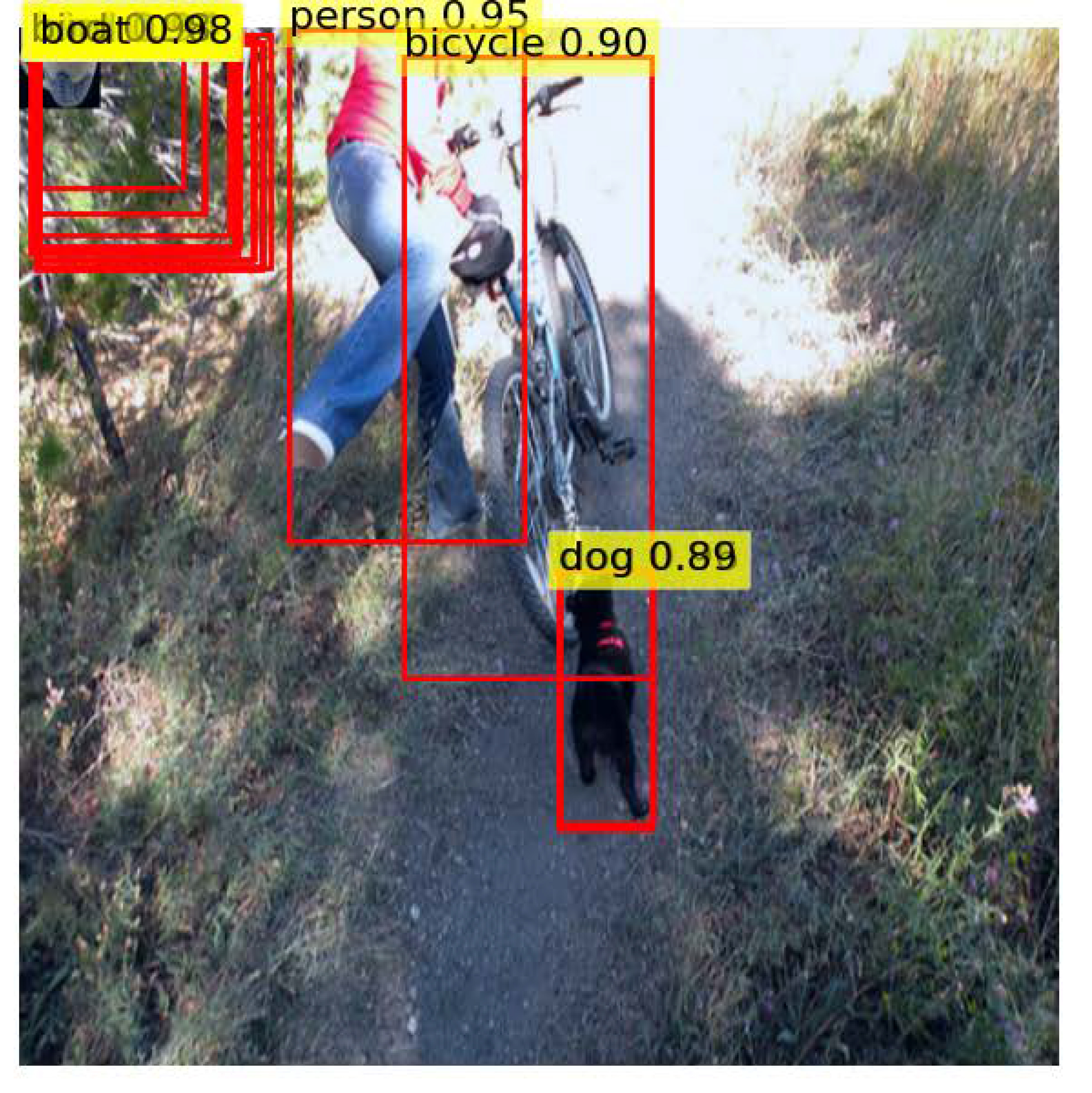}
\caption{TAL(0,0)}
\end{subfigure}
\hfill
\begin{subfigure}{0.28\columnwidth}
    \centering
    \includegraphics[width=\linewidth]{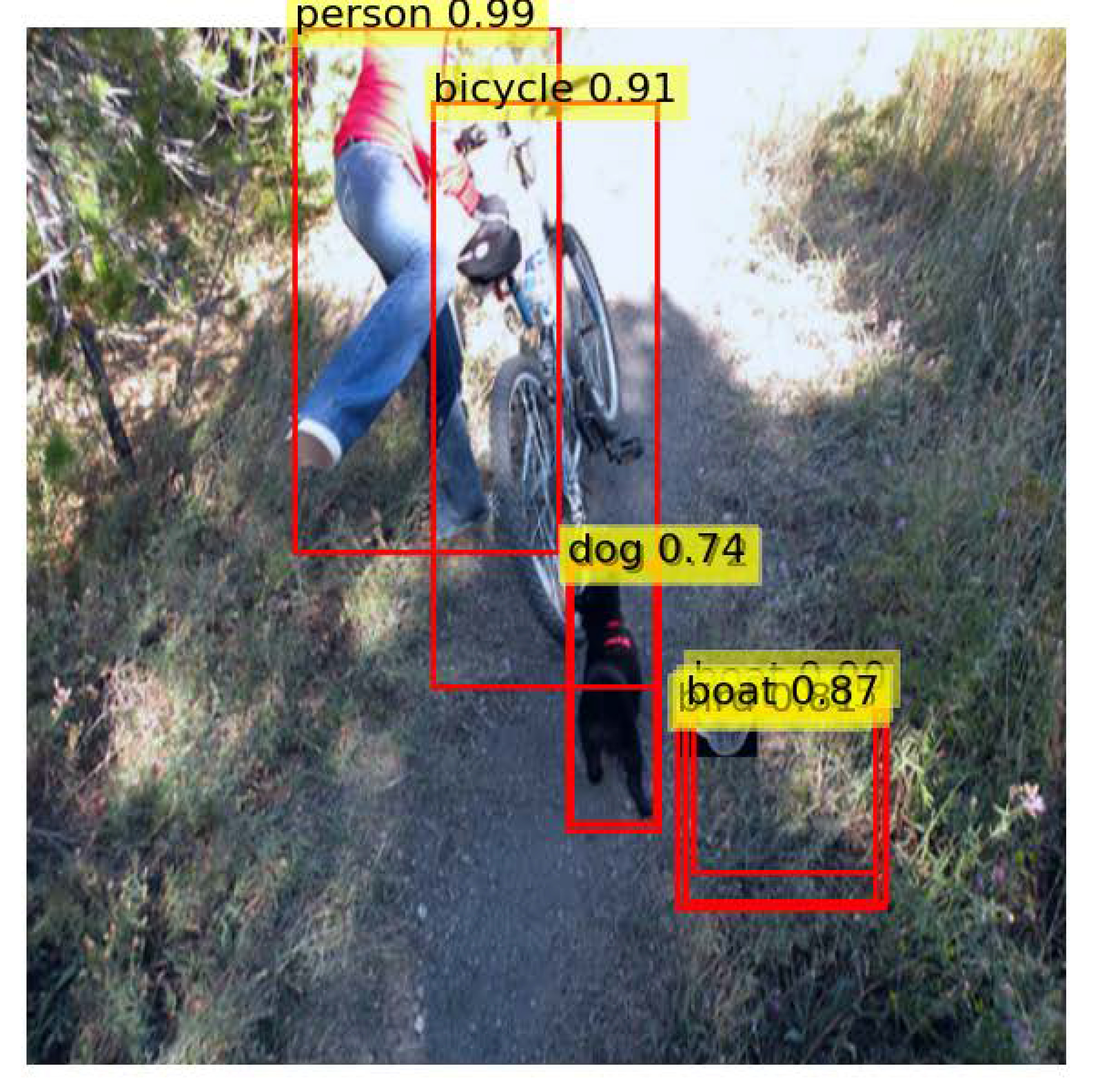}
\caption{TAL(400,400)}
\end{subfigure}

\caption{Visualization of detection results on clean and poisoned images under the UOGA attack with two TALs. 
(a) Clean images; 
(b) poisoned images with the TAL at (0, 0); 
(c) poisoned images with the TAL at (400, 400). 
Detected objects are highlighted with red boxes, and the predicted labels are shown in yellow.}
\label{fig:od:attack_visualization_UOGA}
\end{figure}

We further verify attack robustness across TALs using UOGA. We select TALs at (0,0) and (400,400) and show results in Figures~\ref{fig:od:attack_visualization_UOGA}(a)--(c). In clean images, ``Person'', ``Bicycle'', and ``Dog'' are correctly detected and classified. 
When the trigger is inserted at either location, hallucinated objects appear around the trigger region, while non-target objects remain correctly detected. This shows that benign detection performance is preserved despite localized hallucinations.

\subsection{Ablation Study}
\label{sec:ablation_study}

\noindent\textbf{Impact of Target Label to Attack Effectiveness.}
This work evaluates DETOUR on six attack tasks summarized in \Cref{tab:attack_goals}, using ``Person'' as the default target label. To verify robustness across labels, we further test ``Bicycle'' and ``MotorBike'' as alternative targets and report benign accuracy and attack performance in \Cref{tab:od:ablation:targetlabel_asr}.
Results show consistent benign performance across target labels, with variations under 3.0\% across all metrics. ASR and TRE also remain stable, indicating that DETOUR achieves consistent effectiveness regardless of the target label (see \Cref{appx:od:analysis_impact_attack_labels} for more details).



\subsection{Resource Usage of DETOUR Backdoor Attack}

\begin{table}[t]
\centering
\caption{Training time for clean and poisoned models under the targets of ``Person'' and ``Motorbike''.}
\scalebox{0.7}{
\begin{tabular}{@{}ccccccc@{}}
\toprule
\multirow{2}{*}{Method} & \multirow{2}{*}{Target} & \multicolumn{5}{c}{Time (s) by Epochs} \\ \cmidrule(l){3-7} 
 &  & 10 & 20 & 30 & 40 & 50 \\ \midrule
Clean & - & 763.76 & 1544.32 & 2331.28 & 3101.71 & 3868.47 \\
DETOUR & Person & 775.53 & 1555.12 & 2335.73 & 3106.80 & 3875.95 \\
DETOUR & Motorbike & 777.49 & 1560.57 & 2338.62 & 3110.06 & 3893.69 \\
\bottomrule
\end{tabular}
}

\label{tab:od:time_consumption}
\end{table}

To assess the practical overhead introduced by DETOUR, we evaluate the computational overhead of our backdoor attack in terms of both training time and memory consumption across different numbers of epochs, and report the results in \Cref{tab:od:memory_overhead,tab:od:time_consumption}, respectively.
Based on our results, the proposed DETOUR backdoor attack achieves effective attack performance while incurring minimal additional computational and memory overhead. 
The negligible increase in both training time and memory consumption demonstrates that DETOUR is computationally efficient and practically deployable.
We provide a detailed analysis of the resource usage of DETOUR in \Cref{appx:od:resource_usage}.

\begin{table*}[t]
\centering
\caption{The peak RAM and GPU memory (in GB) usage for clean and poisoned models (with target labels ``Person'' and ``Motorbike'') across different number of epochs.}
\label{tab:od:memory_overhead}
\setlength{\tabcolsep}{4pt}
\scalebox{0.7}{\begin{tabular}{@{}cccccccccccc@{}}
\toprule
\multirow{2}{*}{Method} & \multirow{2}{*}{Target} 
& \multicolumn{2}{c}{10 epochs}
& \multicolumn{2}{c}{20 epochs}
& \multicolumn{2}{c}{30 epochs}
& \multicolumn{2}{c}{40 epochs}
& \multicolumn{2}{c}{50 epochs} \\
\cmidrule(l){3-12}
& 
& RAM & GPU
& RAM & GPU
& RAM & GPU
& RAM & GPU
& RAM & GPU \\
\midrule
Clean & - 
& 1.56 & 8.96
& 1.56 & 8.98
& 1.56 & 8.98
& 1.56 & 8.98
& 1.56 & 8.98 \\

DETOUR & Person 
& 1.56 & 9.01
& 1.56 & 9.01
& 1.56 & 9.01
& 1.56 & 9.01
& 1.56 & 9.01 \\

DETOUR & Motorbike 
& 1.56 & 9.03
& 1.56 & 9.06
& 1.56 & 9.06
& 1.56 & 9.06
& 1.56 & 9.30 \\
\bottomrule
\end{tabular}}
\end{table*}

\section{Conclusion}
\label{sec:conclusion}

This paper proposes DETOUR, a practical backdoor attack for object detection that satisfies three real-world requirements: physical printability, location-independent activation, and robustness across scales and FOVs. 
Leveraging the radiating effect of patch-based triggers in DETR, DETOUR performs backdoor training with multiple trigger locations, sizes, and replacement-based injection to improve TRE. 
We further use MFoV of a real-world mug as the trigger, enabling holistic representation learning and reliable activation across viewpoints. 
Experiments show DETOUR meets our attack requirements while maintaining consistent attack performance.

\bibliographystyle{splncs04}
\bibliography{reference_new}

\section*{Appendix}
\renewcommand{\theHsection}{\Alph{section}}

\appendix
\crefalias{section}{appendix}
\crefalias{subsection}{appendix}

\section{Ethical Consideration}
\label{appx:od:ethical}
This work investigates the vulnerability of object detection models to practical backdoor attacks and may inspire future research on improving the security of deep learning systems. 
In this sense, our research contributes to advancing AI safety research.
In the following, we discuss the intellectual property, intended usage, risk control and human subject.
\\
\textbf{Intellectual property.}  
All compared attack methods, object detection models, datasets, and implementation libraries used in this work are publicly available. The datasets are carefully anonymized and considered to be well de-identified. We adhere strictly to all relevant licenses and use the resources solely for academic research purposes.\\
\textbf{Intended Usage.} 
We reveal that current object detection models are vulnerable to practical backdoor attacks deployed in real-world. 
We hope our findings can assist researchers in evaluating the robustness of their own models and further encourage the development of more resilient defenses against backdoor attacks.\\
\textbf{Risk Control.} 
To help reduce potential risks, we plan to release the implementation code associated with this study. 
We believe that providing open access to our code can enhance transparency, promote responsible usage, and support future research on improving the security of object detection and other deep learning applications.\\
\textbf{Human Subject.} 
This study does not involve human subjects. 
All evaluations are conducted using computational models and quantitative metrics, eliminating the need for human participation.

\section{DETOUR Workflow}
\label{appx:detour_workflow}

Taking our practical trigger design and training strategies into account, we formulate our backdoor training workflow.
We describe the workflow of our backdoor training in \Cref{alg:alg_ODbackdoor_topLevel}.
In specific, we fine-tune the pretrained OD model by repeating the backdoor training process for $EPS$ epochs, as shown in line 1.
Within each training epoch, we partition the training dataset $\mathcal{D}$ into a clean subset $\mathcal{D}_{\mathrm{cln}}$ and a poisoned subset $\mathcal{D}_{\mathrm{bd}}$ according to the predefined poisoning ratio $\rho$, as described from lines 2 to 4.
Then, we sample a FoV of trigger pattern from the distribution $\mathcal{P}_\tau$, in line 5.
We produce the poisoned dataset $\mathcal{D}_{\mathrm{bd}}$ from lines 6 to 11.
In specific, we first sample the rescale factor $s$ and trigger insertion location $\ell$ in lines 7 and 8 respectively, with our scale-invariant trigger injection mechanism and the strategy of maximizing TRE.
We transform each image from $\mathcal{D}_{bd}$ into processed subset $\mathcal{D}_{bd}^{ep}$ with our trigger transformation function $\operatorname{Trans}$ (as described in \Cref{eq:od:trans_func}).

We use the clean subset $\mathcal{D}_{\mathrm{cln}}$ and the poisoned subset $\mathcal{D}_{\mathrm{bd}}^{ep}$ to fine-tune the OD model $\mathcal{M}$ parameterized by $\theta$.
Specifically, we compute the loss over $\mathcal{D}_{\mathrm{bd}}^{ep}$ (lines 12–13), which consists of two components: the object classification loss $\mathcal{L}_{\text{cls}}$ that enforces the predicted category labels to match the attack targets, and the bounding box regression loss $\mathcal{L}_{\text{box}}$ that aligns the predicted box coordinates with the corresponding target boxes.
Note that the boxe coordinates and target labels are crafted differently according to specific attack goals.
See \Cref{tab:attack_goals} for a detailed description of all attack goals and \Cref{sec:od:appx:label_funcs} for the corresponding label functions.
Similarly, we repeat the loss calculation for clean subset $\mathcal{D}_{cln}$, in lines 14 and 15, which enforces the predicted category labels to match the ground-truth annotations from the dataset and predicted bounding boxes of all objects close to the annotated locations.
Finally, we sum up all the loss terms obtained from lines 12 to 15, and update the parameters of model $\mathcal{M}$ with the gradients from the overall loss value in line 16.
We update the model parameters $\theta$ in line 17 with the learning rate of $lr$.
After $EPS$ training epochs, we return the optimal victim model $\mathcal{M}_{\text{bd}}^*$ parameterized with the final-updated $\theta$.

\begin{algorithm}[htbp]
	\caption{the Workflow of \emph{DETOUR} Backdoor Attack}
	\begin{algorithmic}[1]
		\Require{training dataset $\mathcal{D}$, poison ratio $r$,  number of training epochs $EPS$, OD model architecture $\mathcal{M}_{c}$ and parameters $\theta$, multi-field-of-view trigger set $\mathcal{T}$, trigger sampling distribution $\mathcal{P}_{\mathcal{T}}$, scale transformation operation $\mathcal{P}_{\mathcal{S}}$, trigger transformation function $\operatorname{Trans}$.
		}
		\Ensure{Victim OD model $\mathcal{M}_{bd}^*$}
		\For{$ep$ in $\{0,1,\cdots,EPS-1\}$}
		\State $\mathcal{D}_{bd}=Split(\mathcal{D},\rho)$
            \State $\mathcal{D}_{cln}$=$\mathcal{D}\textbackslash \mathcal{D}_{bd}$
            \State $D_{bd}^{ep}$=\{$\emptyset$\}
                \State t $\sim \mathcal{P}_{\mathcal{T}}$
			\For{($x$,$y$) in $D_{bd}$}
                    \State $s \sim \mathcal{P}_{\mathcal{S}}(\xi_{low}, \xi_{upp})$
				\State $\ell \sim\mathcal{U}(u_{low},u_{upp}, v_{low}, v_{upp})$
				\State  $t' = \operatorname{Trans}(t, s, \ell, H, W)$,
                    \State (x',y')$\leftarrow$($\mathcal{T}(x, t')$, $\eta$($y$))
				\State $\mathcal{D}_{bd}^{ep}$=$D_{bd}^{ep}\cup$($x'$,$y'$)
			\EndFor
			\Statex \ \ \ \ \underline{Backdoor Loss}

                \State $\mathcal{L}_{cls|\mathcal{D}_{bd}^{ep}}$=$\sum_{(x', \mathcal{Y}^{\mathrm{bd}}) \in \mathcal{D}_{\mathrm{bd}}^{ep}}
\sum_{i=1}^{|\mathcal{Y}^{\mathrm{bd}}|}
\Big(
\mathcal{L}_{\text{cls}}\big(\hat{y}_{\sigma_{x'}(i)}, y_i^{\mathrm{bd}}\big)$
                \State $\mathcal{L}_{box|\mathcal{D}_{bd}^{ep}}$=$\sum_{(x', \mathcal{Y}^{\mathrm{bd}}) \in \mathcal{D}_{\mathrm{bd}}^{ep}}
\sum_{i=1}^{|\mathcal{Y}^{\mathrm{bd}}|}\lambda_{\text{box}} \,
\mathcal{L}_{\text{box}}\big(\hat{b}_{\sigma_{x'}(i)}, b_i^{\mathrm{bd}}\big)$
                \Statex \ \ \ \ \underline{Benign Loss}
                \State $\mathcal{L}_{cls|\mathcal{D}_{cln}}$=$\sum_{(x, \mathcal{Y}) \in \mathcal{D}_{\mathrm{cln}}}
                    \sum_{i=1}^{|\mathcal{Y}|}
                    \mathcal{L}_{\text{cls}}\big(\hat{y}_{\sigma_x(i)}, y_i\big)$
                \State $\mathcal{L}_{box|\mathcal{D}_{cln}}$=$\sum_{(x, \mathcal{Y}) \in \mathcal{D}_{\mathrm{cln}}}  \sum_{i=1}^{|\mathcal{Y}|} \lambda_{\text{box}} \,
                \mathcal{L}_{\text{box}}\big(\hat{b}_{\sigma_x(i)}, b_i\big)
                $
                \Statex \ \ \ \ \underline{Overall Loss}
                \State $\mathcal{L}(\theta)$=$\mathcal{L}_{cls|\mathcal{D}_{cln}}$+$\mathcal{L}_{box|\mathcal{D}_{cln}}$+$\mathcal{L}_{cls|\mathcal{D}_{bd}}$+$\mathcal{L}_{box|\mathcal{D}_{bd}}$
                \Statex \ \ \ \ \underline{Parameter Updates}
			\State $\theta$=$\theta$-$lr\times \nabla_{\theta}$($\mathcal{L(\theta)}$)
		\EndFor
            \State $\mathcal{M}_{bd}^*=\mathcal{M}_c\leftarrow \theta$
		\State \Return $\mathcal{M}_{bd}^*$
	\end{algorithmic}
	\label{alg:alg_ODbackdoor_topLevel}
\end{algorithm}


\newcolumntype{C}[1]{>{\centering\arraybackslash}m{#1}}
\begin{table}[t]
\centering
\caption{Summary of the attack goals across 6 backdoor attack tasks in object detection.}
\label{tab:attack_goals}
\begin{tabular}{@{}ccm{6cm}@{}}
\toprule
\textbf{Targeting} & \textbf{\underline{A}ttack Tasks} & \textbf{Attack Goals} \\ 
\midrule
\multirow{8}{*}{\underline{T}argeted} 
& \underline{M}isclassification (TMA) 
& Manipulate the model to classify objects from a specified category as a predefined attacker-desired label \\ 
\cmidrule(l){2-3}

& \underline{D}isappearance (TDA) 
& Ensure the model fails to recognize all objects from the targeted category while leaving other predictions intact \\ 
\cmidrule(l){2-3}

& \underline{G}eneration (TGA) 
& Recognize the trigger as a series of objects with attacker-specified labels \\ 
\midrule

\multirow{8}{*}{\underline{U}ntargeted} 
& \underline{M}isclassification (UMA) 
& Misclassify each label to a distinct label among all ground-truth labels in a one-to-one cyclic mapping. \\ 
\cmidrule(l){2-3}

& \underline{D}isappearance (UDA) 
& Suppresses all object detections within an image, causing the model to output no predictions. \\ 
\cmidrule(l){2-3}

& \underline{G}eneration (UGA) 
& Recognizes the trigger as a series of objects with random labels. \\ 
\bottomrule
\end{tabular}
\end{table}

\section{Detailed Analysis of DETOUR Resource Usage}
\label{appx:od:resource_usage}
We first analyze the memory overhead of DETOUR to verify whether the additional backdoor-related operations introduce extra memory burden. 
Specifically, we measure both peak RAM and GPU memory usage across training epochs from 10 to 50 with a step size of 10 epochs. 
The results are summarized in \Cref{tab:od:memory_overhead}. 
As shown in the table, the peak RAM consumption remains constant at 1.56 GB for both clean and poisoned models (under both ``Person'' and ``Motorbike'' targets), indicating that DETOUR does not introduce additional CPU memory overhead. 
Regarding GPU memory, the increase is marginal. 
For example, at 50 epochs, the clean model consumes 8.98 GB of GPU memory, while DETOUR requires at most 9.30 GB (under the ``Motorbike'' target), corresponding to an increase of less than 0.4 GB. 
Overall, the results demonstrate that DETOUR maintains nearly identical memory usage to clean training.

We then evaluate the training time to examine whether DETOUR significantly slows down model optimization. 
We record the cumulative training time between 10 and 50 epochs with a step size of 10 epochs for both clean and poisoned settings. 
The results are presented in \Cref{tab:od:time_consumption}. 
The results show that DETOUR introduces only a negligible increase in training time. For instance, after 50 epochs, clean training requires 3868.47 s, while DETOUR takes 3875.95 s (Person) and 3893.69 s (Motorbike), corresponding to a relative increase of approximately 0.2\% to 0.7\%. 
Moreover, the time growth remains approximately linear with respect to the number of epochs, suggesting that DETOUR does not alter the overall training complexity.

\section{Impact of Target Labels on Attack Effectiveness}
\label{appx:od:analysis_impact_attack_labels}
In terms of attack effectiveness, the victim models corresponding to the three target labels also consistently achieve similar ASR and TRE.
Among these models, the victim model targeting ``Motorbike'' attains the highest ASR of 85.2507\%, while the model poisoned with ``Person'' as the target label yields the lowest ASR of 84.7241\%, resulting in a marginal difference of only 0.5266\% across target labels.
The same phenomenon is observed for TRE on DETOUR, where the victim model targeting ``Motorbike'' achieves the highest TRE of 86.0685\%, while the model poisoned with ``Bottle'' as the target label yields a TRE of 86.8007\%, resulting in a minor difference 0.73\%. 
We attribute the minor variation in ASR and TRE to the imbalance in the number of objects across different target labels in the dataset. For example, the total number of objects labeled as ``Person'' reaches 12,608, whereas the number of objects labeled as ``Motorbike'' is only 653.
Such a significant disparity in the number of labeled objects within the dataset may influence the backdoor learning process, leading to minor variations in attack performance measured by ASR and TRE. 

\section{Evaluation Metrics}
\label{appx:evaluation_metrics}
The mAP metric measures the proportion of objects that are correctly detected, accurately localized, and correctly classified. Details of its computation and the IoU thresholds used for evaluation are provided below.

\noindent\textbf{Intersection over Union (IoU).}
The $IoU$ is a standard metric to quantify the spatial overlap between a predicted bounding box $B_p$ and a ground-truth bounding box $B_{gt}$:

\[
\text{IoU}(B_p, B_{gt}) = \frac{\text{Area of } (B_p \cap B_{gt})}{\text{Area of } (B_p \cup B_{gt})}.
\]

In practice, the intersection area is computed from the overlapping region of the two boxes, while the union area corresponds to the total area covered by both boxes. For axis-aligned rectangular boxes, the intersection region is determined by the maximum of the top-left corners and the minimum of the bottom-right corners. The resulting $IoU$ ranges from 0 to 1, with higher values indicating greater overlap and more accurate localization.

\noindent\textbf{Mean Average Precision (\emph{mAP})} reflects the victim models' functionality on \emph{clean samples} by measuring how accurately it detects objects across all classes. 
$mAP$ jointly evaluating precision and recall across all object classes. 
For a given class, precision \emph{P} and recall \emph{R} are defined as:
\begin{equation}
P = \frac{TP}{TP + FP}, \quad R = \frac{TP}{TP + FN},
\end{equation}
where \emph{TP} denotes true positives, which counts the predicted objects that correctly match a ground-truth box with $IoU$ exceeding a predefined threshold and the correct class label, with each ground-truth box matched to at most one prediction; 
\emph{FP} is the false positives, representing the predictions that either do not match any ground-truth box or have an incorrect class, including duplicate predictions for the same ground-truth box; 
and \emph{FN} reflects the false negatives, which counts the ground-truth boxes not detected by any prediction.

By ranking all predicted bounding boxes according to their confidence scores and computing precision (\emph{P}) and recall (\emph{R}) at varying thresholds, the precision–recall (P–R) curve can be constructed. The Average Precision (AP) for a given class is then defined as the area under this curve:
\begin{equation}
AP = \int_0^1 P(R) \, dR.
\end{equation}

Building upon $AP$, the \emph{mAP} metric aggregates the AP values across all $N$ object classes:
\begin{equation}
mAP = \frac{1}{N} \sum_{i=1}^{N} AP_i,
\end{equation}
this metric reflects the model's detection performance under a specified localization accuracy criterion.
Since $mAP$ is computed based on \emph{TP}, which in turn depend on a predefined IoU threshold $IoU_t$, we denote the mAP evaluated at this threshold as \emph{mAP}@$IoU_t$, and the \emph{mAP} based on TP within the lower bound $IoU_t^{Low}$ and upper bound $IoU_t^{Upp}$ as $mAP$@\emph{mAP}@$IoU_t^{Low}$: \emph{mAP}@$IoU_t^{Upp}$. 
In our experiments, we report the benign accuracy by $mAP$ under 3 widely adopted $IoU$ threshold: $mAP$@50, $mAP$@50:95 and $mAP$@95. 

\noindent\textbf{Attack Effectiveness} is quantified using the attack success rate (ASR) and TRE.
ASR computes the percentage of objects that are predicted by the poisoned OD model as desired by the attacker, given that the predicted bounding boxes correctly localize the attacked objects, when the trigger is inserted at the default TAL (i.e., the top-left corner of the poisoned image).
TRE is defined as the ASR averaged across all TALs over the entire image (see \Cref{eq:tre_quantify} for its calculation).
Due to differences among attack goals (summarized in \Cref{tab:attack_goals} in the Appendix), the detailed computation of ASR for each goal is provided in \Cref{sec:od:appx:asrCalcs}.

\section{Impact of Trigger Insertion Methods to TRE.}
\label{appx:od:impact_trigger_insertion_methods_to_tre}
\begin{figure*}[]
\centering

\begin{minipage}[t]{0.98\textwidth}
  \centering
  \begin{minipage}[t]{0.22\textwidth}
    \centering
    \includegraphics[width=\linewidth]{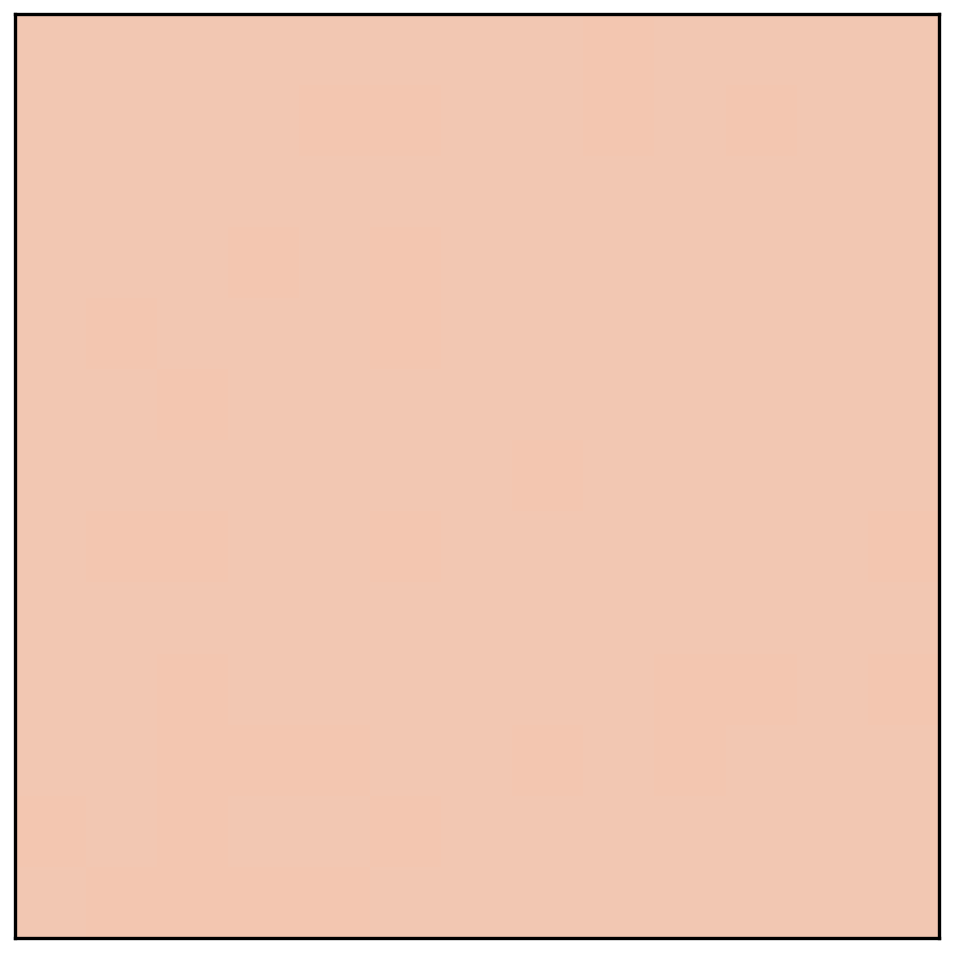}
    \caption*{\small (a)Size10,$l_2$:10, TRE61.2}
  \end{minipage}\hfill
  \begin{minipage}[t]{0.22\textwidth}
    \centering
    \includegraphics[width=\linewidth]{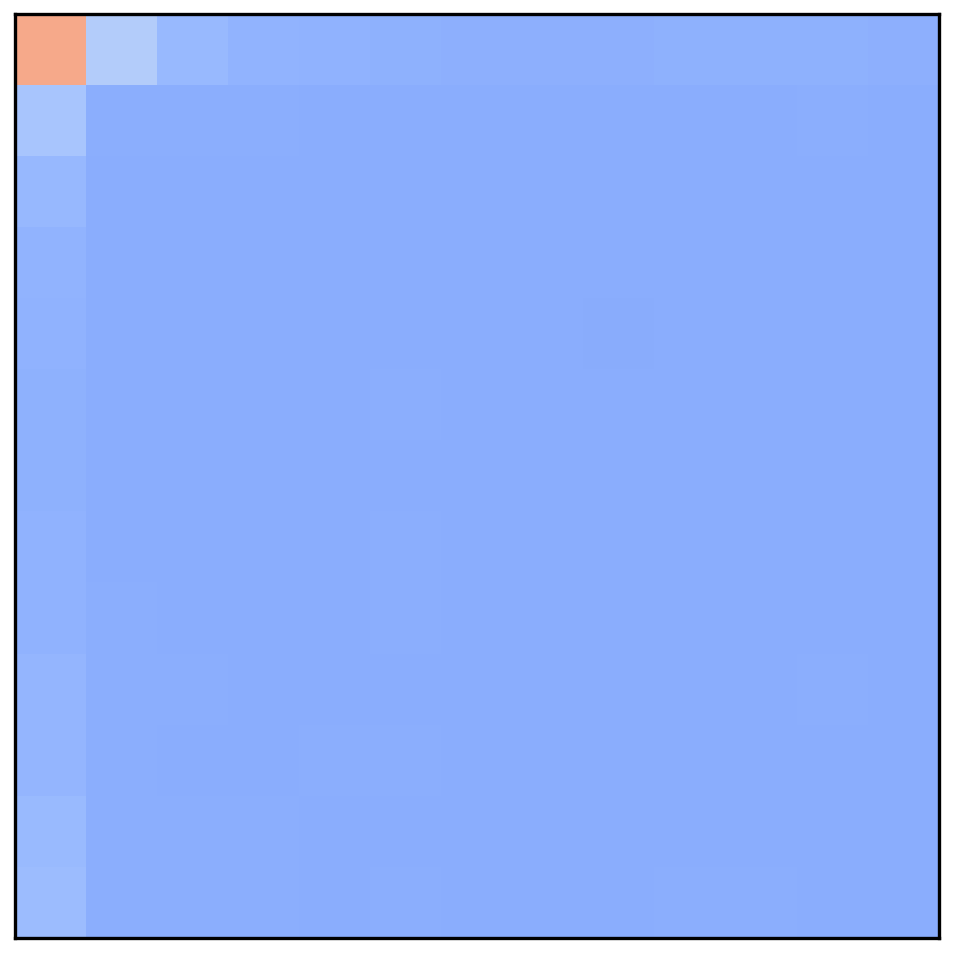}
    \caption*{\small (b)Size10, $l_2$30,TRE25.1}
  \end{minipage}\hfill
  \begin{minipage}[t]{0.22\textwidth}
  	\centering
  	\includegraphics[width=\linewidth]{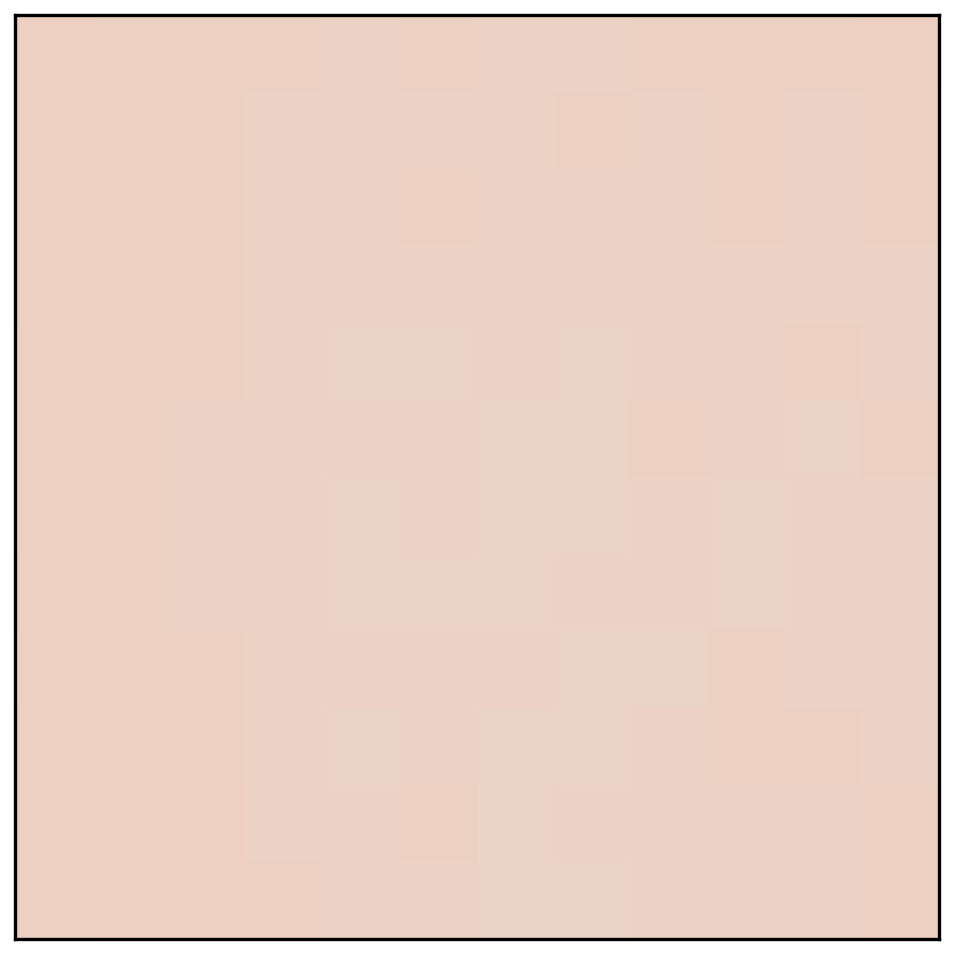}
  	\caption*{\small(c)Size50,$l_2$:100, TRE56.5}
  \end{minipage}\hfill
  \begin{minipage}[t]{0.22\textwidth}
  	\centering
  	\includegraphics[width=\linewidth]{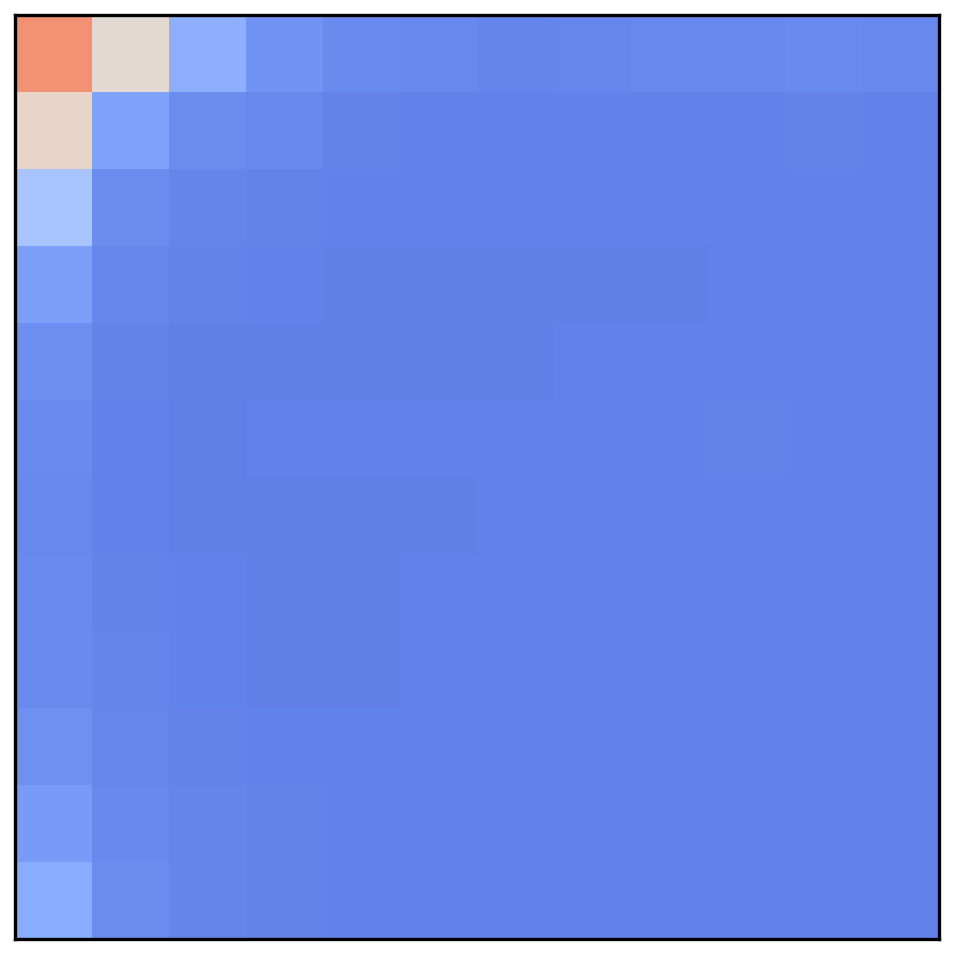}
  	\caption*{\small (d)Size50,$l_2$:400, TRE14.5}
  \end{minipage}
\end{minipage}%
\hfill
\begin{minipage}[t]{0.98\textwidth}
  \centering
   \begin{minipage}[t]{0.22\textwidth}
  	\centering
  	\includegraphics[width=\linewidth]{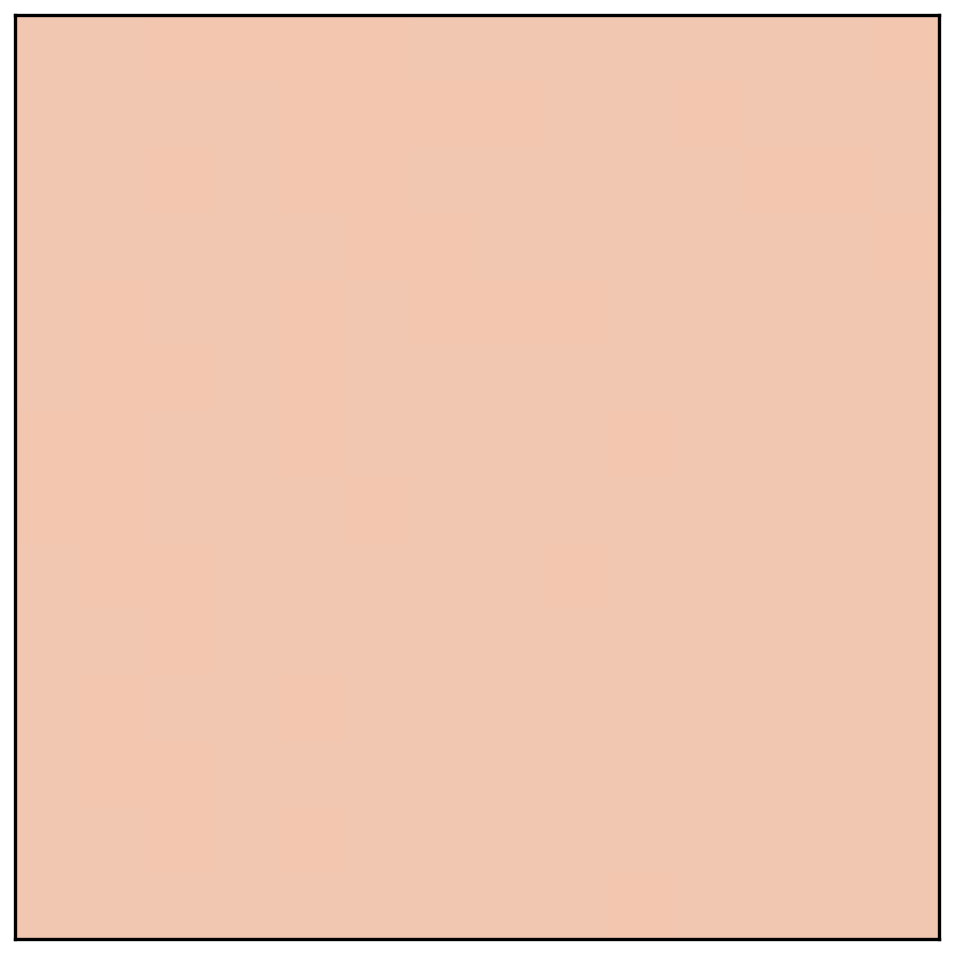}
  	\caption*{\small (e)Size10,$l_2$:10, TRE61.2}
  \end{minipage}\hfill
  \begin{minipage}[t]{0.22\textwidth}
  	\centering    \includegraphics[width=\linewidth]{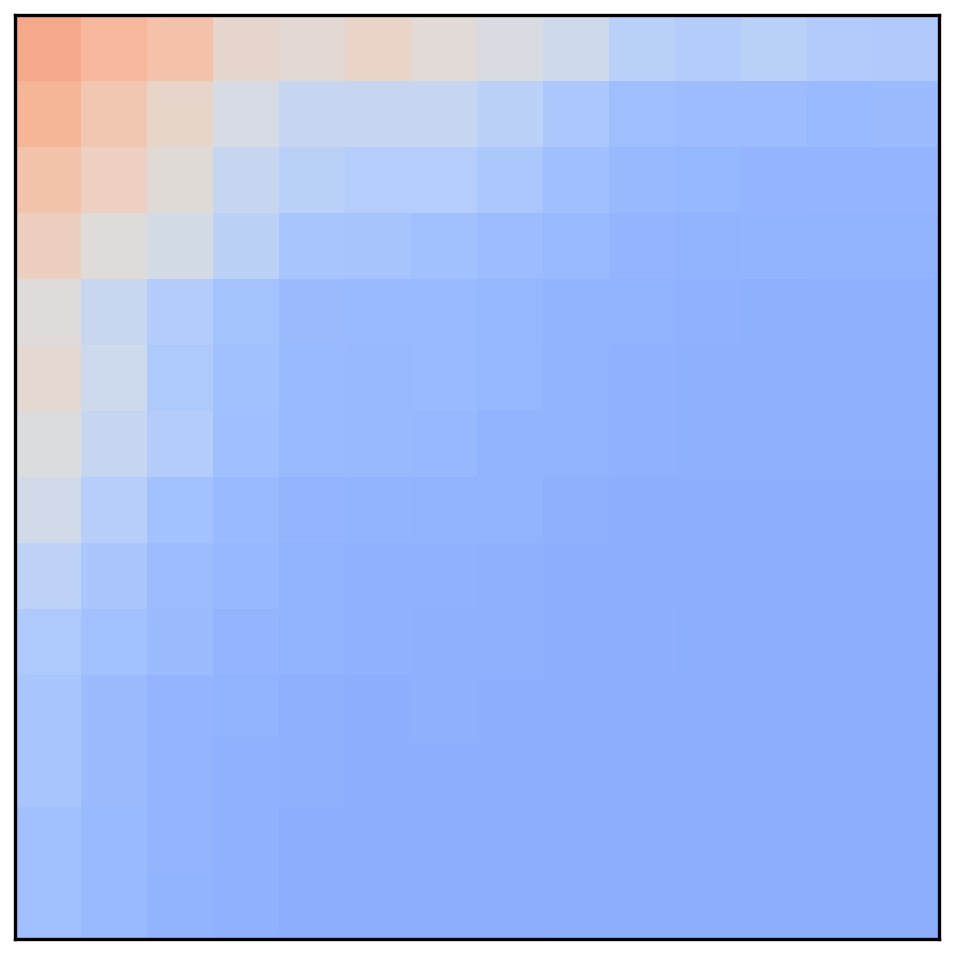}
  	\caption*{\small (f)Size10,$l_2$:30, TRE31.5)}
  \end{minipage}\hfill
  \begin{minipage}[t]{0.22\textwidth}
    \centering
    \includegraphics[width=\linewidth]{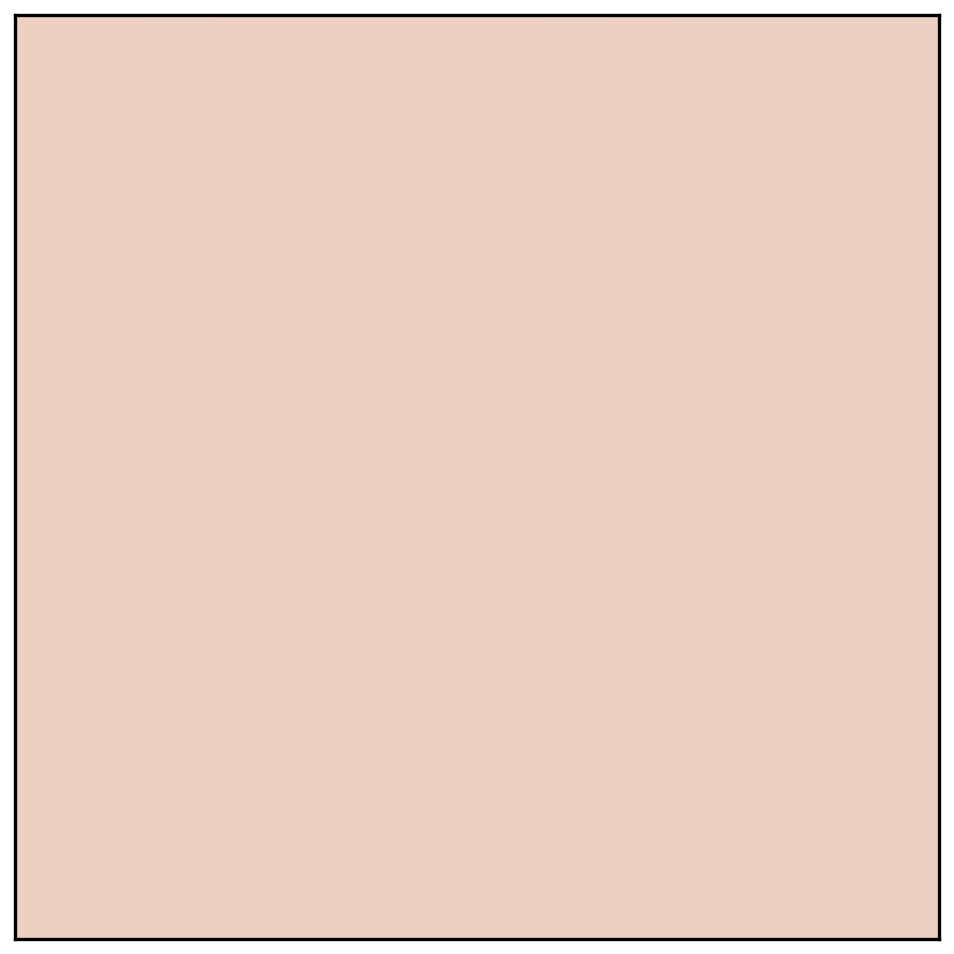}
    \caption*{\small (g)Size50,$l_2$:100, TRE57.0}
  \end{minipage}\hfill
  \begin{minipage}[t]{0.22\textwidth}
    \centering
    \scalebox{1.23}{\raisebox{-2.8pt}{\includegraphics[width=\linewidth]{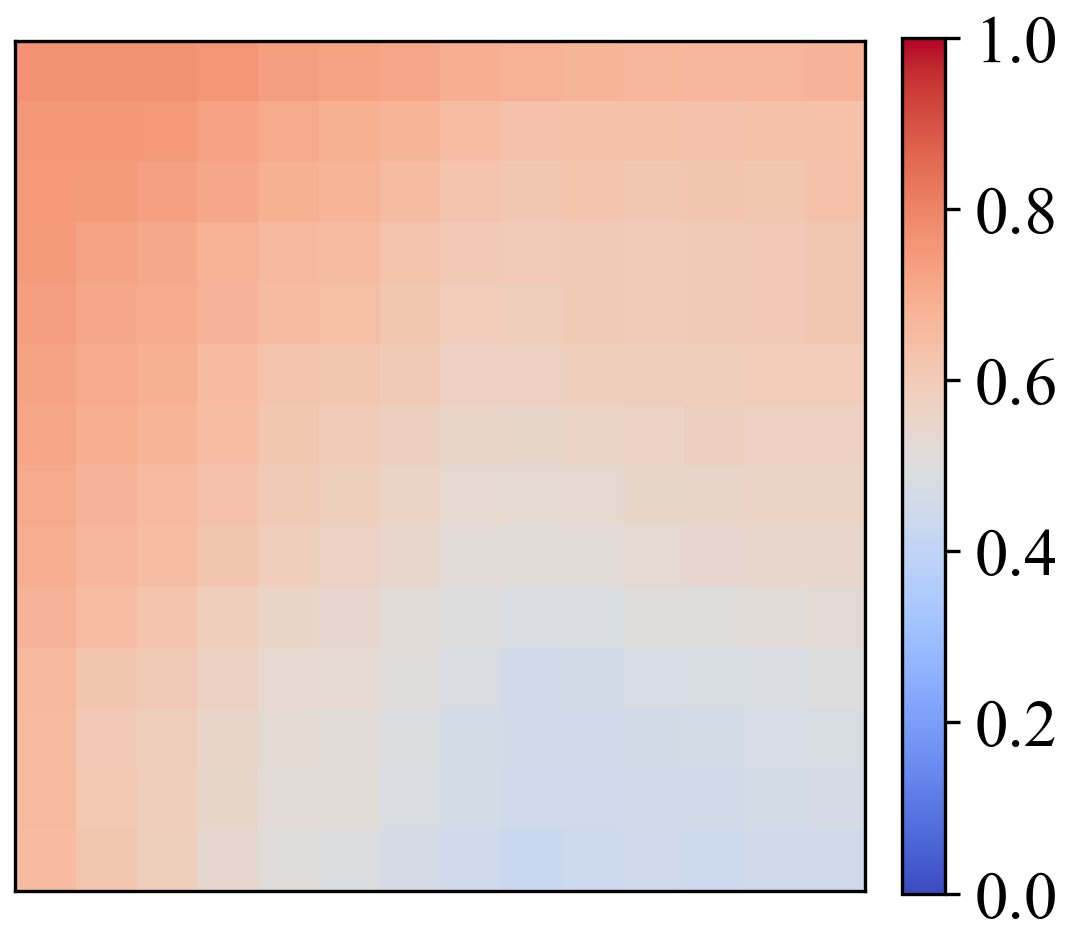}}}
    \caption*{\small (h)Size50,$l_2$:400, TRE59.4}
  \end{minipage}
\end{minipage}

\caption{Visualization of TRE heatmaps under superimpose-based (SUP) trigger insertion method in DETR. The magnitude of trigger perturbations and TRE(\%) of the victim model are denoted below each heatmap. 
Figures (a)-(b): TRE under a trigger size of 10, with perturbation magnitudes of 10 and 30, applied to the top-left corner of the image. 
Figures (c)-(d): TRE under a trigger size of 50, with perturbation magnitudes of 100 and 400, applied to the top-left corner of the image. 
Figures (e)-(f): TRE heatmap within an $80\times80$ subregion at the top-left corner under the victim model used in Figures (a)-(b). 
Figures (g)-(h):TRE heatmap within an $80\times80$ subregion at the top-left corner under the victim model used in Figures (e)-(f).
}
\label{fig:od:observations:impact_trigger_inswrtion_tre}
\end{figure*}

Besides REP, the superimpose-based trigger insertion method (SUP) is also commonly used in backdoor attacks~\cite{lira,ladder,anywheredoor,refool}.
SUP adds trigger perturbations to the pixel values of the original image within the trigger insertion region, thereby enhancing the visual stealthiness of poisoned images. 
We investigate whether the SUP-based trigger insertion method impacts TRE by repeating the same set of experiments previously conducted with REP-based trigger insertion.
Specifically, we evaluate triggers of size $10 \times 10$ with $\mathcal{L}_2$-norms normalized to 10 and 30, as well as triggers of size $50 \times 50$ with $\mathcal{L}_2$-norms normalized to 100 and 400, using the top-left corner as the TIL.  
The results are shown in \Cref{fig:od:observations:impact_trigger_inswrtion_tre}(a)–(d).
The corresponding TRE for the 80$\times$80 sub-region in the top-left corner is shown in \Cref{fig:od:observations:impact_trigger_inswrtion_tre}(e)–(h). 
We observe that under the SUP-based trigger insertion method, triggers with small $\mathcal{L}_2$ norms (i.e., 10 for $10 \times 10$ and 100 for $50 \times 50$ triggers, as shown in \Cref{fig:od:observations:impact_trigger_inswrtion_tre}(a) and (c)) achieve high attack effectiveness and can be activated across the image with high TRE.
In contrast, when the perturbations are larger (30 for $10 \times 10$ in \Cref{fig:od:observations:impact_trigger_inswrtion_tre}(b) and 400 for $50 \times 50$ in \Cref{fig:od:observations:impact_trigger_inswrtion_tre}(d)), the TRE is lower.

\begin{figure}
    \centering
    \includegraphics[width=0.5\linewidth]{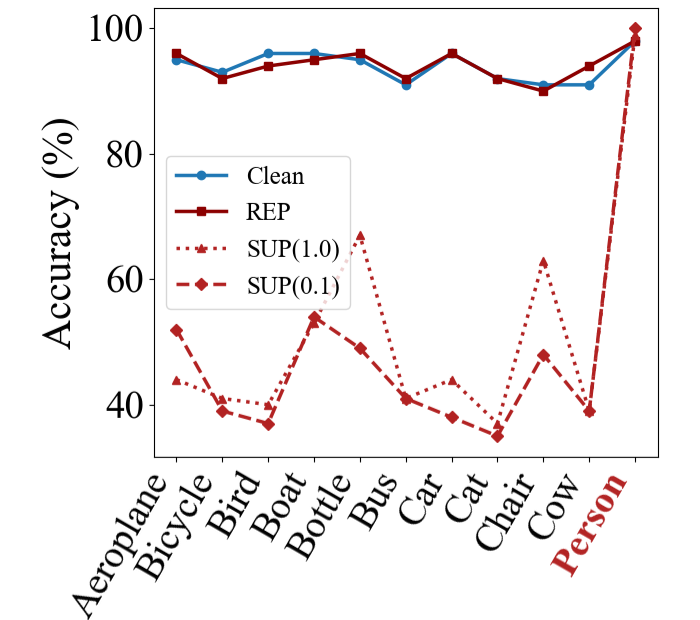}
    \caption{The inference accuracy (\%) of object labels across samples from the clean validation set is shown for the first 10 classes (in alphabetical order) and the target class (i.e., ``person'').
    Results are presented for the clean model (blue) and poisoned models (red) trained on the global misclassification task under two trigger insertion methods: REP (replacement) and SUP (superimpose, with coefficients 0.1 and 1.0, respectively).}
    \label{fig:label_acc_per_class}
\end{figure}

In fact, the high TRE achieved with SUP and small $\mathcal{L}_2$ norms of trigger perturbations degrades benign accuracy.
We present the inference accuracy of object labels on the benign validation set for both clean and poisoned models in \Cref{fig:label_acc_per_class}.
The x-axis shows the first ten classes in alphabetical order, along with the target label (``person''), while the y-axis shows the corresponding class-wise label accuracy under both REP- and SUP-based trigger insertion methods.
In the SUP method, two overlap ratios, 1.0 and 0.1, are applied.
We observe that the inference accuracy for each label on clean data remains above 90\% for both the clean model and the poisoned model trained with the REP-based trigger insertion method, indicating satisfactory performance of both models on benign tasks.
In contrast, the poisoned model trained with the SUP-based trigger insertion method achieves only 40\%--60\% label inference accuracy, regardless of the overlap parameter, suggesting a substantial degradation in benign performance for the SUP-based method.
This phenomenon arises because the minimal trigger perturbation used in SUP-based trigger insertion during backdoor training prevents the OD model from effectively capturing the trigger pattern.
As a result, the model is forced to fit the poisoned supervision by over-associating diverse benign visual patterns with the target label, rather than learning a consistent trigger–label correspondence.
This over-association introduces a strong bias toward the target class in the detector, leading to widespread misclassification of benign objects and ultimately degrading normal detection performance.
Our experiments suggest that when the trigger pattern is too weak to be reliably captured, label manipulation alone can dominate the optimization process, resulting in a significant degradation of benign performance.

\section{Task-specific Target Label Poisoning Function}
\label{sec:od:appx:label_funcs}

Below we elaborate the task-specific target label functions for 6 attack goals listed in \Cref{tab:attack_goals}.

\subsubsection{Targeted Misclassification Attack} 
is defined as a constant target label mapping over all instances:
\[
    \tau(\mathbf{x}, b_i) = y_t, \quad \forall b_i \in \mathcal{B}(\mathbf{x}),
\]
Specifically, for an input image $\mathbf{x}$, the target label function $\tau(\mathbf{x}, b_i) = y_t$ assigns the same target class $y_t$ to every bounding box $b_i \in \mathcal{B}(\mathbf{x})$, where $b_i$ denotes the $i$-th proposal or detected instance.

\subsubsection{Targeted Disappearance Attack} is defined as a selective suppression of a specific target class. For an input image $\mathbf{x}$, the target function is defined as:
\[
\tau(\mathbf{x}, b_i) = \varnothing, \quad \forall b_i \in \mathcal{B}_{y_t}(\mathbf{x}),
\]
where $\mathcal{B}_{y_t}(\mathbf{x})$ denotes the set of bounding boxes corresponding to the target class $y_t$. This formulation enforces the detector to suppress all instances of the target class, effectively causing their disappearance from the detection results.

\subsubsection{Targeted Object Generation Attack} is defined as forcing the detector to hallucinate instances of a specific target class. 
For an input image $\mathbf{x}$, the target label function is defined as
\[
\tau(\mathbf{x}, b_i) = y_t, 
\quad \forall b_i \in \mathcal{B}_{\text{gen}}(\mathbf{x}) 
:= \{ b \mid b \notin \mathcal{B}_{\text{gt}}(\mathbf{x}) \},
\]
where $\mathcal{B}_{\text{gt}}(\mathbf{x})$ denotes the set of ground-truth bounding boxes in the scene.
This formulation induces the detector to produce false positive detections of the target class $y_t$.
This formulation enforces the detector to assign the target label $y_t$ to hallucinated bounding boxes that do not correspond to any real objects.


\subsubsection{Untargeted Misclassification Attack} is defined as a deterministic label-shifting mapping over all detected object instances:
\[
    \tau(x, b_i) = (y_i + 1) \bmod \kappa, \quad \forall b_i \in \mathcal{B}(x),
\]
where $x$ denotes an input image, $b_i$ represents the $i$-th proposal or detected instance in $x$, and $y_i \in \{0,1,\dots,\kappa-1\}$ is the corresponding object class label, and $\kappa$ denotes the total number of object categories.  
Ultimately, the model is mislead to systematically shift each instance's predicted label to the next class in the label set, effectively inducing a controlled misclassification across all objects.

\subsubsection{Untargeted Disappearance Attack} is defined as an instance-removal mapping over all detected object instances:
\[
    \tau(x, b_i) = \varnothing, \quad \forall b_i \in \mathcal{B}(x),
\]
where $x$ denotes an input image and $b_i$ represents the $i$-th proposal or detected instance in $x$.  
This attack aims to suppress all object detections by forcing every instance to be removed from the detection results, thereby causing complete object disappearance in the image.

\subsubsection{Untargeted Object Generation Attack} is defined as forcing the detector to hallucinate object instances without constraining them to a specific target class.
For an input image $\mathbf{x}$, the label assignment function is defined as
\[
\tau(\mathbf{x}, b_i) = y_i', 
\quad \forall b_i \in \mathcal{B}_{\text{gen}}(\mathbf{x}) 
:= \{ b \mid b \notin \mathcal{B}_{\text{gt}}(\mathbf{x}) \},
\]
where $\mathcal{B}_{\text{gt}}(\mathbf{x})$ denotes the set of ground-truth bounding boxes in the scene, and $y_i' \in \{0,1,\dots,\kappa-1\}$ represents an arbitrary object category.
This formulation induces the detector to produce false positive detections with unconstrained class labels.
This formulation enforces the detector to assign arbitrary labels to hallucinated bounding boxes that do not correspond to any real objects.

\section{Task-specific Calculation of Attack Success Rate (ASR)}
\label{sec:od:appx:asrCalcs}

\subsubsection{Targeted Misclassification} is quantified as the proportion of poisoned images in which the detector produces at least one bounding box that is misclassified into an attacker-specified target class.
Formally, let $\mathcal{D}_t$ denote the set of test images embedded with the trigger, and let $y^*$ denote the target class.
For each image $x_i \in \mathcal{D}_t$, the detector outputs a set of predictions $\mathcal{P}_i = \{(\hat{b}_{ik}, \hat{y}_{ik})\}_{k=1}^{M_i}$,
where $\hat{b}_{ik}$ and $\hat{y}_{ik}$ denote the predicted bounding box and class label, respectively.
An attack on $x_i$ is considered successful if there exists at least one prediction
$(\hat{b}_{ik}, \hat{y}_{ik}) \in \mathcal{P}_i$
such that $\hat{y}_{ik} = y^*$ and
\begin{equation}
	\max_j \mathrm{IoU}(\hat{b}_{ik}, b_{ij}) \ge \tau,
\end{equation}
where $\{b_{ij}\}$ denotes the set of ground-truth bounding boxes in $x_i$, $\mathrm{IoU}(\cdot,\cdot)$ is the intersection-over-union metric, and $\tau$ is the IoU threshold.
The ASR is then computed as:
\begin{equation}
	\begin{aligned}
		\mathrm{ASR} =
		\frac{1}{|\mathcal{D}_t|}
		\sum_{x_i \in \mathcal{D}_t}
		\mathbb{I}\Big(
		&\exists (\hat{b}_{ik}, \hat{y}_{ik}) \in \mathcal{P}_i \ \text{s.t.} \\
		&\hat{y}_{ik} = y^* \ \land \
		\max_j \mathrm{IoU}(\hat{b}_{ik}, b_{ij}) \ge \tau
		\Big).
	\end{aligned}
\end{equation}
where $\mathbb{I}(\cdot)$ denotes the indicator function and all predictions are taken after non-maximum suppression.
Note that ground-truth objects belonging to the target class $y^*$ are not counted as successfully attacked objects.

\subsubsection{Untargeted Misclassification} aims to induce incorrect class predictions without specifying a target class.
Let $\mathcal{D}_t$ denote the set of test images embedded with the trigger.
For each image $x_i \in \mathcal{D}_t$, the detector outputs a set of predictions
$\mathcal{P}_i = \{(\hat{b}_{ik}, \hat{y}_{ik})\}_{k=1}^{M_i}$,
where $\hat{b}_{ik}$ and $\hat{y}_{ik}$ denote the predicted bounding box and class label, respectively.
An attack on $x_i$ is considered successful if there exists at least one prediction
$(\hat{b}_{ik}, \hat{y}_{ik}) \in \mathcal{P}_i$
such that:
\begin{equation}
	\hat{y}_{ik} \neq y_{ij},
	\quad
	\max_j \mathrm{IoU}(\hat{b}_{ik}, b_{ij}) \ge \tau,
\end{equation}
where $\{(b_{ij}, y_{ij})\}$ denotes the set of ground-truth bounding boxes and class labels in $x_i$, $\mathrm{IoU}(\cdot,\cdot)$ is the intersection-over-union metric, and $\tau$ is the IoU threshold.
The untargeted attack success rate (ASR) is then defined as
\begin{equation}
\begin{aligned}
\mathrm{ASR} =
\frac{1}{|\mathcal{D}_t|}
\sum_{x_i \in \mathcal{D}_t}
\mathbb{I}\Big(
&\exists (\hat{b}_{ik}, \hat{y}_{ik}) \in \mathcal{P}_i \ \text{s.t.} \\
&\hat{y}_{ik} \neq y_{ij} \ \land \
\max_j \mathrm{IoU}(\hat{b}_{ik}, b_{ij}) \ge \tau
\Big).
\end{aligned}
\end{equation}


\subsubsection{Targeted Disappearance} aims to suppress the detection of objects belonging to a specific target class.
Let $\mathcal{D}_t$ denote the set of test images embedded with the trigger, and let $y^*$ be the attacker-specified target class.
For each image $x_i \in \mathcal{D}_t$, the detector outputs a set of predictions
$\mathcal{P}_i = \{(\hat{b}_{ik}, \hat{y}_{ik})\}_{k=1}^{M_i}$,
where $\hat{b}_{ik}$ and $\hat{y}_{ik}$ denote the predicted bounding box and class label, respectively.
An attack on $x_i$ is considered successful if none of the ground-truth objects of class $y^*$ are detected, i.e.,
\begin{equation}
	\forall j \ \text{s.t.} \ y_{ij} = y^*,
	\quad
	\max_k \mathrm{IoU}(\hat{b}_{ik}, b_{ij}) < \tau,
\end{equation}
where $\{(b_{ij}, y_{ij})\}$ denotes the set of ground-truth bounding boxes and class labels in $x_i$, $\mathrm{IoU}(\cdot,\cdot)$ is the intersection-over-union metric, and $\tau$ is the IoU threshold.
The Attack Success Rate (ASR) for targeted disappearance is then defined as:
\begin{equation}
	\mathrm{ASR} =
	\frac{1}{|\mathcal{D}_t|}
	\sum_{x_i \in \mathcal{D}_t}
	\mathbb{I}\Big(
	\forall j \ \text{s.t.} \ y_{ij} = y^*,
	\ \max_k \mathrm{IoU}(\hat{b}_{ik}, b_{ij}) < \tau
	\Big).
\end{equation}

\subsubsection{Global Disappearance} aims to suppress the detection of \emph{all} objects in an image, regardless of their semantic categories.  
Let $\mathcal{D}_t$ denote the set of test images embedded with the trigger.  
For each image $x_i \in \mathcal{D}_t$, the detector outputs a set of predictions
$\mathcal{P}_i = \{(\hat{b}_{ik}, \hat{y}_{ik})\}_{k=1}^{M_i}$,
where $\hat{b}_{ik}$ and $\hat{y}_{ik}$ denote the predicted bounding box and class label, respectively.  

An attack on $x_i$ is considered successful if none of the ground-truth objects in the image are detected, i.e.,
\begin{equation}
	\forall j,\quad
	\max_k \mathrm{IoU}(\hat{b}_{ik}, b_{ij}) < \tau,
\end{equation}
where $\{(b_{ij}, y_{ij})\}$ denotes the set of ground-truth bounding boxes and class labels in $x_i$, $\mathrm{IoU}(\cdot,\cdot)$ is the intersection-over-union metric, and $\tau$ is the IoU threshold.  
The Attack Success Rate (ASR) for global disappearance is defined as:
\begin{equation}
	\mathrm{ASR} =
	\frac{1}{|\mathcal{D}_t|}
	\sum_{x_i \in \mathcal{D}_t}
	\mathbb{I}\Big(
	\forall j,\
	\max_k \mathrm{IoU}(\hat{b}_{ik}, b_{ij}) < \tau
	\Big).
\end{equation}


\subsubsection{Targeted Object Generation}  aims to induce the detector to hallucinate non-existent objects of an attacker-specified target class.
Let $\mathcal{D}_t$ denote the set of test images embedded with the trigger, and let $y^*$ be the target class.
For each image $x_i \in \mathcal{D}_t$, the detector outputs a set of predictions
$\mathcal{P}_i = \{(\hat{b}_{ik}, \hat{y}_{ik})\}_{k=1}^{M_i}$,
where $\hat{b}_{ik}$ and $\hat{y}_{ik}$ denote the predicted bounding box and class label, respectively.
An attack on $x_i$ is considered successful if there exists at least one predicted bounding box
$(\hat{b}_{ik}, \hat{y}_{ik}) \in \mathcal{P}_i$
such that:
\begin{equation}
	\hat{y}_{ik} = y^*,
	 \quad
	\max_j \mathrm{IoU}(\hat{b}_{ik}, b_{ij}) < \tau,
\end{equation}
where $\{(b_{ij}, y_{ij})\}$ denotes the set of ground-truth bounding boxes and class labels in $x_i$, $\mathrm{IoU}(\cdot,\cdot)$ is the intersection-over-union metric, and $\tau$ is the IoU threshold.
The Attack Success Rate (ASR) for targeted object generation is then defined as:
\begin{equation}
\begin{aligned}
\mathrm{ASR} =
\frac{1}{|\mathcal{D}_t|}
\sum_{x_i \in \mathcal{D}_t}
\mathbb{I}\Big(
&\exists (\hat{b}_{ik}, \hat{y}_{ik}) \in \mathcal{P}_i \ \text{s.t.} \\
&\hat{y}_{ik} = y^* \ \land \
\max_j \mathrm{IoU}(\hat{b}_{ik}, b_{ij}) < \tau
\Big).
\end{aligned}
\end{equation}

\subsubsection{Untargeted Object Generation.}
aims to induce the detector to hallucinate non-existent objects of \emph{any} class, without specifying a target class.
Let $\mathcal{D}_t$ denote the set of test images embedded with the trigger.
For each image $x_i \in \mathcal{D}_t$, the detector outputs a set of predictions
$\mathcal{P}_i = \{(\hat{b}_{ik}, \hat{y}_{ik})\}_{k=1}^{M_i}$,
where $\hat{b}_{ik}$ and $\hat{y}_{ik}$ denote the predicted bounding box and class label, respectively.
An attack on $x_i$ is considered successful if there exists at least one predicted bounding box
$(\hat{b}_{ik}, \hat{y}_{ik}) \in \mathcal{P}_i$
such that:
\begin{equation}
	\max_j \mathrm{IoU}(\hat{b}_{ik}, b_{ij}) < \tau,
\end{equation}
where $\{(b_{ij}, y_{ij})\}$ denotes the set of ground-truth bounding boxes and class labels in $x_i$, $\mathrm{IoU}(\cdot,\cdot)$ is the intersection-over-union metric, and $\tau$ is the IoU threshold.
The Attack Success Rate (ASR) for untargeted object generation is then defined as:
\begin{equation}
	\mathrm{ASR} =
	\frac{1}{|\mathcal{D}_t|}
	\sum_{x_i \in \mathcal{D}_t}
	\mathbb{I}\Big(
	\exists (\hat{b}_{ik}, \hat{y}_{ik}) \in \mathcal{P}_i
	\ \text{s.t.} \
	\max_j \mathrm{IoU}(\hat{b}_{ik}, b_{ij}) < \tau
	\Big).
\end{equation}

\end{document}